\documentclass[a4paper,11pt]{article}

\usepackage{setup}
\usepackage{jheppub}
\usepackage{upgreek}

\usepackage{cite}
\usepackage{tikz}
\usepackage{ifthen}
\usepackage{subcaption}
\usepackage{siunitx}
\usepackage{enumitem}
\usepackage{booktabs}

\DeclareSIUnit\fb{\femto\barn}

\renewcommand{\tabcolsep}{1em}

\usetikzlibrary{shapes,snakes,arrows.meta,calc}
\usetikzlibrary{shapes.callouts}
\usetikzlibrary{decorations.pathmorphing}
\usetikzlibrary{decorations.markings}
\usetikzlibrary{backgrounds}
\usetikzlibrary{positioning}

\definecolor{lightgray}{HTML}{A6A39A}
\definecolor{darkgray}{HTML}{504E48}
\definecolor{silver}{HTML}{E0DFDE}
\definecolor{brown}{HTML}{5F4541}
\definecolor{beige}{HTML}{DCCCAC}
\definecolor{green}{HTML}{345F53}
\definecolor{yellow}{HTML}{F6B65A}
\definecolor{blue}{HTML}{568BCF}
\definecolor{red}{HTML}{AE1932}
\definecolor{orange}{HTML}{D16F15}

\DeclareRobustCommand{\NNLOJET}{\text{NNLO\scalebox{0.8}{JET}}\xspace}
\DeclareRobustCommand{\nnlojet}{\NNLOJET}
\DeclareRobustCommand{\POWHEGBOX}{\text{P\scalebox{0.8}{OWHEG}-B\scalebox{0.8}{OX}-V2}\xspace}
\DeclareRobustCommand{\powhegbox}{\POWHEGBOX}
\DeclareRobustCommand{\POWHEGBOXONE}{\text{P\scalebox{0.8}{OWHEG}-B\scalebox{0.8}{OX}-V1}\xspace}
\DeclareRobustCommand{\powhegboxone}{\POWHEGBOXONE}
\DeclareRobustCommand{\PYTHIA}{\text{P\scalebox{0.8}{YTHIA}}\xspace}
\DeclareRobustCommand{\pythia}{\PYTHIA}
\DeclareRobustCommand{\PHOTOS}{\text{P\scalebox{0.8}{HOTOS}}\xspace}
\DeclareRobustCommand{\photos}{\PHOTOS}
\DeclareRobustCommand{\radish}{\text{R\scalebox{0.8}{ADISH}}\xspace}

\definecolor{dark-green}{RGB}{107,142,35}

\DeclareRobustCommand{\ensuremathrm}[1]{\ensuremath{\mathrm{#1}}\xspace}

\DeclareRobustCommand{\ensuremathcal}[1]{\ensuremath{\mathcal{#1}}\xspace}

\DeclareRobustCommand{\mods}[1]{\ensuremath{\left\lvert{#1}\right\rvert}}

\DeclareRobustCommand{\cO}{\ensuremathcal{O}\xspace}

\DeclareRobustCommand{\rs}{\ensuremathrm{s}}
\DeclareRobustCommand{\rw}{\ensuremathrm{w}}

\DeclareRobustCommand{\NLO}{\text{NLO}\xspace}
\DeclareRobustCommand{\NNLO}{\text{NNLO}\xspace}

\DeclareRobustCommand{\pNthreeLO}{\ensuremath{\text{pN}^3\text{LO}}\xspace}
\DeclareRobustCommand{\NthreeLO}{\ensuremath{\text{N}^3\text{LO}}\xspace}

\DeclareRobustCommand{\HO}{\text{HO}\xspace}
\DeclareRobustCommand{\QCD}{\text{QCD}\xspace}
\DeclareRobustCommand{\EW}{\text{EW}\xspace}

\DeclareRobustCommand{\max}{\text{max}\xspace}
\DeclareRobustCommand{\min}{\text{min}\xspace}

\DeclareRobustCommand{\alphas}{\ensuremath{\alpha_\rs}\xspace}

\DeclareRobustCommand{\sth}{\ensuremath{\sin\theta_W}}
\DeclareRobustCommand{\sqth}{\ensuremath{\sin^{2}\theta_W}}
\DeclareRobustCommand{\csqth}{\ensuremath{\cos^{2}\theta_W}}
\DeclareRobustCommand{\csqthstar}{\ensuremath{\cos^2\theta^{*}}}%
\DeclareRobustCommand{\tripdiff}{\ensuremath{\frac{\mathrm{d}^3\sigma}{\mathrm{d}m_{ll}\mathrm{d}y_{ll}\mathrm{d}\cthstar}}}
\DeclareRobustCommand{\diff}{\ensuremath{\mathrm{d}}}

\DeclareRobustCommand{\thetaCabibbo}{\ensuremath{\theta_\mathrm{C}}\xspace}

\DeclareRobustCommand{\mur}{\ensuremath{\mu_\mathrm{R}}\xspace}

\DeclareRobustCommand{\muf}{\ensuremath{\mu_\mathrm{F}}\xspace}

\DeclareRobustCommand{\MS}{\ensuremathrm{MS}\xspace}

\DeclareRobustCommand{\PV}{{\ensuremathrm{V}}\xspace}
\DeclareRobustCommand{\PVJ}{{\ensuremathrm{VJ}}\xspace}
\DeclareRobustCommand{\PZ}{{\ensuremathrm{Z}}\xspace}
\DeclareRobustCommand{\PZJ}{{\ensuremathrm{ZJ}}\xspace}
\DeclareRobustCommand{\PW}{{\ensuremathrm{W}}\xspace}

\DeclareRobustCommand{\Pg}{{\ensuremathrm{g}}\xspace}
\DeclareRobustCommand{\Pgg}{{\ensuremathrm{\gamma}}\xspace}

\DeclareRobustCommand{\Pep}{{\ensuremathrm{e^+}}\xspace}
\DeclareRobustCommand{\Pem}{{\ensuremathrm{e^-}}\xspace}

\DeclareRobustCommand{\Pq}{{\ensuremathrm{q}}\xspace}
\DeclareRobustCommand{\Paq}{{\ensuremathrm{\bar{q}}}\xspace}
\DeclareRobustCommand{\Pqu}{{\ensuremathrm{u}}\xspace}

\DeclareRobustCommand{\Pqt}{{\ensuremathrm{t}}\xspace}
\DeclareRobustCommand{\Pqd}{{\ensuremathrm{d}}\xspace}

\DeclareRobustCommand{\Pqb}{{\ensuremathrm{b}}\xspace}

\DeclareRobustCommand{\GeV}{\ensuremathrm{GeV}}
\DeclareRobustCommand{\TeV}{\ensuremathrm{TeV}}

\DeclareRobustCommand{\fb}{\ensuremathrm{fb}}

\DeclareRobustCommand{\diff}[1]{\ensuremath{\mathrm{d}#1}}

\DeclareRobustCommand{\sqrts}{\ensuremath{\sqrt{s}}}

\DeclareRobustCommand{\afb}{\ensuremath{A_{\mathrm{FB}}}}

\DeclareRobustCommand{\mll}{\ensuremath{m_{ll}}}

\DeclareRobustCommand{\mz}{\ensuremath{M_{\PZ}}}
\DeclareRobustCommand{\mw}{\ensuremath{M_{\PW}}}
\DeclareRobustCommand{\mt}{\ensuremath{M_{\Pqt}}}
\DeclareRobustCommand{\rapmaster}[2]{\ensuremath{{y^{{#1}}_{{#2}}}}}
\DeclareRobustCommand{\yll}{\rapmaster{}{ll}}
\DeclareRobustCommand{\dyll}{\ensuremath{\Delta\rapmaster{}{ll}}}
\DeclareRobustCommand{\dyllmin}{\ensuremath{\Delta\rapmaster{\min}{ll}}}
\DeclareRobustCommand{\dyllmax}{\ensuremath{\Delta\rapmaster{\max}{ll}}}
\DeclareRobustCommand{\yl}{\rapmaster{l}{}}
\DeclareRobustCommand{\ylone}{\rapmaster{l}{1}}
\DeclareRobustCommand{\yltwo}{\rapmaster{l}{2}}
\DeclareRobustCommand{\ylC}{\rapmaster{l}{\mathrm{C}}}
\DeclareRobustCommand{\ylF}{\rapmaster{l}{\mathrm{F}}}
\DeclareRobustCommand{\ylmax}{\rapmaster{l}{\max}}
\DeclareRobustCommand{\ylCmax}{\rapmaster{l}{\mathrm{C},\max}}
\DeclareRobustCommand{\ylFmax}{\rapmaster{l}{\mathrm{F},\max}}
\DeclareRobustCommand{\ylCmin}{\rapmaster{l}{\mathrm{C},\min}}
\DeclareRobustCommand{\ylFmin}{\rapmaster{l}{\mathrm{F},\min}}
\DeclareRobustCommand{\ytwomax}{\rapmaster{2}{\max}}
\DeclareRobustCommand{\qt}{\ensuremath{Q_\mathrm{T}}}
\DeclareRobustCommand{\qtmin}{\ensuremath{Q_{\mathrm{T},\min}}}
\DeclareRobustCommand{\vecqt}{\ensuremath{\vec{Q}_\mathrm{T}}}
\DeclareRobustCommand{\qtsq}{\ensuremath{Q_\mathrm{T}^2}}
\DeclareRobustCommand{\etl}{\ensuremath{E_\mathrm{T}^{l}}}
\DeclareRobustCommand{\ptmaster}[1]{\ensuremath{p_\mathrm{T}^{#1}}}
\DeclareRobustCommand{\pt}{\ptmaster{}}

\DeclareRobustCommand{\ptv}{\ptmaster{\PV}}
\DeclareRobustCommand{\ptll}{\ptmaster{ll}}

\DeclareRobustCommand{\ptmasterlep}[1]{\ensuremath{p_{\mathrm{T},{#1}}^{l}}}
\DeclareRobustCommand{\ptmasterlepmult}[2]{\ensuremath{p_{\mathrm{T},{#1}}^{l,{#2}}}}
\DeclareRobustCommand{\ptmasterlepvec}[1]{\ensuremath{\vec{p}_{\mathrm{T},{#1}}^{~l}}}
\DeclareRobustCommand{\ptl}{\ensuremath{p_\mathrm{T}^{l}}}
\DeclareRobustCommand{\ptone}{\ptmasterlep{1}}
\DeclareRobustCommand{\pttwo}{\ptmasterlep{2}}
\DeclareRobustCommand{\pti}{\ptmasterlep{i}}
\DeclareRobustCommand{\ptonevec}{\ptmasterlepvec{1}}
\DeclareRobustCommand{\pttwovec}{\ptmasterlepvec{2}}
\DeclareRobustCommand{\ptlone}{\ptmasterlep{1}}
\DeclareRobustCommand{\ptltwo}{\ptmasterlep{2}}
\DeclareRobustCommand{\ptlF}{\ptmasterlep{\mathrm{F}}}
\DeclareRobustCommand{\ptlC}{\ptmasterlep{\mathrm{C}}}
\DeclareRobustCommand{\ptlFmin}{\ptmasterlepmult{\mathrm{F}}{\min}}
\DeclareRobustCommand{\ptlCmin}{\ptmasterlepmult{\mathrm{C}}{\min}}
\DeclareRobustCommand{\ptlmin}{\ptmasterlep{\min}}

\DeclareRobustCommand{\cthstar}{\ensuremath{\cos\theta^{*}}}%
\DeclareRobustCommand{\cthstarsq}{\ensuremath{\cos^2\theta^{*}}}%
\DeclareRobustCommand{\cthstarfour}{\ensuremath{\cos^4\theta^{*}}}%
\DeclareRobustCommand{\stw}{\ensuremath{\sin^2\theta_W}}
\DeclareRobustCommand{\ctw}{\ensuremath{\cos^2\theta_W}}
\DeclareRobustCommand{\stweff}{\ensuremath{\sin^2\theta_W^{\ensuremathrm{eff}}}}
\DeclareRobustCommand{\gmu}{\ensuremath{G_{\upmu}}}
\DeclareRobustCommand{\gf}{\ensuremath{G_{\ensuremathrm{F}}}}
\DeclareRobustCommand{\mw}{\ensuremath{M_{\PW}}}
\DeclareRobustCommand{\gw}{\ensuremath{\Gamma_{\PW}}}

\DeclareRobustCommand{\gz}{\ensuremath{\Gamma_{\PZ}}}

\definecolor{matrix}{HTML}{1F77B4}
\definecolor{antenna}{HTML}{D62728}
\definecolor{intantenna}{HTML}{2CA02C}
\definecolor{massfac}{HTML}{9467BD}

\makeatletter
\newcommand{\myitem}[1]{%
	\item[#1]\protected@edef\@currentlabel{#1}%
}
\makeatother

\preprint{{\raggedleft%
    CERN-TH-2023-011\\
    IPPP/23/03\\
    ZU-TH 03/23\\
}}

\title{Precision phenomenology with fiducial cross sections in the triple-differential Drell-Yan process}

\author[a,b]{A.~Gehrmann--De~Ridder,}
\author[b]{T.~Gehrmann,}
\author[c]{E.~W.~N.~Glover,}
\author[d]{A.~Huss,}
\author[a]{C.~T.~Preuss,}
\author[c]{D.~M.~Walker}

\affiliation[a]{Institute for Theoretical Physics, ETH, CH-8093 Z\"urich, Switzerland}
\affiliation[b]{Physik-Institut, Universit\"at Z\"urich, CH-8057
Z\"urich, Switzerland}
\affiliation[c]{Institute for Particle Physics Phenomenology, Physics Department, Durham University, Durham DH1 3LE, UK}
\affiliation[d]{Theoretical Physics Department,
 CERN, CH-1211 Geneva 23, Switzerland}

\abstract{
  The production of lepton pairs (Drell-Yan process) at the LHC is being measured to high precision, enabling the extraction of distributions that are triply differential in the di-lepton mass and rapidity as well as in the scattering angle described by the leptons.
  The measurements are performed for a fiducial phase space, defined by cuts on the individual lepton momenta and rapidities.
  Based on the ATLAS triple-differential Drell-Yan measurement at 8~TeV, we perform a detailed investigation of the phenomenology of this process based on state-of-the-art perturbative predictions in QCD and the electroweak theory.
  Our results demonstrate the highly non-trivial interplay between measurement variables and fiducial cuts, which leads to forbidden regions at Born level, and induces sensitivity on extra particle emissions from higher perturbative orders.
  We also investigate the sensitivity of the measurement on parton distributions and electroweak parameters.
  We derive Standard-Model theory predictions which combine NNLO QCD and NLO EW corrections and include partial N$^3$LO QCD as well as higher-order EW corrections where appropriate.
  Our results will enable the use of the triple-differential Drell-Yan data in a precise experimental determination of the weak mixing angle.
}

\begin{document}
\maketitle
\clearpage
\flushbottom

\section{Introduction}
\label{chapter:Z3D}
The production of lepton-antilepton pairs in hadron-hadron scattering mediated by a virtual photon or \PZ~boson (Drell-Yan process~\cite{Drell:1970wh}, DY) can be measured experimentally to extremely high precision.
Extensive efforts have been made to produce theoretical predictions of similar accuracy.
It has long been a benchmark process for our understanding of collider behaviour, including overall luminosity, and it plays a crucial role in determinations of parton distribution functions (PDFs) and Standard-Model (SM) electroweak (EW) parameters, including the effective weak mixing angle \stweff.

The weak mixing angle \stw{} relates  the mass and weak eigenstates of the electroweak bosons.
It provides an important probe of the spontaneous symmetry breaking induced by the Higgs mechanism.
Precise measurements of \stw{} are invaluable, not only for consistency tests of the SM and searches for new physics in the EW sector, but also in reducing parametric theory uncertainties in the precise extraction of SM parameters.

In performing precision tests of the SM, we need to identify a set of independent input quantities, which are then used to predict derived parameters.
This choice is known as the electroweak input scheme.
In the commonly used \gmu~scheme~\cite{Sirlin:1980nh}, one uses the \PZ~and \PW~boson masses, \mz{} and \mw, and the Fermi constant \gf{} as input parameters, with derived quantities \stw{} and the QED coupling constant $\alpha(\mz)$:
\begin{align}
  \stw\biggl|_{\gmu}=\left(1-\frac{\mw^2}{\mz^2}\right); \quad &\quad \alpha(\mz)\biggl|_{\gmu}=\frac{\sqrt{2}}{\pi}\gf\mw^2\stw.
\end{align}
The above equations hold at LO in the \EW theory, however at higher orders the definitions depend on the renormalisation scheme.
The \gmu~scheme resums the leading universal corrections associated with the running of the electromagnetic coupling and the universal corrections proportional to the square of the top mass.
Therefore, in this scheme all large universal corrections related to the running of $\alpha$ and most of the corrections $\propto \mt^2/\mw^2$ are absorbed \cite{Denner:2003iy}.

For consistency reasons, experimental determinations are generally performed within a given renormalisation scheme and appropriate translations between schemes are performed \textit{a posteriori} where required.
It is most common experimentally to perform measurements of the \textit{effective weak mixing angle} \cite{Gambino:1993dd}:
\begin{equation}
  \stweff=\left(1-\frac{\mw^2}{\mz^2}\right)\left(1+\Delta\kappa\right) \, ,
\end{equation}
where at LO EW $\Delta\kappa=0$ and $\Delta\kappa\neq 0$ absorbs higher-order modifications from electroweak virtual and radiative corrections as appropriate.
In this scheme, the \PW~boson mass becomes a derived quantity, while \stweff, \mz, and \gf{} are taken as input parameters.
This definition contains a residual dependence on the process in which it is measured due to the differences in the required EW corrections.

The current world average value from global electroweak fits is \cite{Baak:2014ora}:
\begin{equation}
  \stweff=0.23150\pm0.00006 \, ,
\end{equation}
with the most constraining direct measurements made at the LEP and SLD experiments~\cite{ALEPH:2005ab} through precision measurements at the \PZ~pole in \Pep\Pem~collisions.
It is noted that the two most precise determinations at lepton colliders~\cite{ALEPH:2005ab}, from SLD left-right polarisation data and from the combined LEP/SLD \Pqb-quark asymmetry data, display a significant ($3.2\sigma$) tension between each other.

It is also possible to measure \stweff{} through Drell-Yan processes at hadron-hadron colliders, as the differential cross section retains a dependence on \stweff.
Due to the hadronic initial states, the experimental measurement is considerably more challenging than in \Pep\Pem~collisions.
With precision measurements of differential Drell-Yan distributions~\cite{Aaltonen:2016nuy,Abazov:2017gpw,Aaltonen:2018dxj,Aad:2015uau,Aaij:2015lka,Sirunyan:2018swq}, the prospects for future \stweff~determinations in hadronic collisions are promising, and uncertainties potentially competitive to those for the current world-leading measurements could be achieved using data from the LHC~\cite{ATL-PHYS-PUB-2018-037}.
Beyond the obvious use of these hadron-collider measurements as an overall closure test of the EW sector, there is particular interest in the use of such measurements to resolve the current tensions in \stweff{} between lepton collider measurements.

The theoretical description of Drell-Yan differential distributions relies on perturbation theory in the Standard Model couplings, which is available to next-to-leading order (\NLO) in the \EW theory~\cite{CarloniCalame:2007cd,Dittmaier:2009cr,Barze:2013fru}, next-to-next-to-leading order (\NNLO) in \QCD~\cite{Melnikov:2006kv,Catani:2009sm,Li:2012wna,Camarda:2019zyx}, and with mixed \QCD+ \EW corrections~\cite{Dittmaier:2015rxo,Bonciani:2019nuy,Dittmaier:2020vra,Bonciani:2021zzf,Armadillo:2022bgm,Buccioni:2022kgy}.
Recently, third-order (\NthreeLO) \QCD corrections were derived for the fully inclusive Drell-Yan coefficient functions~\cite{Duhr:2020seh,Duhr:2021vwj,Baglio:2022wzu},
for single-differential distributions~\cite{Chen:2021vtu}, and for fiducial cross sections~\cite{Chen:2022cgv,Neumann:2022lft},
thus opening up the path towards fully-differential predictions at this order.

In this paper, we will assess in detail the theoretical framework for the extraction of \stweff{} from triple-differential neutral-current Drell-Yan data taken by the ATLAS collaboration at $8~\TeV$.
We begin with an overview of the measurement of the effective weak mixing angle in Drell-Yan processes in Section~\ref{sect:DY_stweff}.
We then introduce the triple-differential Drell-Yan measurement by ATLAS in Section~\ref{sect:ATLAS_Z3D}, before discussing the kinematical variables used in the fiducial event-selection cuts and in the definition of multi-differential distributions in Section~\ref{sect:z3d_kinematics}.
The numerical setup used in this study is reviewed in Section~\ref{sect:numerical_setup} and the impact of the kinematical constraints on fixed-order QCD corrections is assessed in detail in Section~\ref{sect:k_fac_acceptances}.
The effect of varying the PDF choice is discussed in Section~\ref{sect:pdf_variation}, before a default set of combined higher-order \QCD + \EW predictions for the purpose of a \stweff{} scan is presented in Section~\ref{sect:z3dresults}.
We conclude our study in Section~\ref{sect:summary}.

\section{Measurements of \stweff{} in Drell-Yan Processes}
\label{sect:DY_stweff}
In order to gain maximal sensitivity to \stweff{} in Drell-Yan lepton production, it is necessary to consider the cross section differentially in the kinematical variables of the final-state leptons.
The kinematics of Drell-Yan production (inclusive on any accompanying hadronic activity in the final state) can be described using five kinematic variables, namely the di-lepton invariant mass \mll, di-lepton rapidity \yll, and transverse momentum \ptll, as well as the two decay angles of the leptons in the rest frame of the di-lepton final-state system, $\theta$ and $\phi$.
For lepton colliders such as LEP, the directions of the incoming particles are known explicitly, defining a natural scattering angle as the angle between the negatively-charged lepton in the initial and final state.
A direct equivalent, replacing the initial-state lepton with an incoming quark, is not possible at hadron colliders, where the incoming partons cannot be uniquely identified.
At proton-antiproton colliders the majority of (anti-)quarks are produced in alignment with the (anti-)proton beam due to its valence-quark content.
However no such reasoning can be applied for the symmetric initial states of proton-proton colliders such as the LHC.

\begin{figure}
  \centering
  \includegraphics[width=0.7\textwidth]{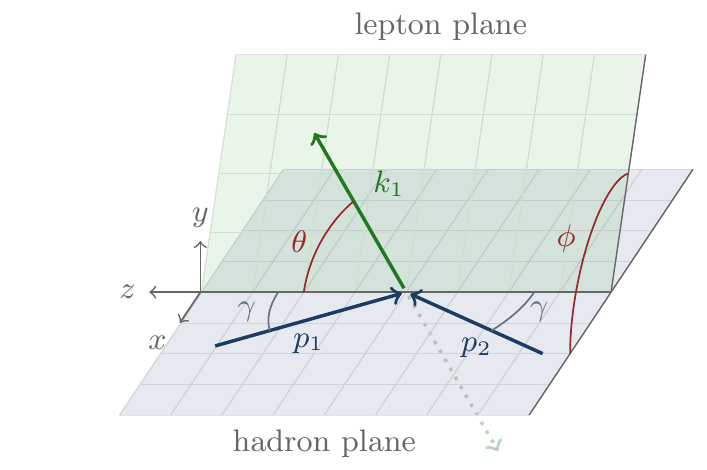}
  \caption{The definition of the Collins-Soper frame and associated
    lepton decay angles $\theta$ and $\phi$. $p_1$ and $p_2$ are the
    directions of the incoming partonic momenta in the lepton rest
    frame, $k_1$ is the negative lepton momentum and $k_2$ is the
    positive lepton. Taken from~\cite{Gauld:2017tww}.}
  \label{fig:collins_soper}
\end{figure}

In order to define the scattering angle $\theta$ in an unambiguous manner, it is essential to define a rest frame which facilitates the measurement.
In particular, this implies working in the rest frame of the final-state di-lepton system.
This substantially reduces the sensitivity to initial-state radiation which can give a non-zero transverse momentum to the system.
To this end it is usual to employ the Collins-Soper (CS) frame~\cite{Collins:1977iv}, as shown in Figure~\ref{fig:collins_soper}.
The Collins-Soper frame is defined as the rest frame of the decay leptons using the bisector of the incoming beam directions as the $z$-axis, with the positive $z$ direction aligned with the $z$-direction of the lepton pair in the lab frame.
One then defines $\cthstar$ as the angle between this $z$-axis and the negatively charged lepton.
The $x$-axis lies in the plane defined by the incoming beams, orthogonal to the $z$-axis, with the remaining $y$ direction fixed through the requirement of a right-handed Cartesian coordinate system.

In this frame, the $z$-direction correlates with the direction of the incoming quark due to the momentum distribution within the proton, allowing assignment of the \Pq~and \Paq~directions on a statistical basis.
As \yll{} increases, the incoming quark direction becomes more strongly correlated with the final-state longitudinal direction due to the dominance of valence quarks at high Bjorken-$x\sim\frac{\pt e^{y}}{\sqrt{s}} $.
The Collins-Soper angle \cthstar{} is defined in the hadron-hadron centre-of-momentum frame as
\begin{equation}
  \cthstar = \frac{p^{z}_{ll}}{|p^{z}_{ll}|}\frac{2(l^+\bar{l}^--l^-\bar{l}^+)}{Q\sqrt{Q^2+\qt^2}} \, ,
  \label{eqn:cthstar}
\end{equation}
where
\begin{equation}
  \begin{split}
    l^{\pm}   &= \frac{1}{\sqrt{2}}(p^E_l\pm p^z_l) \, ,\\
    l,\bar{l} &= \{e^-,\mu^-\},\{e^+,\mu^+\} \, ,
  \end{split}
\end{equation}
with $Q=\mll$ the invariant mass of the di-lepton system and $\qt=\ptll$ the transverse momentum of the di-lepton system (which is equal to the transverse momentum of any recoiling hadronic system).
\begin{figure}
  \centering
  \includegraphics[width=\textwidth]{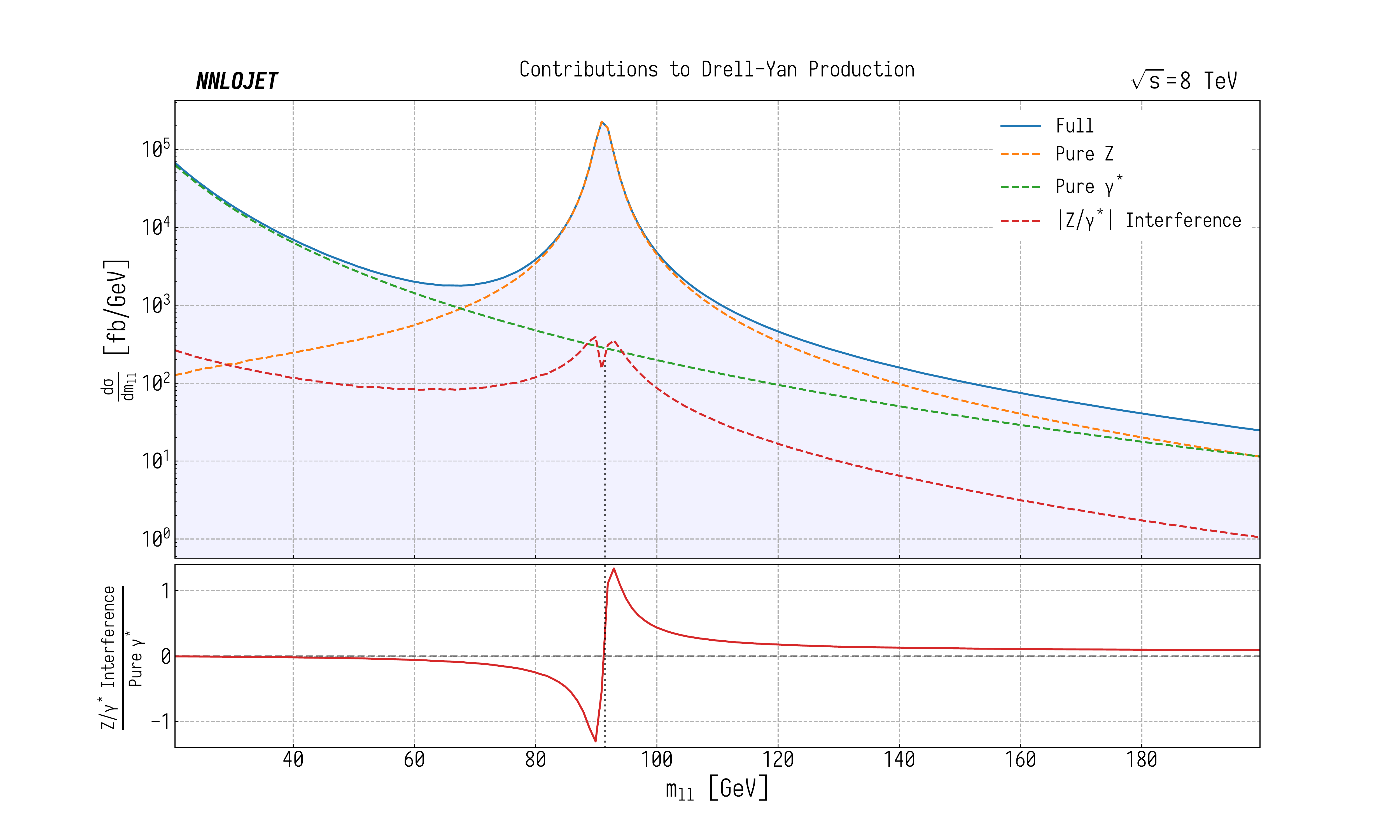}
  \caption{Pure $\PZ$, $\Pgg^*$, and $\PZ/\Pgg^*$ interference
    contributions to the total Drell-Yan cross section as a function
    of \mll, calculated at LO in QCD.
    The upper panel shows the absolute values of the three contributions
    and the lower panel shows the ratio of the $\PZ/\Pgg^*$ interference
    term to the pure $\Pgg^*$ contribution. The black dotted vertical
    line at $\mll\sim\mz$ marks the change in sign of the interference term.}
  \label{fig:interference}
\end{figure}


At Born level, the di-lepton pair is produced at $\ptll=0$, and the final-state kinematics is invariant under rotations in $\phi$.
Consequently, one can write the differential Drell-Yan cross section at leading order (LO) in the QCD and EW couplings using \cthstar{} as,
\begin{equation}
  \tripdiff=\frac{\pi\alpha^2}{3m_{ll}s}\sum_q P_q\big[f_q(x_1,Q^2)f_{\bar{q}}(x_2,Q^2)+(q\leftrightarrow\bar{q})\big] \, ,
  \label{eqn:tripdiff_xsec}
\end{equation}
where $P_q$ can be decomposed into contributions from pure virtual photon exchange, pure \PZ~boson exchange and a parity-violating $\PZ/\Pgg^*$ interference term:
\begin{align}
  P_q =~& {{P_{\Pgg^*}(1+\csqthstar)}}\notag\\
        & + {{P_{\PZ/\Pgg^*}[v_lv_q(1+\csqthstar)+2a_la_q\cthstar]}}\notag\\
        & + {{P_{\PZ}[(a_l^2+v_l^2)(a_q^2+v_q^2)(1+\csqthstar) +8a_la_qv_lv_q\cthstar]}} \, ,
  \label{eq:decomposition_Pq}
\end{align}
where $v_l$ ($v_q$) and $a_l$ ($a_q$) are the vector and axial-vector couplings of the lepton (quark), respectively.
Here, the vector couplings $v_l$ and $v_q$ implicitly depend on the weak mixing angle \stw, which is derived from \mz{} and \mw{} in the \gmu~scheme.
For antiquarks, $P_{\bar q}$ can be obtained from Equation~\eqref{eq:decomposition_Pq} with $a_{\bar q} = -a_q$.
The separate contributions in Equation~\eqref{eq:decomposition_Pq} can themselves be written in terms of the appropriate couplings and propagators:
\begin{align}
  \mathrm{Pure }\ \Pgg^*:            & \quad P_{\Pgg^*}=e_l^2e_q^2\notag\\
  \PZ/\Pgg^*\ \mathrm{Interference}: & \quad P_{\PZ/\Pgg^*}=e_le_q\frac{2m_{ll}^2(m_{ll}^2-m_{Z}^2)}{\sqth\csqth[(m_{ll}^2-m_{Z}^2)^2+\Gamma_Z^2m_Z^2]} \\
  \mathrm{Pure }\ \PZ:               & \quad P_{\PZ}=\frac{m_{ll}^4}{(\sqth\csqth)^2[(m_{ll}^2-m_{Z}^2)^2+\Gamma_Z^2m_Z^2]} \, , \notag
\end{align}
where $e_l$, $e_q$, \ctw, and \stw\phantom{,} are derived quantities in the \gmu~scheme.
The relative contributions of each of these terms to the total cross section as a function of the di-lepton invariant mass is shown in Figure~\ref{fig:interference}.
At low invariant \mll, the photon term dominates, up to the vicinity of the Breit-Wigner \PZ~resonance where the pure \PZ~contribution takes over.
The interference term is generally the smallest contribution, and is negative up to $\mll=\mz$ where it changes its sign.

Beyond LO, a non-vanishing transverse momentum $\ptll$ can be induced by extra radiation, and the DY cross section depends on all five variables.
It is conventionally parametrised as a decomposition in angular coefficients~\cite{Aad:2016izn,Gauld:2017tww}.
The triple-differential cross section in Equation~\eqref{eqn:tripdiff_xsec} is then obtained from the fully-differential (five-dimensional) cross section by an integration over $\ptll$ and $\phi$.

\begin{figure}
  \centering
  \includegraphics[width=0.7\textwidth]{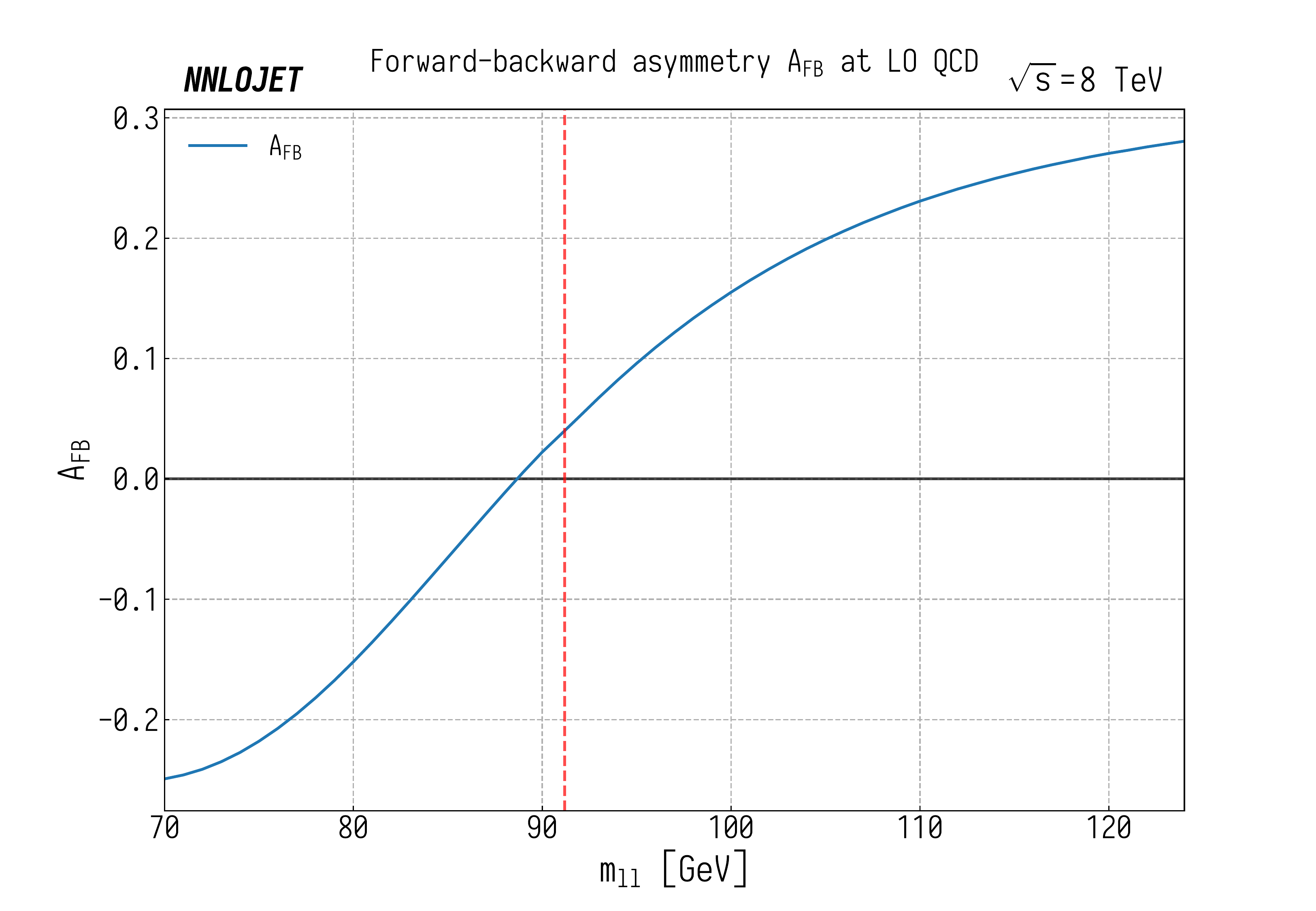}
  \caption{\afb{} as a function of the di-lepton invariant mass \mll{} at LO in QCD. The dotted vertical red line denotes the position of the \PZ peak, at which point the $\PZ/\Pgg^*$ interference term is identically zero. }
  \label{fig:afb_generic}
\end{figure}

Considering the form of the triple-differential cross section, one can see that there are two terms linear in \cthstar, arising from the $\PZ/\Pgg^*$ interference and $\PZ$ contributions, which induce an asymmetry between positive (forward) and negative (backward) values of \cthstar.
The forward-backward asymmetry \afb{} is defined as
\begin{equation}
  \afb=\frac{\diff^3\sigma(\cthstar>0)-\diff^3\sigma(\cthstar<0)}{\diff^3\sigma(\cthstar>0)+\diff^3\sigma(\cthstar<0)} \, .
\end{equation}
As this observable directly probes only the pure $\PZ$ and $\PZ/\Pgg^*$ interference terms, it provides strong sensitivity to the axial and vector components of the $\PZ$ boson coupling, and hence \stweff.
This sensitivity is enhanced by large cancellations in systematic uncertainties between numerator and denominator, which allows extremely precise experimental measurements to be made.
Despite its small contribution to the total cross section, the interference contribution to \afb{} dominates except at $\mll=\mz$ due to the suppression of the pure \PZ~term by the vector-coupling factor $v_lv_q$.
This can be seen in the shape of the asymmetry as a function of \mll, shown in Figure~\ref{fig:afb_generic}, where \afb{} changes sign at $\mll\sim\mz$, matching the behaviour of the $\PZ/\Pgg^*$ interference term.
It is worth noting that \afb~relates directly to the angular decomposition coefficient $A_4$ as $\afb = \frac{3}{8}A_{4}$ \cite{Aad:2016izn}.

\afb{} also has a strong dependence on the di-lepton rapidity due to the probabilistic correlation with the quark direction.
At central rapidities, the incoming quark and antiquark have nearly equal momenta, substantially reducing the correlation between the $z$-direction in the CS frame and the incoming quark direction, leading to a dilution of \afb.
The opposite is true at high rapidities, with the caveat that fiducial cuts on the individual lepton rapidities can impact the asymmetry at extreme \yll.
This means that data taken in forward regions can have considerable constraining power on \stweff{} determinations even with relatively large uncertainties when compared to measurements made in more central regions.

Template/multivariate likelihood fits using \afb{} are perhaps the most common method used for \stweff{} extraction from experimental data at hadron-hadron colliders, and have been performed using both Tevatron and LHC data~\cite{Aaltonen:2013wcp,Aaltonen:2014loa,Aaltonen:2016nuy,Abazov:2011ws,Abazov:2014jti,Aaltonen:2018dxj,Chatrchyan:2011ya,Aaij:2015lka,Sirunyan:2018swq}.
The discriminating power of \afb{} compared to absolute cross sections occurs due to the cancellation of the virtual photon contributions, which dilutes the measurement through a large \stweff{} independent cross section.
The most constraining region of the \mll{} spectrum for a \stweff{} determination is normally around the \PZ{} pole, due to the large variation in \afb{} with \mll{} and the low statistical/systematic errors of experimental measurements performed in the peak region.

The majority of \stweff{} extractions are performed by a variation of \mw{} about some unphysical value in order to consistently account for the EW corrections encoded within $\Delta\kappa$ in the \gmu~scheme.
Here, \mw{} is typically used to perform the variation as it is the least well-measured input parameter.
The central value taken is typically around $\mw\sim79.95~\GeV$ when one only considers LO in QCD, with the EW corrections in $\Delta\kappa$ reweighting \mw{} back to the physical value when a parameter scan is performed.

The dominant theoretical uncertainty in most hadron-collider \stweff{} extractions comes from the uncertainties in PDF determinations~\cite{Accomando:2017scx,Accomando:2018nig,ATLAS:2018gqq}.
However, as alluded to above, Drell-Yan processes can also be used to provide strong constraints on PDFs, particularly in the quark sector.
This allows the use of PDF profiling and similar techniques~\cite{Camarda:2015zba,Bodek:2016olg} to systematically reduce PDF uncertainties when measuring electroweak parameters, in particular when the cross sections differential in \cthstar{} are directly used in the fit as opposed to the reconstructed \afb.
If we consider the cross section differential in \mll, \yll, and \cthstar{} as introduced in Equation~\eqref{eqn:tripdiff_xsec}, the PDF sensitivity is again enhanced with respect to simply measuring $\diff\sigma/\diff\cthstar$, as each of these observables allows us to probe different aspects of the PDF content:
\begin{itemize}
\item \yll{} has a strong sensitivity to the Bjorken-$x$ values of the partons
\item \mll{} probes the $\Pqu/\Pqd$ quark ratio as the relative
  $\PZ$ and $\Pgg^*$ contributions vary with \mll{} through the mass dependence of the $\PZ$ propagator, giving a considerable variation in relative $\Pqu$- and $\Pqd$-type contributions across the \mll{} spectrum
\item Higher-order \QCD terms modify the $\cthstar$ decay-angle dependence through $\Pq\Pg$-, $\Paq\Pg$-, and $\Pg\Pg$-initiated channels which open up at NLO and NNLO and give the measurement sensitivity to gluon and sea-quark PDF contributions
\end{itemize}
The direct use of such differential cross-section data rather than \afb{} provides substantial sensitivity to PDFs and allows a competitive determination of \stweff{} to be made.
Beyond this, standalone differential cross section data provides important input data for future PDF fits.
In the following sections, we will introduce such a triple-differential measurement and consider some of the associated theoretical challenges with the goal of producing consistent \NNLO \QCD corrections for an associated \stweff{} fit.

\begin{figure}
  \centering
  \includegraphics[width=0.6\textwidth]{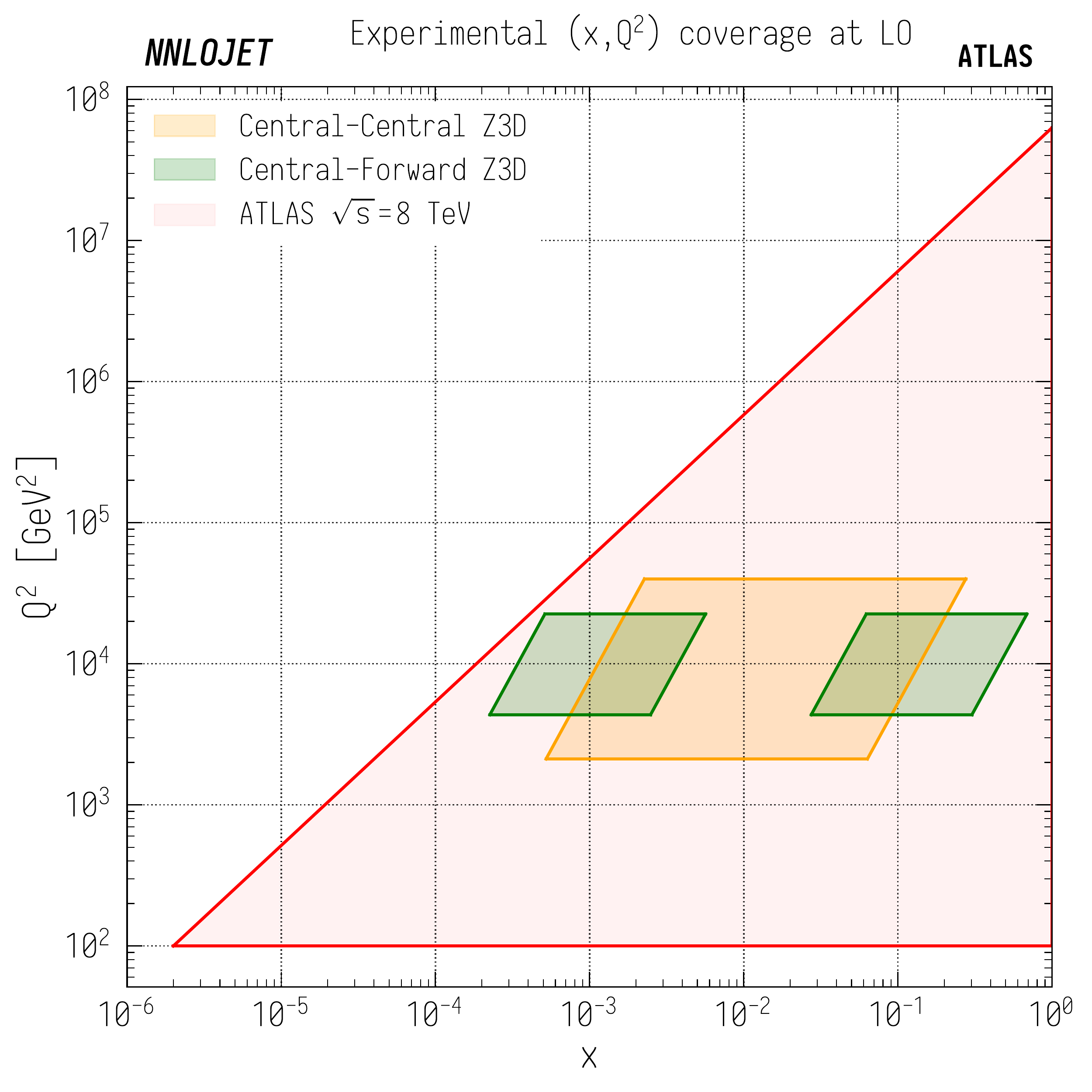}
  \caption{The kinematic regions in the $(x,Q^2)$ plane associated with the central-central (yellow) and central-forward (green) fiducial selections of the Z3D measurement.
    The total kinematic reach of the ATLAS detector at $\sqrt{s}=8~\TeV$ is shown in red.}
  \label{fig:Z3D_kinematic_region}
\end{figure}

\section{ATLAS Triple-Differential Drell-Yan Measurement}
\label{sect:ATLAS_Z3D}
The ATLAS collaboration performed a measurement of the inclusive Drell-Yan process at $\sqrts=8~\TeV$~\cite{Aaboud:2017ffb}, based on $20.2~\fb^{-1}$ of data taken in 2012 using combined electron and muon decay channels.%
\footnote{We will henceforth refer to this measurement as Z3D in order to distinguish this from the complementary DY angular analysis also performed by ATLAS on $8~\TeV$ data~\cite{Aad:2016izn}.}
The results are triply differential in the di-lepton invariant mass \mll, the di-lepton rapidity \yll, and the scattering angle in the Collins-Soper frame \cthstar.
Depending on the rapidities of the individual leptons, the measurement is divided into two regions,  defined by different selection criteria; a central-central (CC) region where both leptons are observed in the central rapidity region of the ATLAS detector, and a central-forward (CF) region where one lepton is found in the central region whilst the other is measured in the forward-detector region.
The full fiducial cuts and binnings are summarised in Tables~\ref{table:Z3Dfidcuts} and \ref{table:Z3Dfidbins}.
Furthermore, the kinematic regions covered by each of the central-central and central-forward regions can be seen in Figure~\ref{fig:Z3D_kinematic_region}, where one sees the distinctive ``split'' kinematic region associated with the central-forward selection.

\begin{table}
  \centering
  \begin{tabular}{c | c c }
    Central-Central & \multicolumn{2}{c}{Central-Forward} \\
    \hline \\[-1em]
    $\ptl > 20~\GeV$      & $\ptlF > 20~\GeV$ & $\ptlC > 25~\GeV$ \\
    $\vert\yl\vert < 2.4$ & $2.5<|\ylF|<4.9$  & $\vert\ylC\vert <2.4$ \\
    $ 46~\GeV<\mll<200~\GeV$ & \multicolumn{2}{c}{$66~\GeV<\mll<150~\GeV$}
  \end{tabular}
  \caption{Selection criteria for the central-central and central-forward fiducial regions in the ATLAS measurement of \cite{Aaboud:2017ffb}.}
  \label{table:Z3Dfidcuts}
\end{table}

The original measurement in~\cite{Aaboud:2017ffb} was presented alongside theoretical results generated at NLO QCD using \powhegboxone~\cite{Nason:2004rx,Frixione:2007vw,Alioli:2008gx,Alioli:2010xd} in conjunction with \pythia~8.1~\cite{Sjostrand:2007gs} to model parton-shower, hadronisation, and underlying-event effects as well as \photos~\cite{Golonka:2005pn} to include QED photon radiation.
The distributions were then corrected using a set of NNLO QCD + NLO EW $k$-factors differential only in the invariant di-lepton mass \mll{} generated using FEWZ 3.1~\cite{Li:2012wna}, which varied from $1.035$ for the lowest \mll{} bin to $1.025$ in the highest bin.
Preliminary results for a  fit of \stweff{} to the data by the ATLAS collaboration were  presented in~\cite{ATLAS:2018gqq}.

\begin{table}[t]
  \centering
  \begin{tabular}{c | c | c}
    Observable & Central-Central & Central-Forward \\
    \hline & & \\[-1em]
    $\mll~[\GeV]$    & [46,\,66,\,80,\,91,\,102,\,116,\,150,\,200]   & [66,\,80,\,91,\,102,\,116,\,150]             \\
    $\vert\yll\vert$ & [0,\,0.2,\,0.4,\,0.6,\,0.8,\,1,\,1.2,\,       & [1.2,\,1.6,\,2,\,2.4,\,2.8,\,3.6]            \\
                     &  1.4,\,1.6,\,1.8,\,2,\,2.2,\,2.4]             &                                              \\
    $\cthstar$       & [$-1$,\,$-0.7$,\,$-0.4$,\,0,\,0.4,\,0.7,\,1]  & [$-1$,\,$-0.7$,\,$-0.4$,\,0,\,0.4,\,0.7,\,1] \\
    \hline & & \\[-1em]
    Total Bin Count  & 504                                           & 150
  \end{tabular}
  \caption{Binnings for the central-central and central-forward fiducial regions in the ATLAS measurement of~\cite{Aaboud:2017ffb}.}
  \label{table:Z3Dfidbins}
\end{table}

The definition of the fiducial cut on individual lepton momenta contrasts with the use of di-lepton variables in the definition of the triple-differential cross section.
This interplay of kinematical variables leads to an intricate structure of the measurement regions, potentially implying non-trivial acceptance effects and an enhanced sensitivity to extra radiation from higher-order corrections.
\QCD predictions must therefore be complemented by appropriate EW corrections for multiple values of \stweff{} in order for a scan of \stweff{} to be performed, which requires careful attention to avoid consistency issues between the two theory inputs.

Whilst differential \NNLO \QCD results for the Drell-Yan process have been known for almost two decades and many codes are available for these calculations (see e.g.~\cite{Li:2012wna,Catani:2009sm,Grazzini:2017mhc,Boughezal:2016wmq}), accurate and exclusive results typically require substantial computing resources.
This is particularly true when producing multi-differential results, and it is for this reason that generating accurate predictions for the 654 separate bins of the Z3D analysis remains technically challenging.
These issues are multiplied when producing results for a parameter fit, where multiple sets of such results are required for parameter variation, uncertainty estimation, and closure tests.
As a result, one can consider the numerical demands of producing such predictions to be more comparable to those required for \PVJ~production than in the more standard single- or double-differential inclusive Drell-Yan distributions.

\section{Kinematics of the Z3D Measurement}
\label{sect:z3d_kinematics}
Inclusive Drell-Yan production contains a rich kinematic structure, which becomes increasingly apparent in multi-differential measurements.
This is particularly true when considering results differential in both \cthstar{} and \yll, where indirect kinematic constraints from fiducial cuts restrict the available phase space for the di-lepton system.
As shown in detail in the following, these constraints occur naturally as a consequence of rapidity cuts both in the Born phase space and beyond.

\subsection{Born-Level Kinematics}
At Born level, the kinematics are particularly simple, and serve as a good illustration of how phase-space constraints can be induced by fiducial cuts.
We begin with the definition of \cthstar{} from \eqref{eqn:cthstar} and use the standard momentum parameterisation of a four-vector in terms of rapidity and \pt{} for each of the outgoing leptons:
\begin{align}
  p^{\mu}_{l}&=(E_l,p^x_l,p^y_l,p^z_l) \notag\\
  &=(\etl\cosh(\yl),\ptl\cos\theta,\ptl\sin\theta,\etl\sinh(\yl)) \, .
  \label{eq:massless_momentum_parameterisation}
\end{align}
From this we can construct the separate component parts of \cthstar, noting that for massless leptons, $\etl=\ptl$:
\begin{align}
  l_i^{\pm} &= \frac{1}{\sqrt{2}}\pti \exp(\pm y_i)\,, \notag\\
  2l_i^{+}l_i^{-} &=  (\pti)^2\,, \notag\\
  l_1^+l_2^--l_1^-l_2^+ &= \ptone \pttwo \sinh(\dyll)\,, \notag\\
  \Delta  \yll & =\ylone- \yltwo \,, \notag\\
  Q^2 &= E_{12}^2-(p^z_{12})^2-{\qt}^2 = \mll^2  \,, \notag\\
  Q^2+{\qt}^2 &= E_{12}^2-(p^z_{12})^2\,, \notag\\
  &=2(l_1^++l_2^+)(l_1^-+l_2^-)\,, \notag\\
  &=2l_1^{+}l_1^{-}+2l_2^{+}l_2^{-}+2l_1^{+}l_2^{-}+2l_2^{+}l_1^{-}\,, \notag\\
  &=({\ptone })^2+({\pttwo })^2+2\ptone \pttwo \cosh(\dyll)\,. \label{eqn:Z3D_kinematic_components}
\end{align}
At Born level, $\ptone =\pttwo =\ptl$ and $\qt=0$ as there is no recoiling system and
\begin{equation}
  \yll =\frac{\ylone+\yltwo}{2} \,. \notag\\
\end{equation}
We can thus directly reconstruct Equation~\eqref{eqn:cthstar}:
\begin{equation}
  \cthstar = \frac{\sinh(\dyll)}{1+\cosh(\dyll)}=\tanh\left(\frac{\dyll}{2}\right) \,.
  \label{eqn:cthstar_qt0}
\end{equation}
This immediately allows one to derive constraints on \cthstar{} which are induced through constraints on \dyll.

\subsubsection{Central-Central (CC) Region}
For the case of rapidity cuts symmetric between the leptons and about the origin, as is the case in the central-central region of the Z3D measurement, this procedure is particularly simple.
If we note that the minimal value of $\vert\dyll\vert$ from the cuts is $0$, and that Equation~\eqref{eqn:cthstar_qt0} is symmetric in \dyll, we can see that constraints on \cthstar{} come from the upper bounds on $\vert\dyll\vert$.
For a given \yll{} value with lepton rapidity cut $\vert\yl\vert < \vert\ylmax\vert$, the greatest value of \dyll{} permitted by the cuts is $2(\ylmax-\vert\yll\vert)$, which leads to the constraint
\begin{equation}
  \cthstar \leq\frac{\sinh (2(\ylmax-|\yll|))}{1+\cosh(2(\ylmax-|\yll|))} \, .
\end{equation}
This defines a region in ($|\yll|$, $\cthstar$)-space which is forbidden at LO in QCD.

\begin{figure}
  \centering
  \includegraphics[width=0.8\textwidth]{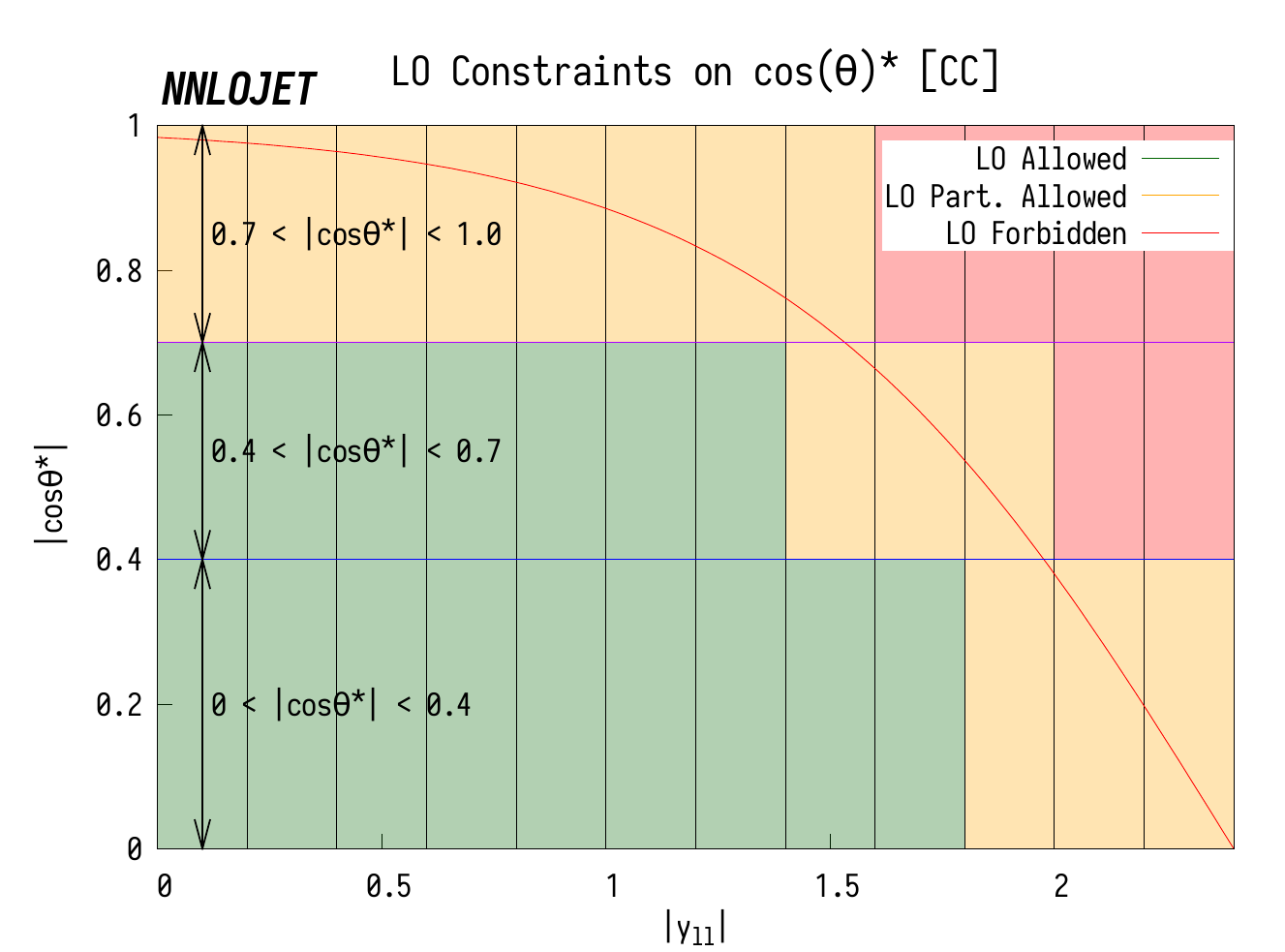}
  \caption{Bin classifications at LO for the central-central Z3D
    Drell-Yan fiducial region in the ($|\yll|$, \cthstar)
    plane. Overlaid are the measurement bins, integrated over \mll.}
  \label{fig:bin_classifications_CC}
\end{figure}

One can then use this constraint to classify the measurement bins in ($|\yll|$, $\cthstar$)-space into three categories, depending on whether the associated fiducial regions can be fully accessed, partially accessed, or are completely forbidden at LO.
The bin classifications for the central-central region are shown in Figure~\ref{fig:bin_classifications_CC} where, as one would expect, the majority of bins in the central region ($\yll\sim 0$) are fully allowed, while beyond $|\yll|=1.4$ the restrictions take effect.

One important corollary of this is that the forbidden bins, shown in red, will be described at best at \NLO{} accuracy within the full \NNLO{} calculation.
These bins can only be populated starting at $\mathcal{O}(\alphas)$ for a full fixed-order \NNLO{} Drell-Yan calculation as is also the case for the vector-boson transverse momentum distribution.
In effect, the cuts of the lepton rapidities have induced an indirect transverse momentum cut which becomes exposed when one is simultaneously differential in \cthstar{} and \yll.

\subsubsection{Central-Forward (CF) Region}

\begin{table}[t]
  \centering
  \begin{tabular}{c | c | c | c}
    & Midpoint   & Region 1 & Region 2 \\
    Bound &$|\yll|^{\mathrm{mid}}$ & $|\yll|<|\yll|^{\mathrm{mid}}$ & $|\yll|>|\yll|^{\mathrm{mid}}$\\[0.1em]
    \hline
    &&\\[-1em]
    Upper &$\frac{1}{2}(\ylFmax+\ylCmin)$& $\dyllmax=2(\yll-\ylCmin)$ & $\dyllmax=2(\ylFmax-\yll)$\\[0.1em]
    Lower &$\frac{1}{2}(\ylCmax+\ylFmin)$ & $\dyllmin=2(\ylFmin-\yll)$ & $\dyllmin=2(\yll-\ylCmax)$\\[0.1em]
  \end{tabular}
  \caption{Maximum/minimum values of \yll~permitted in different regions of phase space for the CF Z3D selection.}
  \label{tab:cf_regions}
\end{table}

A similar procedure can be undertaken for the central-forward region of the Z3D measurement, where one lepton is emitted in the forward region.
In order to extract the LO constraints on \dyll{} in the asymmetric case, it is easiest to first divide the phase space into regions by introducing mid-points of the lepton cuts, detailed in Table~\ref{tab:cf_regions}, which are then used for the construction of each of the upper and lower bounds.
These regions correspond to values of $|\yll|$ for which a particular lepton rapidity cut provides the limiting value of \dyll, and hence extremal values for \cthstar.

\begin{figure}
  \centering
  \includegraphics[width=0.8\textwidth]{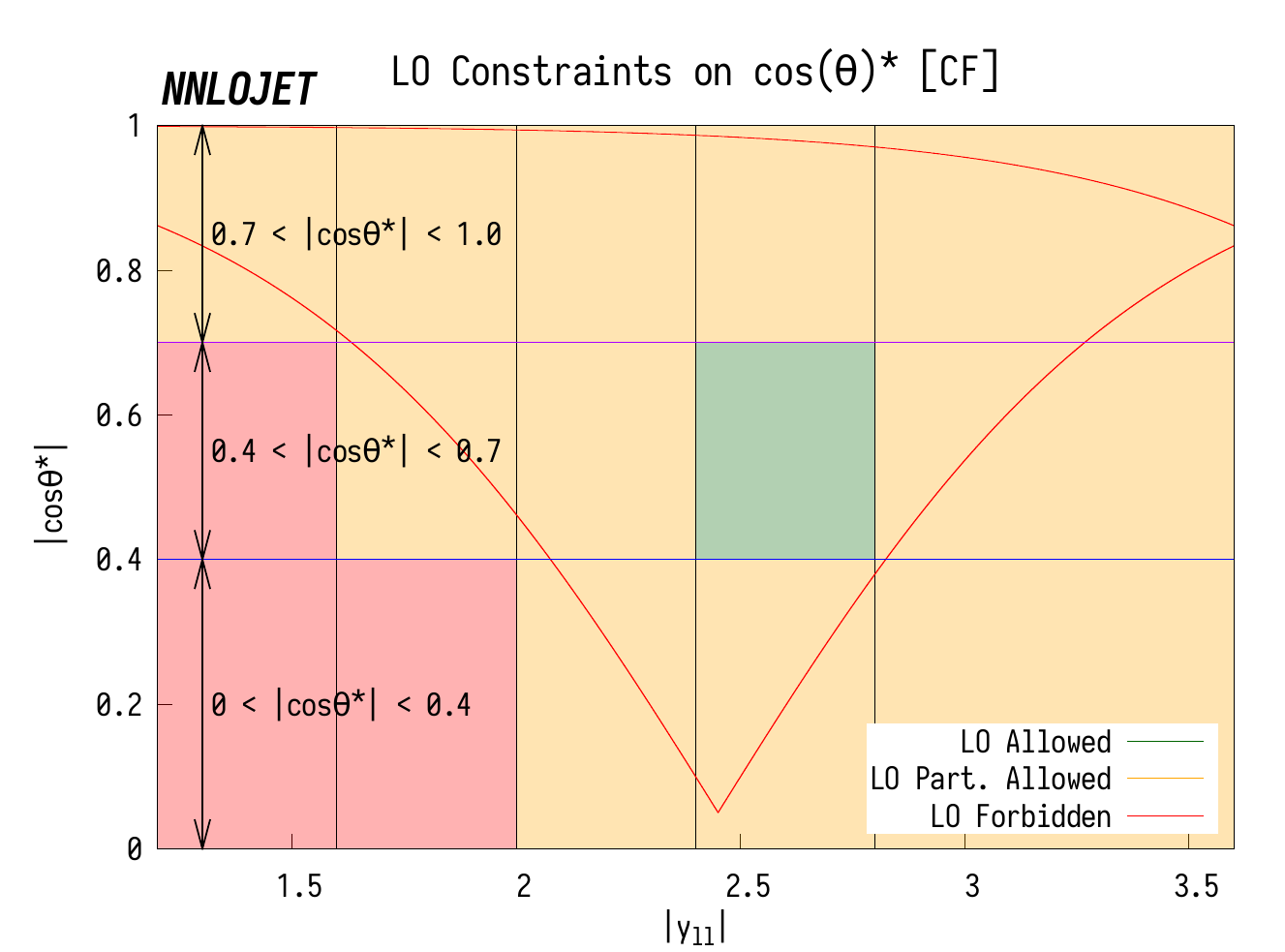}
  \caption{Bin classifications at LO for the central-forward Z3D
    Drell-Yan fiducial region in the ($|\yll|$, $\cthstar$)
    plane. Overlaid are the measurement bins, integrated over \mll.}
  \label{fig:bin_classifications_CF}
\end{figure}

The associated phase space and bin classifications are shown in Figure~\ref{fig:bin_classifications_CF}, where one can see that (unlike the CC region), there is a bias in the allowed phase space towards non-central values of $|\cthstar|$.
In the context of a \sth{} fit this is particularly interesting, as the cuts imply that the distribution of the cross section is biased towards larger values of \yll.
The number of LO-forbidden bins is also greatly increased with respect to the CC-case.
This means that inclusive \NNLO{} QCD predictions for Drell-Yan production are less effective in describing the bulk of the data than for the CC region.
As the forward-backward asymmetry can be measured more precisely at large values of rapidity, since the incoming quark and antiquark are better defined in the Collins-Soper frame, one could consider exploiting this in order to construct an experimental binning in which a maximum number of bins are fully accessible at LO (statistical and detector constraints notwithstanding).

\subsection{Constraints Beyond Born Level}
\label{sect:better_than_born}
The above derivations of kinematical constraints at Born level rely on the vanishing of the transverse momentum of the di-lepton pair, leading to an exact balance of the transverse momenta of both leptons.
At higher orders in perturbation theory, these constraints are lifted since the di-lepton system acquires a non-zero transverse momentum through recoil against some partonic radiation.
In the following, we are looking to evaluate the minimum transverse momentum required to populate the bins forbidden at LO rather than to evaluate the bounds of this region.

From the kinematical relations of Equation~\eqref{eqn:Z3D_kinematic_components}, one obtains the general form of \cthstar{} by considering the \pt{} dependence of the \PZ~boson, using $Q = \sqrt{(Q^2+\qtsq)-\qtsq}$:
\begin{align}
  \cthstar =& \frac{2 \ptone \pttwo \sinh(\dyll)}{\sqrt{(\ptone )^2+(\pttwo )^2+2\ptone \pttwo \cosh(\dyll)}} \notag\\
  &\times \frac{1}{\sqrt{(\ptone )^2+(\pttwo )^2+2\ptone \pttwo \cosh(\dyll)-\qtsq}} \, .
  \label{eqn:cthstar_full_kin}
\end{align}
Directly deriving constraints from Equation~\eqref{eqn:cthstar_full_kin} is considerably less trivial than in the Born case, given the complex interplay of the different fiducial cuts, but can still be performed for the two fiducial regions of the Z3D measurements under certain approximations.

\begin{figure}
  \centering
  \includegraphics[width=\textwidth]{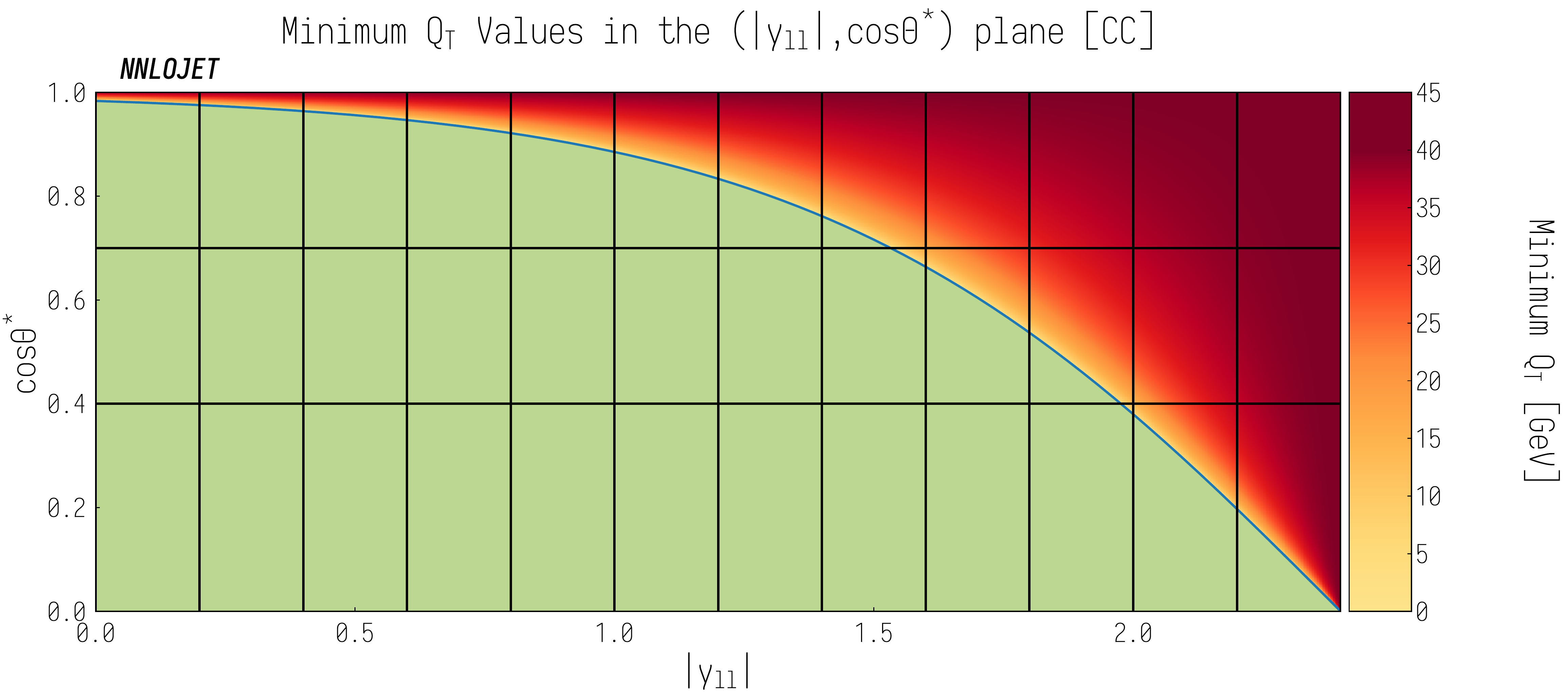}
  \caption{Minimum $\qt$ values required for the different regions of the ($|\yll|$, $\cthstar$) plane in the central-central region, obtained with the simplifying assumption of
    equal transverse momenta of both leptons. Overlaid are the Z3D measurement bins, integrated over \mll.}
  \label{fig:CC_heatmap}
\end{figure}

For the central-central region, the minimum value of \qt{} at a given point in the ($|\yll|$, $\cthstar$) plane can be obtained simply by rearranging for \qtsq, and substituting in the minimum lepton transverse momenta \ptlmin{} and maximum rapidity difference $\dyllmax$ permitted by the cuts:
\begin{equation}
  \qtsq \geq 2\bigl(\ptlmin\bigr)^2\biggl(1+\cosh(\dyllmax) -\frac{\sinh^2(\dyllmax)}{(1+\cosh\left(\dyllmax)\right)\cthstarsq}\biggr) \, .
  \label{eqn:cthstar_qt}
\end{equation}
This constraint means that for a given \qt{} value, there is a region in \cthstar{} space that cannot be populated, determined by the accessible lepton $\pt$ and $\yl$ values given by the cuts.
These \qt{} constraints are shown in Figure~\ref{fig:CC_heatmap} for the central-central Z3D region under the additional assumption that
\begin{equation}
  \yll=\frac{\ylone+\yltwo}{2} \, ,
  \label{eqn:av_yll_assumption}
\end{equation}
which holds only when the leptons have equal transverse momentum (the general expression of $\yll$ will be given below).
We see that the constraints, whilst significant, are not strong enough to prevent population of the forbidden bins beyond Born level for centre-of-mass energies one typically observes at the Tevatron or the LHC.

This effective \qt{} cut can be employed to extend the accuracy of the fixed-order theoretical predictions for the Born-forbidden bins by using calculations of \PZJ{} in a similar manner as for the \ptv{} spectrum~\cite{Ridder:2016nkl,Gehrmann-DeRidder:2016jns,Gehrmann-DeRidder:2017mvr}, since the requirement for a non-zero \qt{} implicitly enforces partonic radiation.
Fully-differential results  for the \PZJ{} process are known to $\alphas^3$, corresponding to the NNLO QCD corrections~\cite{Boughezal:2015ded,Ridder:2015dxa}. 
Most recently, the $\alphas^3$ corrections were also computed to differential DY production~\cite{Chen:2021vtu,Chen:2022cgv,Neumann:2022lft}, 
where they correspond to \NthreeLO QCD. However, these \NthreeLO 
calculations are not yet sufficiently efficient to yield the multi-differential distributions as considered here. By restricting the $\alphas^3$ contributions to the forbidden bins, 
where the triple-virtual contribution which lies in the Born phase space does not contribute, this represents an improvement equivalent to extending the predictions to \NthreeLO accuracy in these bins.

\begin{figure}
  \centering
  \includegraphics[width=\textwidth]{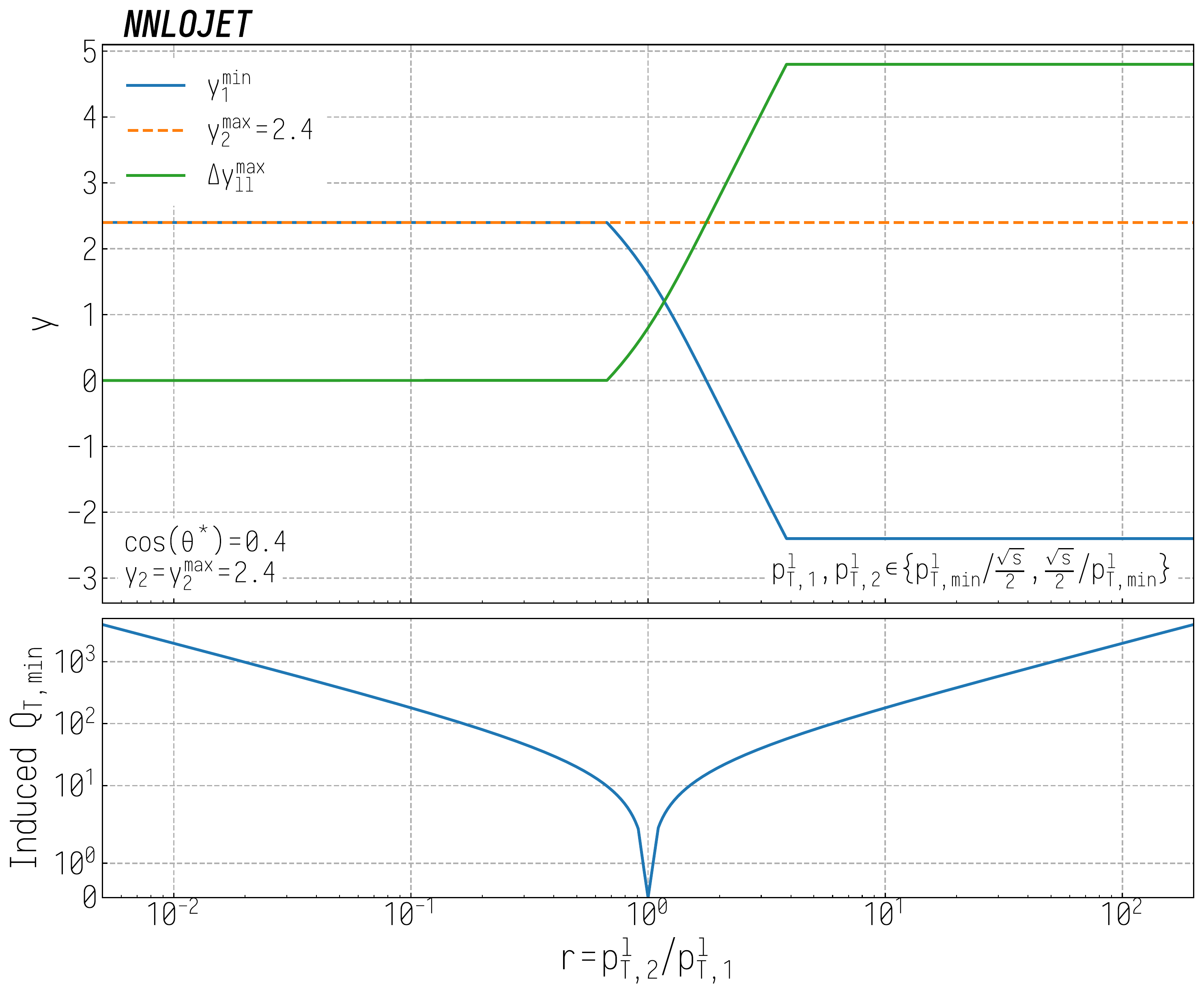}
  \caption{The upper panel shows the minimum values of $\ylone$ and maximum values of
    $\dyllmax$ as a function of the lepton transverse momentum
    ratio $r$. The lower panel shows the resultant minimum transverse
    momentum induced by the momentum ratio $r$.
    For this we consider the forbidden phase-space
    point at $\yll=2$, $\cthstar=0.4$ in the Z3D central-central
    fiducial region, and take $\yltwo=\ytwomax=2.4$ in order to
    maximise the rapidity difference. }
  \label{fig:yll_cc}
\end{figure}

However, the assumption made in Equation~\eqref{eqn:av_yll_assumption} is not strictly true.
When we relax the Born phase-space constraints, the leptons have different transverse momenta which alters the relationship between the lepton rapidities $\ylone$, $\yltwo$, and $\yll$.
Taking the definition of rapidity:
\begin{align}
  \yll=&\frac{1}{2}\log\left(\frac{\ptone \left(\cosh \ylone+\sinh \ylone\right)+\pttwo \left(\cosh \yltwo+\sinh \yltwo\right)}{\ptone \left(\cosh \ylone-\sinh \ylone\right)+\pttwo \left(\cosh \yltwo-\sinh \yltwo\right)}\right)\notag\\
      =&\frac{1}{2}\log\left(\frac{e^{\ylone}+\pttwo /\ptone e^{\yltwo}}{e^{-\ylone}+\pttwo /\ptone e^{-\yltwo}}\right)\notag\\
      =&\frac{1}{2}\left(\ylone+\yltwo\right)+\frac{1}{2}\log\left(r\cdot\left[\frac{1+r\cdot e^{\yltwo-\ylone}}{1+1/r\cdot e^{\yltwo-\ylone}}\right]\right) \,,
      \label{eqn:yll_non_equal_transverse}
\end{align}
where we have introduced the ratio of lepton transverse momenta as
\begin{align}
  r=\pttwo /\ptone \, .
\end{align}
Equation~\eqref{eqn:yll_non_equal_transverse} is explicitly dependent on both the \pt{} ordering of the leptons as well as the rapidity ordering, and we observe that it reduces to the correct form in the Born limit $r\rightarrow1$.
The new term weakens the constraints on \qt{} and allows $\dyllmax$ to take a larger range of values for a given \yll{} value by introducing a large transverse momentum imbalance between the leptons (up to the maximum directly permitted by the lepton cuts).
Evaluating the full \qt~dependence for a given \yll, \cthstar{} value requires an iterative numerical solution.
Since
\begin{equation}
  r\in\left\{\sim \ptlmin/\frac{\sqrt{s}}{2},\sim\frac{\sqrt{s}}{2}/\ptlmin\right\}
\end{equation}
(which for the central-central region in the Z3D~measurement is equivalent to the approximate range $r \in [0.005, 200]$), this would allow one to effectively produce any $\dyllmax$ given the correct conditions.
This is shown in Figure~\ref{fig:yll_cc}, where we show the minimum values of $\ylone$ and maximum values of $\dyllmax$ as a function of $r$ for the Born-level forbidden phase-space point $\yll=2$, $\cthstar=0.4$, setting $\yltwo$ to the maximum permitted value in order to maximise $\dyllmax$.

Here we see that for final states with values of $r\sim 8$, one could imagine that it is possible to generate any lepton rapidity separation and thus any value of \qt.
However, hidden within $r$, there is a second restriction on \qt, governed by the requirement for transverse momentum conservation:
\begin{align}
  \ptonevec+\pttwovec+\vecqt=0 \, .
\end{align}
From here, one can see that the minimal value for \qt{} is given when the lepton \pt~values are back to back, which allows us to conclude that there is a second competing \qt~bound at
\begin{equation}
  \ptlmin(r-1)=\qtmin \, .
\end{equation}
which has no effect for $r=1$, but gives a minimum $\qt$ of $140~\GeV$ for $r=8$ and $\ptlmin=20~\GeV$.
Somewhat counter-intuitively, one can in effect \textit{decrease} the required \qt{} for a given point in ($|\yll|$, $\cthstar$) space by \textit{generating} a lepton \pt{} imbalance through a non-zero \qt.
The lower panel of Figure~\ref{fig:yll_cc} shows the variation of this \qtmin~value with $r$ for the selected Born-level forbidden phase-space point $\yll=2$, $\cthstar=0.4$.

\begin{figure}
  \centering
  \includegraphics[width=\textwidth]{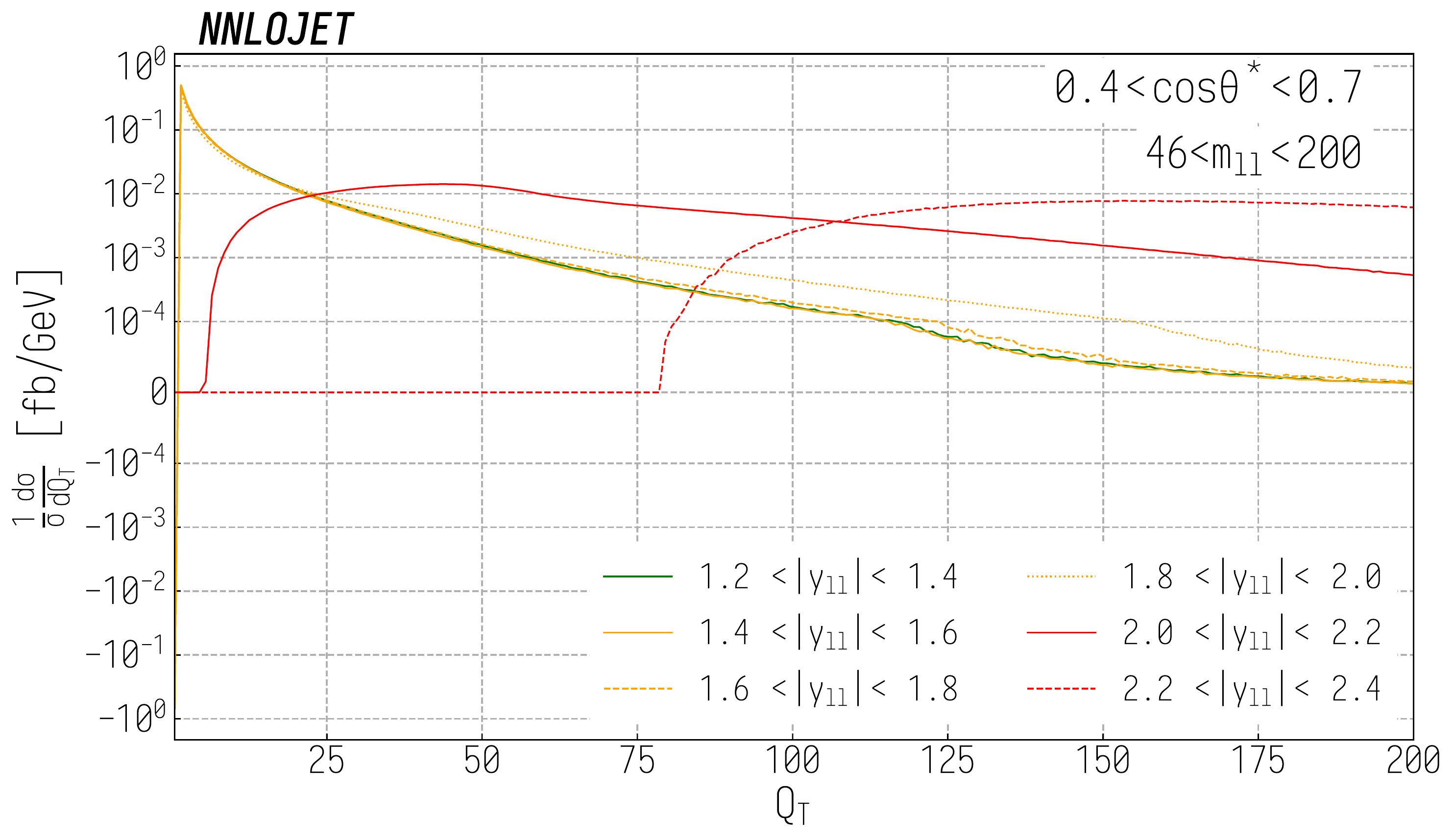}
  \caption{The normalised $\qt$ spectrum for the $0.4<\cthstar<0.7$,
    $46<\mll<200$ region for each rapidity bin of the Z3D
    central-central measurement region
    between $\yll=1.2$ and $\yll=2.4$. The results are produced to
    $\mathcal{O}(\alphas)$, with the colours labelling as before the
    allowed (green), mixed (orange) and forbidden (red) bins. The kink
    observed at $1/\sigma\cdot\diff\sigma/\diff \qt=10^{-4}$ is a consequence of the
    linearisation of the axes between $\qt\pm10^{-4}$ to allow the negative
    contribution at $\qt=0$ to be shown.}
  \label{fig:CC_qt_forbidden}
\end{figure}

Given that we intend to compute fixed-order predictions for these Born-level forbidden bins, it is instructive to determine the \qt~spectrum within a given bin to assess the potential impact of large logarithms in $\mll/\qt$.
If present, such logarithms could in principle be resummed to $\text{N}^3\text{LL}$ accuracy using tools such as \radish~\cite{Bizon:2017rah,Bizon:2018foh,Bizon:2019zgf}.
In Figure~\ref{fig:CC_qt_forbidden} we show the normalised \qt~spectra at $\mathcal{O}(\alphas)$ for each \yll{} bin in the $0.4<\cthstar<0.7$, $46~\GeV<\mll<200~\GeV$, $1.2<|\yll|<2.4$ region of the Z3D analysis.
One clearly observes the evolution of the \pt~spectrum with $\yll$ as one passes from fully allowed bins at $1.2<|\yll|<1.4$ (green) through the mixed region $1.4<|\yll|<2.0$ (yellow) to the fully forbidden region $|\yll|>2.0$ (red).
For the forbidden bins we observe that no logarithmically divergent behaviour is present at low \qt, with no \qt~values below $5~\GeV$.
This can be understood as the volume of the phase space in which low \qt~production is permitted decreases at a faster rate than the matrix element diverges as $\qt\rightarrow 0$.

Particularly when one considers cross sections integrated over \qt{} as in the Z3D measurement, any residual breakdown in the perturbative series will be largely suppressed.
Large logarithms in \qt{} emerge from a kinematical mismatch between real and virtual contributions at low \qt, which is cured by the \qt~integration.
The kinematic suppression towards low values of \qt{} in the Born-level forbidden bins can be understood in this context as a direct result of the lack of $\qt=0$ virtual contributions.
This is visible in Figure~\ref{fig:CC_qt_forbidden}, where the peaks in the \qt~distributions for the two forbidden bins occur at $40~\GeV$ and $160~\GeV$, while the kinematical limits are much lower at $5~\GeV$ and $80~\GeV$ respectively.
This is in stark contrast to the mixed and allowed bins, where such logarithms give a large enhancement to the low \qt~cross section.
These distributions are only rendered finite when one includes the $\qt=0$ contribution as is the case when one integrates out \qt{} to form the inclusive cross section.
As a consequence of this kinematic suppression, we can conclude that fixed-order results are indeed reliable in the Born-level forbidden regions of phase space.

At this point, we can consider the contributions as being intrinsically \PZJ{} in nature due to the implicit \qt~requirement.
As a consequence, it becomes possible in these bins to extend the results to $\mathcal{O}(\alphas^3)$ through the use of a \PZJ~calculation, which gives exactly the contributions that would be found in a full calculation of the inclusive Drell-Yan cross section to \NthreeLO.
This will have the impact of enhancing the accuracy of these predictions in the high-\yll~region, where the asymmetry \afb{} is largest.
The same cannot be said for the partially forbidden, partially allowed `mixed' bins, where one will encounter the divergence at the boundary between the two regions and where logarithmically divergent $\qt=0$ contributions are present.
In these bins, one is restricted to $\mathcal{O}(\alphas^2)$, which is \NNLO for the inclusive Drell-Yan cross section.

In principle, the above discussion would allow one to adjust the experimental bin edges in order to maximise the precision of the available theory in future experimental measurements.
Were one to construct bins with edges that align with the kinematic boundary, one could consider using the inclusive \NNLO \PZ~calculation in the allowed region, and solely using the \NNLO \PZJ{} calculation (potentially with resummation) in the forbidden region, as it would amount to a systematic removal of the mixed bins.

Since the current discussion is related only to kinematics, similar observations can be made for higher-order EW corrections to the DY process.
For the associated real contributions, the transverse momentum required can be created through photon emission, such that there are two contributions in the forbidden regions; one of QCD-type at $\mathcal{O}(\alphas)$ and one of EW-type at $\mathcal{O}(\alpha)$, where due to the relative size of the coupling constants, the QCD contribution will dominate.

A similar reasoning can be repeated for the CF case to generate the \qt~dependence of the phase-space constraints.
For the region of phase space above $\cthstar\sim 0.9$, limited by the maximum \dyll{} permitted by the cuts, one can proceed simply by rearranging Equation~\eqref{eqn:cthstar_full_kin}, this time retaining the full \ptl~dependence.
Substituting the $\dyllmax$ values from Table~\ref{tab:cf_regions} for the appropriate $|\yll|$ region, alongside minimum \ptl{} values permitted by the cuts, this gives the minimum \qt~dependence of the upward Born-level forbidden region as
\begin{align}
  \qt^2\geq & \ ({\ptlCmin})^2+({\ptlFmin})^2+2 \ptlCmin \ptlFmin \cosh(\dyllmax)\notag\\
  & - \frac{\bigl(2 {\ptlCmin} {\ptlFmin} \sinh(\dyllmax)\bigr)^2}{\cthstarsq\bigl(({\ptlCmin})^2+({\ptlFmin})^2+2 \ptlCmin \ptlFmin \cosh(\dyllmax)\bigr)} \, .
  \label{eq:qt_full_pt}
\end{align}

We can now consider the new case in which the bound is given by $\dyllmin$, in the lower region of the central-forward (\yll,\cthstar) plane.
Taking the transverse components from Equation~\eqref{eq:massless_momentum_parameterisation}, one can write
\begin{equation}
  \qt^2=({\ptlF})^2+({\ptlC})^2 + 2\ptlF\ptlC\cos(\Delta\theta_\mathrm{FC}) \, ,
\end{equation}
where the angular separation of the two leptons is $\Delta\theta_\mathrm{FC}=|\theta_\mathrm{F}-\theta_\mathrm{C}|$.
Using this in conjunction with Equation~\eqref{eq:qt_full_pt}, one can then identify
\begin{align}
  \cos(\Delta\theta_\mathrm{FC})^{\min}\geq&\cosh(\dyllmax)\\
  &- \frac{ 2{\ptlCmin}{\ptlFmin} \sinh^2(\dyllmax)}{\cthstarsq\bigl(({\ptlCmin})^2+({\ptlFmin})^2+2 \ptlCmin \ptlFmin \cosh(\dyllmax)\bigr)},\notag
\end{align}
such that when $\qt^2$ is minimised, so is $\cos(\Delta\theta_\mathrm{FC})$.
However, for the constraints derived from minimal values $\dyllmin$, there is not an equivalent meaningful lower bound on $\cos(\Delta\theta_\mathrm{FC})$, as it first saturates at $\cos(\Delta\theta_\mathrm{FC})=-1$.
It is this saturation that complicates the picture when considering the minimum values of \qt{} in the region below $\dyllmin$, corresponding to $\cthstar\lesssim 0.9$, as one cannot rely on $\cos(\Delta\theta_\mathrm{FC})$ being minimised to some value by \dyll.

\begin{figure}
  \centering
  \includegraphics[width=\textwidth]{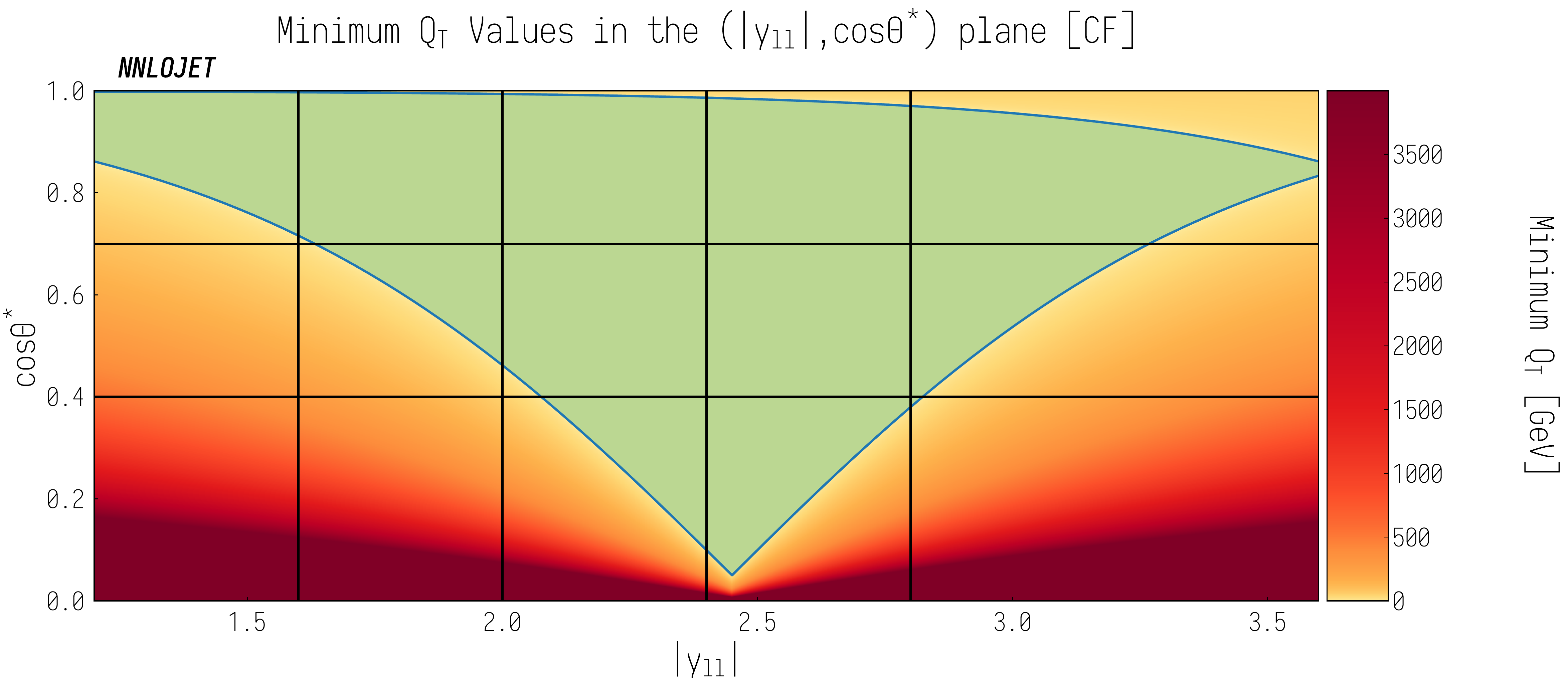}
  \caption{Minimum \qt{} values required for the different regions of the ($|\yll|$, $\cthstar$) plane in the central-forward region. Overlaid are the Z3D measurement bins, integrated over \mll.}
  \label{fig:CF_heatmap}
\end{figure}

In order to minimise \qt{} for a given (\yll,\cthstar) point in this region, one must find values of \ptlF{} and \ptlC{} that are consistent with $\cos(\Delta\theta_\mathrm{FC})=-1$.
This is straightforward if one enforces that $\Delta\theta_\mathrm{FC}=\pi$, $\cos(\Delta\theta_\mathrm{FC})=-1$ in order to minimise \qt, such that (for arbitrary lepton ordering)
\begin{equation}
  \ptlone = \qt+\ptltwo = \qt+\ptl \, ,
  \label{eqn:pt_min_bounds} 
\end{equation}
which ensures that one lepton is parallel to $\vec{Q}_\mathrm{T}$ in the $\vec{p}_\mathrm{T}$ plane.
One can then substitute this directly into Equation~\eqref{eqn:cthstar_full_kin} and solve for \qt{} to find:
\begin{align}
  \qt^2\geq &\frac{2{\ptl}^2(1 + \cosh(\dyllmin))^2(\cthstarsq - \tanh(\dyllmin)^2)}{\cthstarfour}\notag\\
  &\hspace{0.5cm}\times\biggl[\frac{\cthstarsq\cosh(\dyllmin)}{1 + \cosh(\dyllmin)} - \tanh(\dyllmin)^2\notag\\
    &\hspace{1cm}+ \biggl(\frac{{\cthstarsq - \tanh(\dyllmin)^2}}{(1 + \cosh(\dyllmin))}(\cthstarsq(\cosh(\dyllmin)-1) \notag\\
    &\hspace{1cm} - (1 + \cosh(\dyllmin))\tanh(\dyllmin)^2)\biggr)^{\frac{1}{2}}\biggr] \, .
\end{align}
It is straightforward to see that this is minimised for the smallest value of \ptl{} accessible to \textit{both} \ptlC{} and \ptlF{} permitted by the cuts
\begin{equation}
  \ptl = \ptlmin = \max(\ptlCmin,\ptlFmin) \, ,
\end{equation}
which for the Z3D measurement gives $\ptlmin=25~\GeV$.
With this in place, the minimum values of \qt{} required across the (\yll, \cthstar) plane can be calculated, which is shown in Figure~\ref{fig:CF_heatmap} in the $\yll=(\ylone+\yltwo)/2$ approximation.

One interesting effect in the CF region is the existence of an ultra-forbidden region due to the constraints induced by \dyllmin, which is excluded to all orders in perturbation theory.
This is present as $\cthstar\rightarrow 0$, and is defined by the region where $\qt^{\min}>\frac{\sqrt{s}}{2}$, such that there can never be enough energy present in the event to overcome the minimum \qt{} and allow an event to occur.



\section{Numerical Setup}
\label{sect:numerical_setup}
In this section, we describe the numerical setup used in this study.

\NNLO and partial \NthreeLO QCD predictions are calculated using existing implementations of the \PZ{} and \PZJ{} processes in \nnlojet.
We employ the \gmu~EW scheme $(\gf, \mz, \mw)$ including running-width effects, with the following on-shell input parameters:
\begin{align}
  \mz &= 91.1876~\GeV \, , & \quad \mw &= 79.939~\GeV \, ,  \notag\\
  \gz &= 2.4952~\GeV  \, , & \quad \gw &= 2.085~\GeV  \, ,  \\
  \gf &= 1.663787\times10^{-5}~\GeV^{-2} \, , && \notag
\end{align}
where the unphysical value of \mw{} corresponds to that required to set $\stw=\stweff$ in the \gmu~scheme at LO EW.
From this, the following values of \stw{} and \alpha are derived:
\begin{equation}
  \stw = 0.23148638 \,, \quad \alpha(\mz) = 0.0077601936 \,.
  \label{eq:derived_quantities}
\end{equation}
For central predictions, we use the NNPDF3.1 parton distributions~\cite{Ball:2017nwa} with $\alpha_S(M_Z)=0.118$ and estimate theory uncertainties by a seven-point variation of renormalisation and factorisation scales within a factor 2 around a central value of $\mu^2=\mll^2$.

Fixed-order electroweak corrections are computed using a modified version of the \verb|Z_ew-BMNNPV| code~\cite{Barze:2013fru} in \powhegbox~\cite{Nason:2004rx,Frixione:2007vw,Alioli:2008gx,Alioli:2010xd}.
The computation is performed at NLO EW, including the leading higher-order (\HO) corrections to $\Delta\rho$ and $\Delta\alpha$ described in \cite{Dittmaier:2009cr}.
For \powhegbox we use the same settings as for \nnlojet, however using a derived \gmu~EW scheme proposed in~\cite{Chiesa:2019nqb}, where the input parameters are $(\gf, \mz, \stweff)$ with $\stweff$ as in Equation~\eqref{eq:derived_quantities}:
\begin{equation}
  \stweff = 0.2318638 \, .
\end{equation}
In this scheme \stweff{} has the same definition used at LEP and SLD and reabsorbs part of the higher-order corrections to \afb.
We use the constant-width scheme and treat singularities associated with the unstable nature of the \PW/\PZ vector bosons circulating in the loops according to the factorisation scheme~\cite{Dittmaier:2001ay,Dittmaier:2009cr}.
As the data is already corrected for the effect of QED photon radiation, we consider only genuine weak loop corrections.

\begin{figure}[ht]
  \centering
  \includegraphics[width=1.0\textwidth]{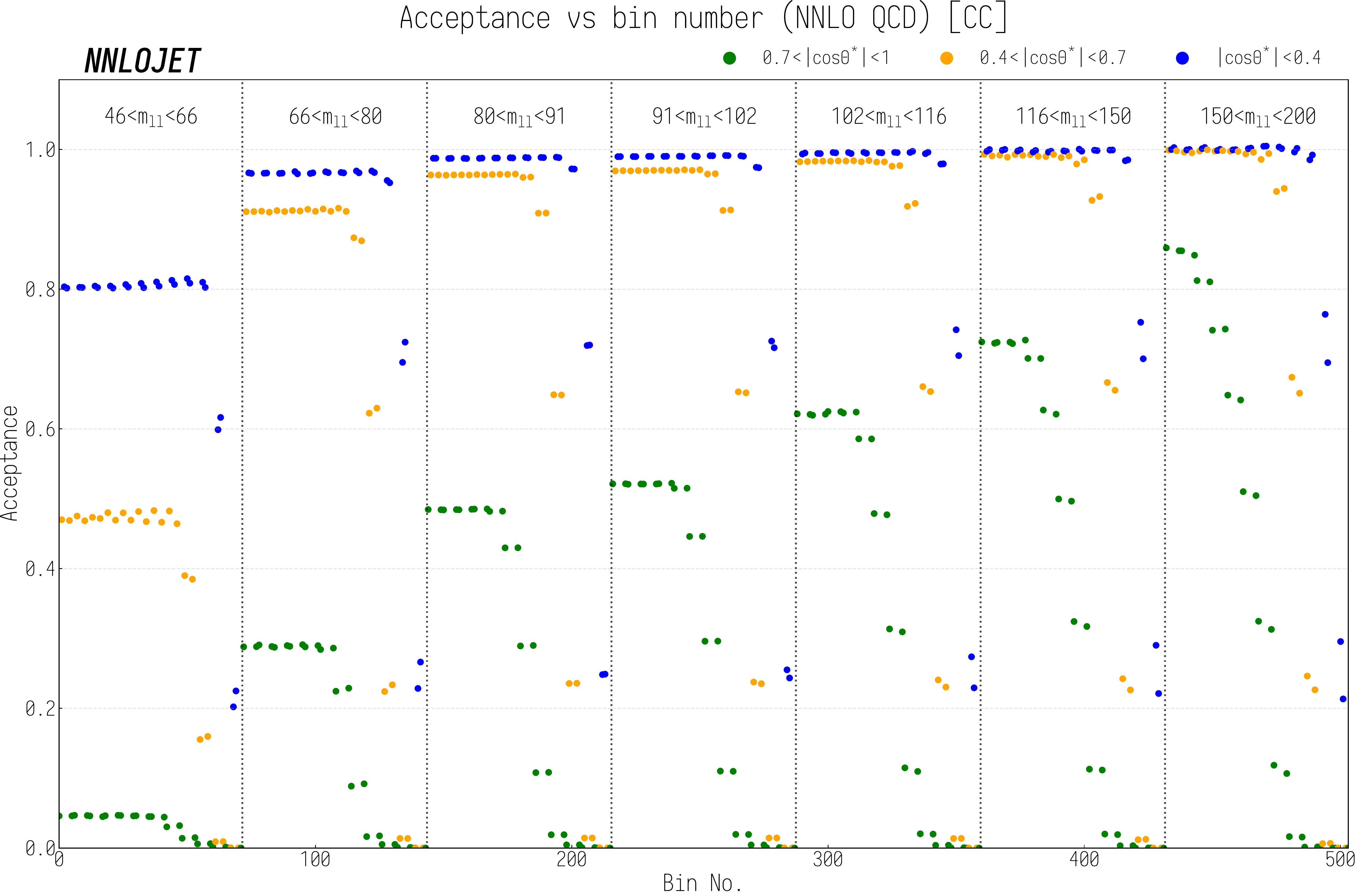}
  \caption{Acceptances for the CC Z3D fiducial region. The bin
    number is as defined in Equation~\eqref{eqn:bin_idx_one}, such that the
    major \mll~bins are divided into 12 \yll~sub-bins from $0-2.4$
    (left to right) which are in turn divided into 6
    \cthstar~sub-bins from $-1$ to $1$ (left to right). The different
    \cthstar~values are denoted by the central colour of each
    point.}
  \label{fig:Z3D_acceptances}
\end{figure}

\section{Acceptances and $k$-factors}
\label{sect:k_fac_acceptances}
The kinematical considerations in Section~\ref{sect:z3d_kinematics} 
have unravelled a highly non-trivial interplay between the variables that define the Z3D cross section and the fiducial cuts.
These fiducial cuts lead to a rejection of events that would normally be contained in specific kinematical bins of the Z3D measurement.
On a bin-by-bin basis, this effect can be quantified through an acceptance factor, which is the ratio of the bin-integrated cross section computed with application of the fiducial cuts to the same bin-integrated cross section without fiducial cuts.
A high acceptance close to unity indicates that the measurement is relatively independent of the cuts, and the bulk of the total cross section contribution is contained within the fiducial region.

Figure~\ref{fig:Z3D_acceptances} displays the acceptances for all bins in the CC region using \NNLO QCD predictions.
To display the full triple-differential structure, we define the bin index using the index of each observable $\mathcal{O}^{\mathrm{idx}}$ from low to high as
\begin{equation}
  \mathrm{Bin\ No.} = 72\cdot\mll^{\mathrm{idx}}+6\cdot\yll^{\mathrm{idx}}+{\cthstar}^{\mathrm{idx}} \, ,
  \label{eqn:bin_idx_one}
\end{equation}
such that the major \mll~bins are divided into 5 \yll~sub-bins from $0$ to $2.4$ (left to right), which are in turn divided into 6 \cthstar~sub-bins from $-1$ to $1$ (left to right).
We use this as our $x$-axis.

One can see a sharp decline in acceptance within each invariant mass bin as one moves into the partially allowed and then forbidden regions with increasing \yll.
This is expected, as the \qt~restriction for a given point in (\yll,\cthstar) space is caused entirely by the lepton rapidity cuts.
The bulk of the fixed-order Drell-Yan cross section lies at low \qt, meaning that a \qt~cut greatly decreases the acceptance.
The structure between invariant mass bins, where the acceptance increases with \mll{} independently of the values of \cthstar~and \yll, are an indirect result of the fiducial lepton \pt~cuts.
Events at lower invariant di-lepton mass are less likely to have the lepton transverse momenta required to pass the fiducial cuts.
As a result, more of the cross section lies outside of the fiducial region, resulting in a reduced acceptance in the low-\mll{} phase-space regions.
\begin{figure}[ht]
  \centering
  \includegraphics[width=1.0\textwidth]{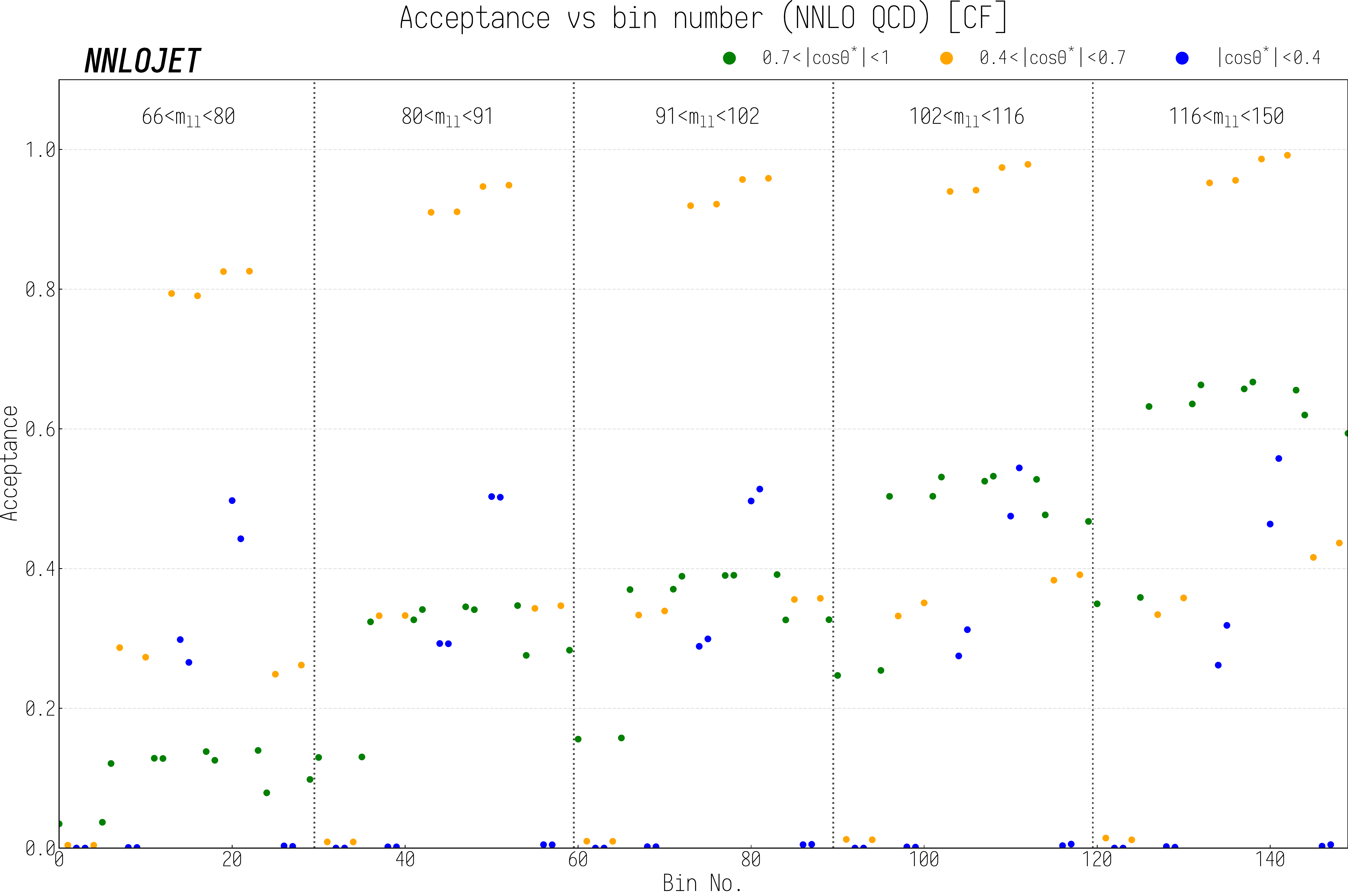}
  \caption{Acceptances for the CF Z3D fiducial region. The bin
    number is as defined in Equation~\eqref{eqn:bin_idx_CF}, such that the
    major \mll~bins are divided into 5 \yll~sub-bins from $1.2$ to $3.6$
    (left to right) which are in turn divided into 6
    \cthstar~sub-bins from $-1$ to $1$ (left to right). The different
    \cthstar~values are denoted by the central colour of each
    point.}
  \label{fig:Z3D_acceptances_CF}
\end{figure}

In the central-forward (CF) region, bins with low acceptance are much more prevalent than in the CC region, as can be seen in Figure~\ref{fig:Z3D_acceptances_CF}, where the bin index is computed as
\begin{equation}
  \mathrm{Bin\ No.}= 30\cdot\mll^{\mathrm{idx}}+6\cdot\yll^{\mathrm{idx}}+{\cthstar}^{\mathrm{idx}},\label{eqn:bin_idx_CF}
\end{equation}
such that the major \mll~bins are divided into 5 \yll~sub-bins from $1.2$ to $3.6$ (left to right) which are in turn divided into 6 \cthstar~sub-bins from $-1$ to $1$ (left to right).
As anticipated from Figure~\ref{fig:bin_classifications_CF}, there are only few bins with acceptances close to unity, all within the central range of $0.4<|\cthstar|<0.7$.
Especially for low $|\cthstar|$, some of the bins have very low acceptances, indicating that they are kinematically disfavoured even at higher orders. 
\begin{figure}
  \centering
  \includegraphics[width=1.0\textwidth]{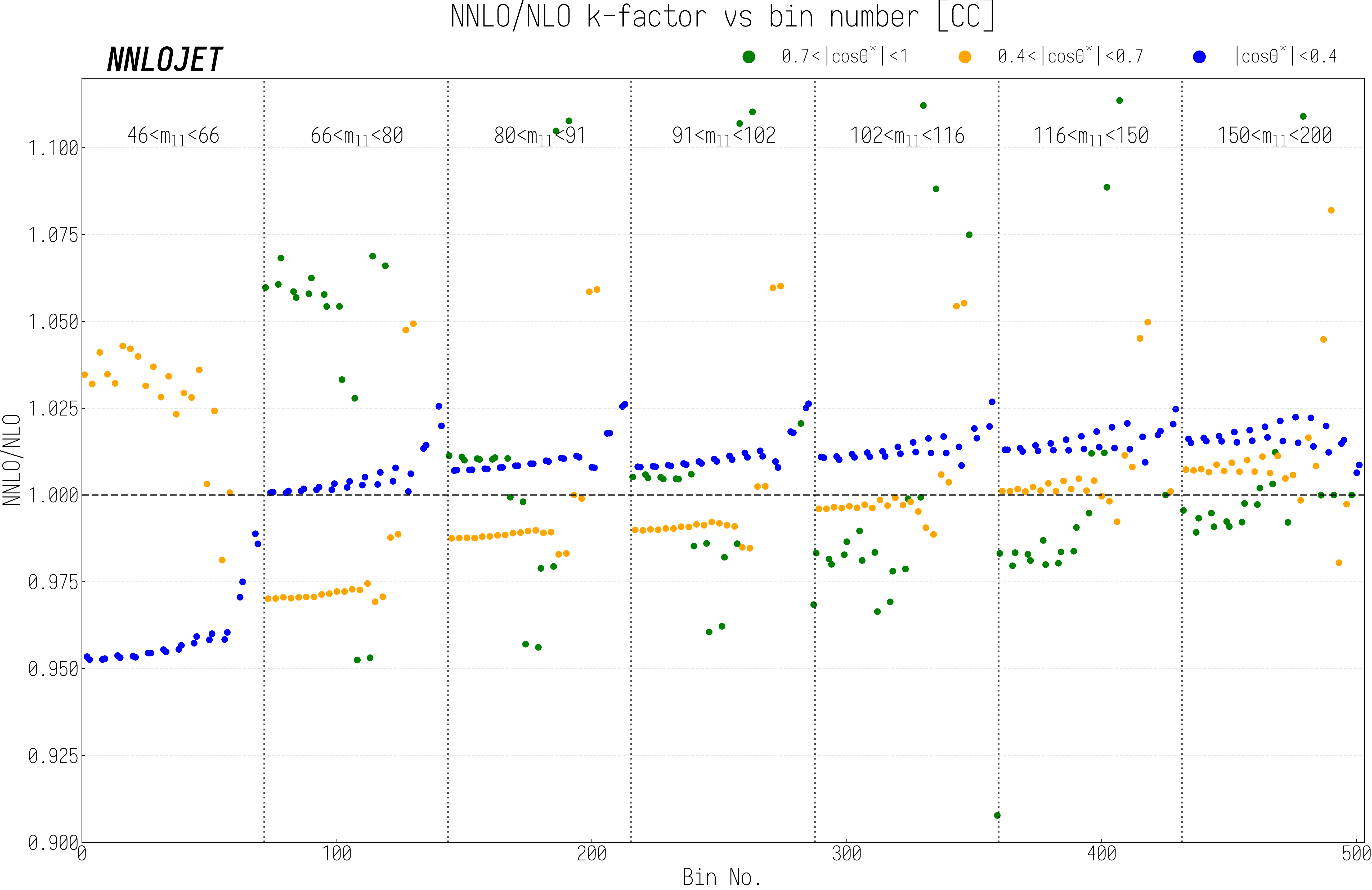}
  \includegraphics[width=1.0\textwidth]{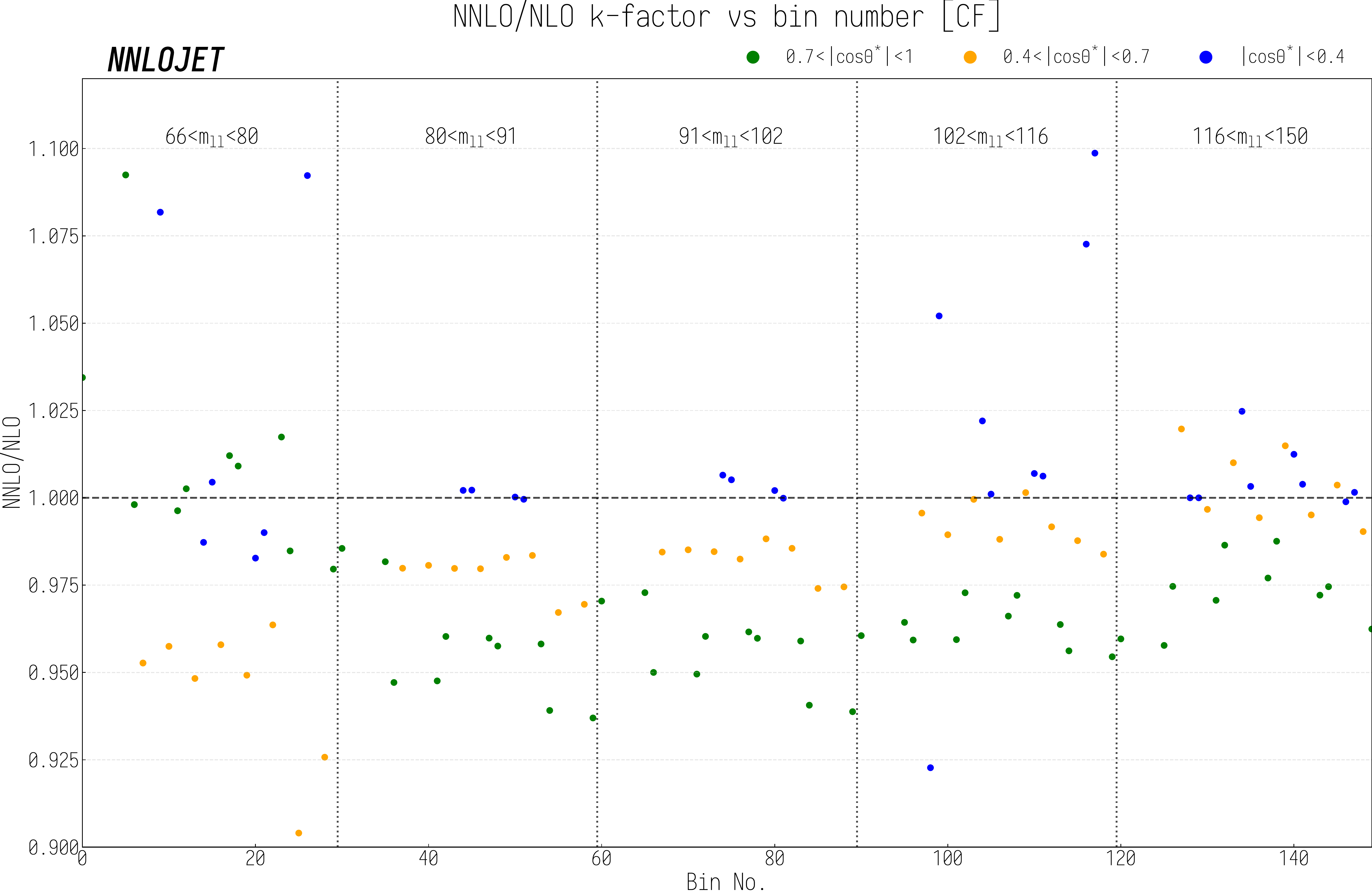}
  \caption{$\NNLO/\NLO$ $k$-factors for the CC and CF Z3D fiducial region. The bin number is
    as defined in Equation~\eqref{eqn:bin_idx_one} for the CC region and
    Equation~\eqref{eqn:bin_idx_CF} for the CF region. The majority of
    $k$-factors for the forbidden region are outside the bounds of the plot.}
  \label{fig:Z3D_kfactors}
\end{figure}

The very substantial variation of the acceptance in the kinematical range of the Z3D measurement reflects the large impact of the cuts on the cross section across phase space.
When using data for extractions of \stweff{} it is important that fiducial cuts do not introduce any systematic bias into the final results.
Thus, it is crucial to be aware of any limitations that low-acceptance regions of phase space might have.
A possibility to mitigate the impact of fiducial cuts in the \stweff~extraction is to impose a cut in the acceptance below which multi-differential asymmetry \afb{} values are constructed.
This strategy considerably reduces the dependence of the result on the definition of the fiducial region due to the relative independence of the cross-section ratios and angular coefficients from the cuts.
This occurs at the cost of a decreased PDF sensitivity as one loses the ability to fit directly at the cross-section level.

To reduce this dependence on the fiducial cuts whilst still keeping the PDF sensitivity, the procedure followed in the ATLAS Z3D analysis is to use an acceptance cut of $\mathcal{O}(95\%)$ whereby cross sections are fitted directly for high acceptance, while for lower acceptance differential \afb~asymmetries are constructed and then fitted.
One can then vary the exact value of the acceptance cut to ensure that the extracted value of \stweff{} is independent of this cut to a predefined level.
From the kinematics, we can then understand the majority of below-cut bins as lying within the forbidden and mixed regions of phase space where they suffer kinematic suppression directly as a result of the cuts applied.

Differential asymmetries are computed as
\begin{equation}
  A(\mll,\yll,|\cthstar|) = \frac{\sigma(\mll,\yll,{\cthstar}_+) - \sigma(\mll,\yll,{\cthstar}_-) }{ \sigma(\mll,\yll,{\cthstar}_+)+ \sigma(\mll,\yll,{\cthstar}_-)} \,,
  \label{eqn:asym}
\end{equation}
where $\sigma(\mll,\yll,{\cthstar}_+)$ and $\sigma(\mll,\yll,{\cthstar}_-)$ are differential cross sections with $\cthstar>0$ and $\cthstar<0$ measured at the same absolute value of $\cthstar$.
The uncertainties on $A(\mll,\yll,|\cthstar|)$ are estimated separately for all correlated and uncorrelated sources following the standard error propagation procedure.
The correlation information among asymmetry and cross-section measurements is preserved.
As a result of this procedure, using the default acceptance cut of $95\%$, the data in the central-central region are represented as $188$ cross section and $128$ asymmetry measurements.
All central-forward data are converted to $56$ asymmetry measurements.
When presented in figures, the differential asymmetries follow the convention defined by Equations~\eqref{eqn:bin_idx_one} and \eqref{eqn:bin_idx_CF}, with the bin number given by the cross section with the negative $\cthstar$.

The use of an acceptance cut also has a secondary effect of ensuring that the theoretical predictions are robust and relatively insensitive to higher-order corrections.
Low-acceptance regions are strongly correlated with large $k$-factors in the theory predictions as the phase-space restrictions become relaxed at higher orders.
This occurs as a result of partonic radiation which generates kinematic configurations inaccessible at lower orders and can be seen in Figure~\ref{fig:Z3D_kfactors}, which shows the $\NNLO/\NLO$ QCD $k$-factors for both the CC and the CF region.

The effect is particularly evident in the forbidden region, where the LO contribution is identically zero, such that the perturbative series effectively begins at $\mathcal{O}(\alphas)$.
As a result, the $\NNLO/\NLO$ ($\mathcal{O}(\alphas^2)/\mathcal{O}(\alphas)$) $k$-factor in these regions only captures the inclusion of the first additional perturbative order, which typically gives $\mathcal{O}(20\%)$ corrections for processes in which a single vector boson is produced.
In Figure~\ref{fig:Z3D_kfactors} this is the case, where the majority of forbidden bins lie outside of the $y$-axis range, corresponding to corrections of a magnitude larger than $\pm 10\%$.
Another problematic feature in the forbidden regions, which will be detailed below in Section~\ref{sect:z3dresults}, is that the theory uncertainty of the \NNLO prediction is considerably larger than for predictions in allowed and partially allowed bins at this order.
\begin{figure}
  \centering
  \includegraphics[width=1\textwidth]{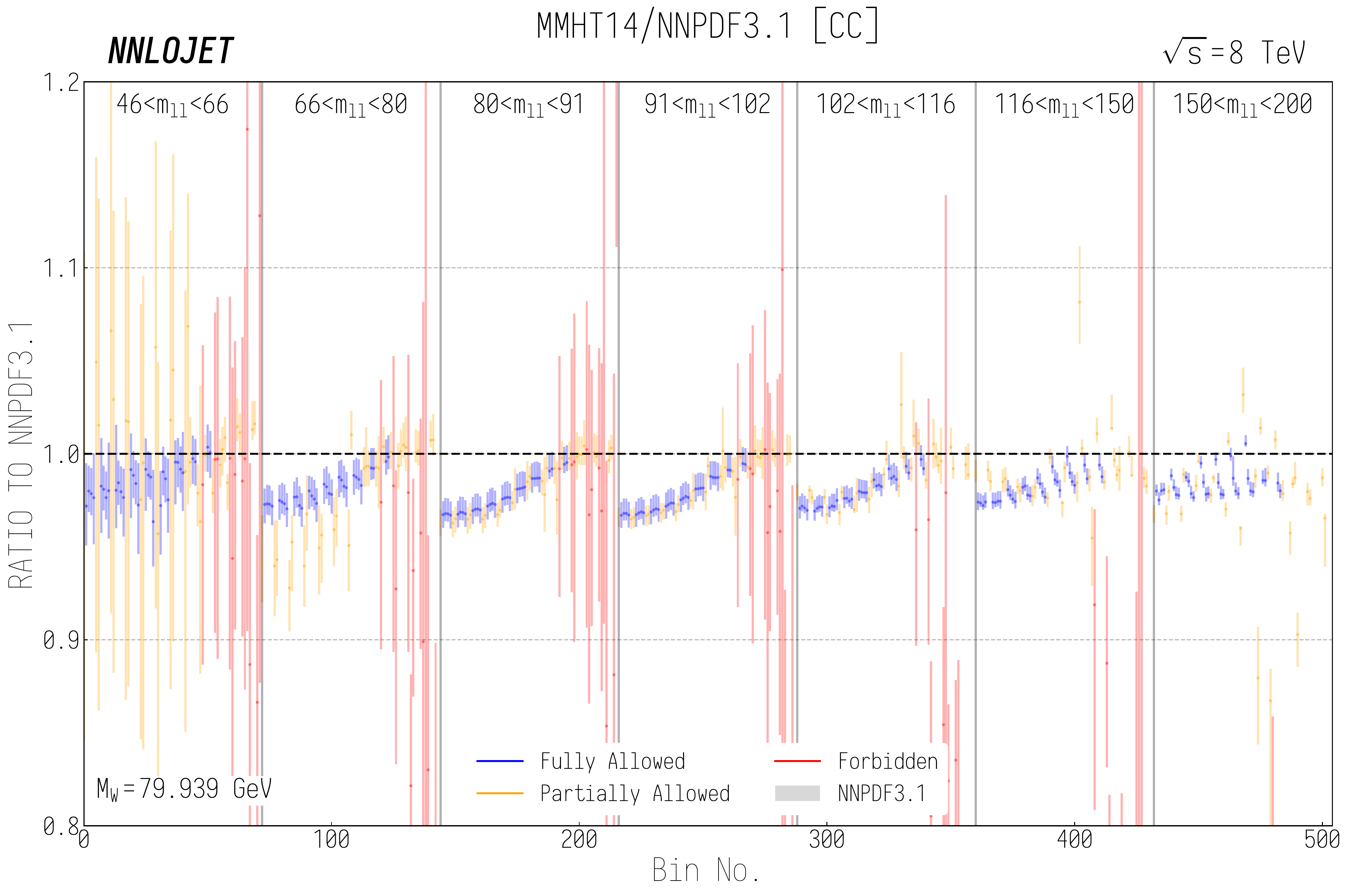}\\
  \includegraphics[width=1\textwidth]{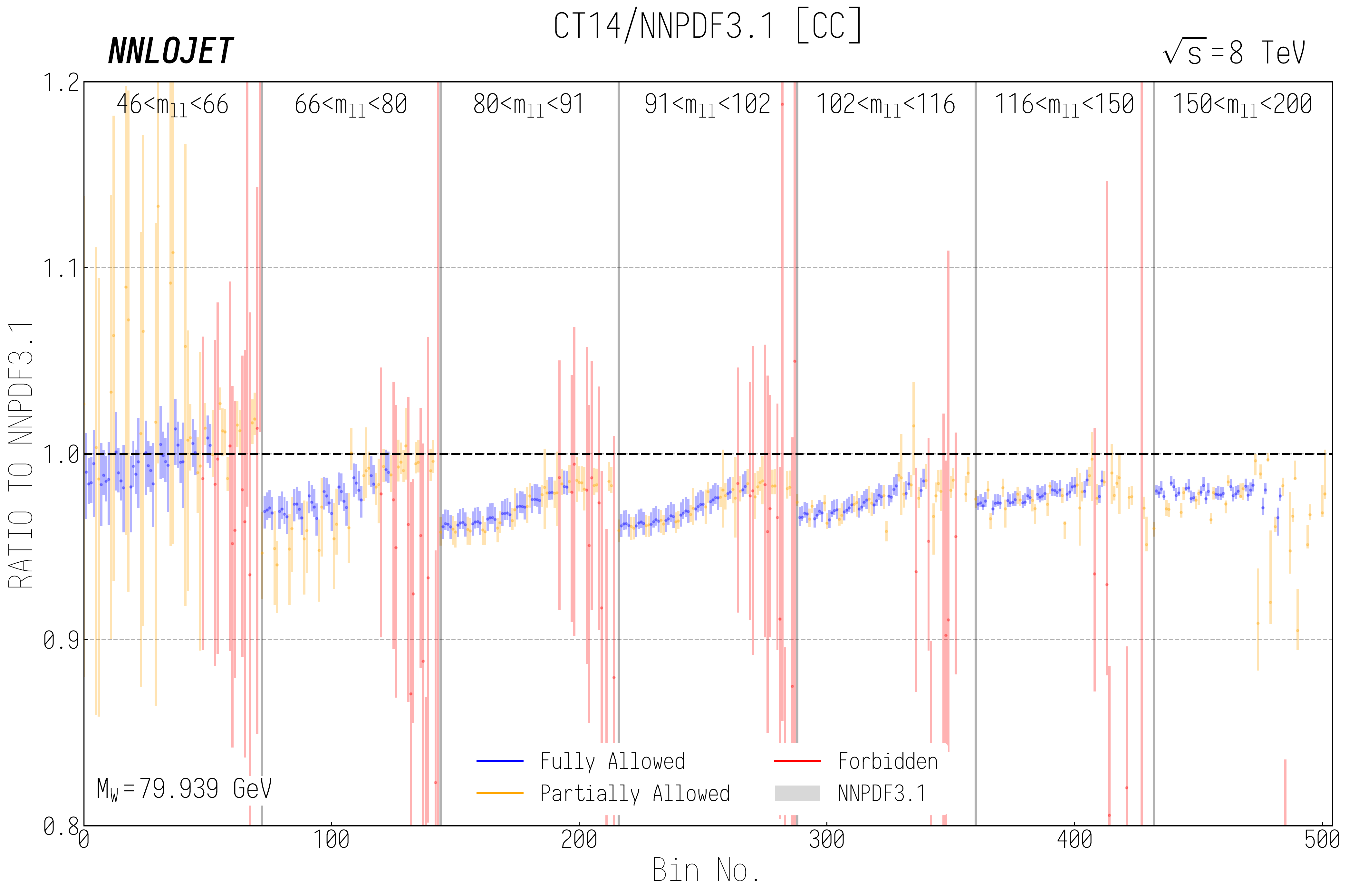}
  \caption{Ratio of the central predictions of the \texttt{MMHT14} (top panel) and \texttt{CT14} (bottom panel) PDF sets to the predictions of \texttt{NNPDF3.1} data in the central-central region of the Z3D analysis.
    Both plots show the predictions in all bins to \NNLO.
    Light error bars on the theory predictions correspond to the scale-variation uncertainty.
    The bin number is as defined in Equation~\eqref{eqn:bin_idx_one}.}
  \label{fig:Z3D_PDF_variation}
\end{figure}

\begin{figure}
  \centering
  \includegraphics[width=0.98\textwidth]{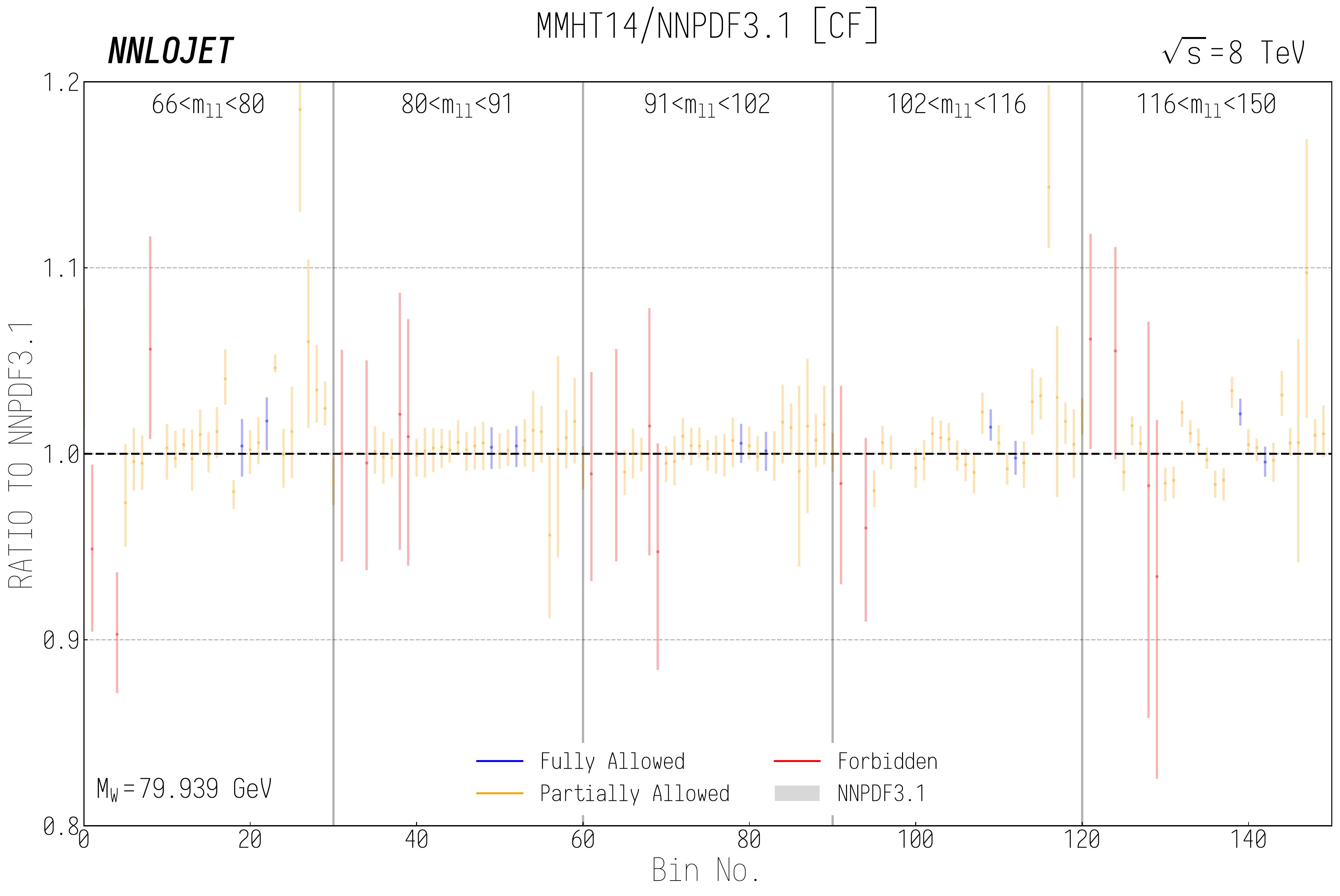}\\
  \includegraphics[width=0.98\textwidth]{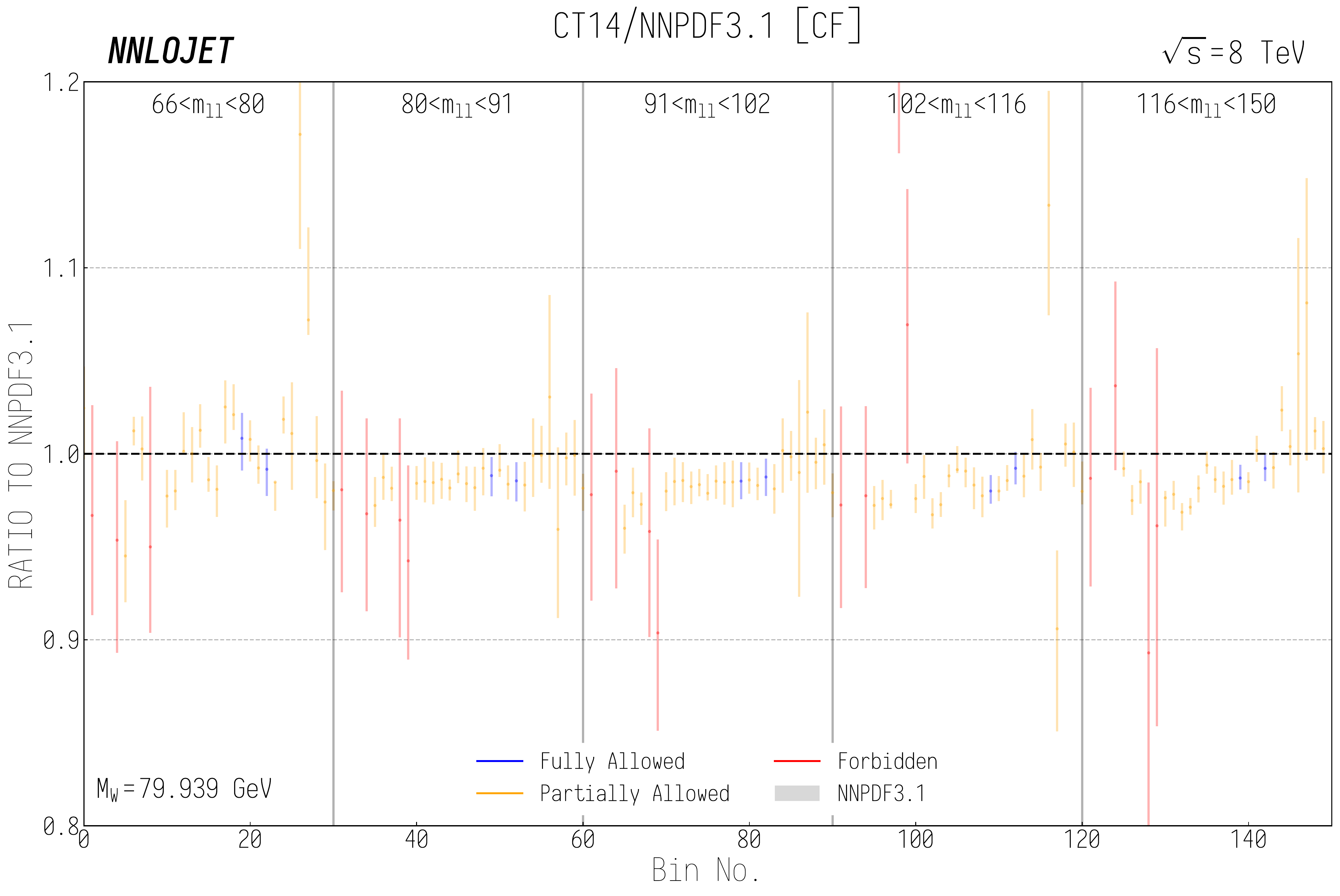}
  \caption{The ratio of the central predictions of the \texttt{MMHT14} (top panel) and \texttt{CT14} (bottom panel) PDF sets to the predictions of \texttt{NNPDF3.1} data in the central-forward region of the Z3D analysis.
    Both plots show the predictions in all bins to \NNLO.
    Light error bars on the theory predictions correspond to the scale-variation uncertainty.
    The bin number is as defined in Equation~\eqref{eqn:bin_idx_CF}.}
  \label{fig:Z3D_PDF_variation_CF}
\end{figure}

\section{PDF Variation}
\label{sect:pdf_variation}
Before discussing complete \NNLO predictions and comparing it to experimental data, we turn to the effects of varying the PDF choice.
PDFs constitute the dominant theoretical uncertainty in the \stweff~extraction, but it is also interesting to study the constraining power of the ATLAS data on the PDFs themselves.
We show in Figures~\ref{fig:Z3D_PDF_variation}--\ref{fig:Z3D_PDF_variation_CF} the ratios of the central members of the \texttt{MMHT14}~\cite{Harland-Lang:2014zoa} and \texttt{CT14}~\cite{Dulat:2015mca} PDF sets to our benchmark \texttt{NNPDF3.1}~\cite{Ball:2017nwa} results in the central-central region and central-forward regions.
The comparison between different PDF sets is primarily representative of methodological differences between the PDF fitting collaborations, incorporating effects due to fitting procedures, parametrisations, experimental data sets, input theory, and so on.
The PDF variation is largely uniform in \mll, such that the characteristic \mll-dependence of the \stweff~variation can potentially be exploited to disentangle the PDF effects in its measurement~\cite{Bodek:2016olg}.

\begin{figure}
  \centering
  \includegraphics[width=1\textwidth]{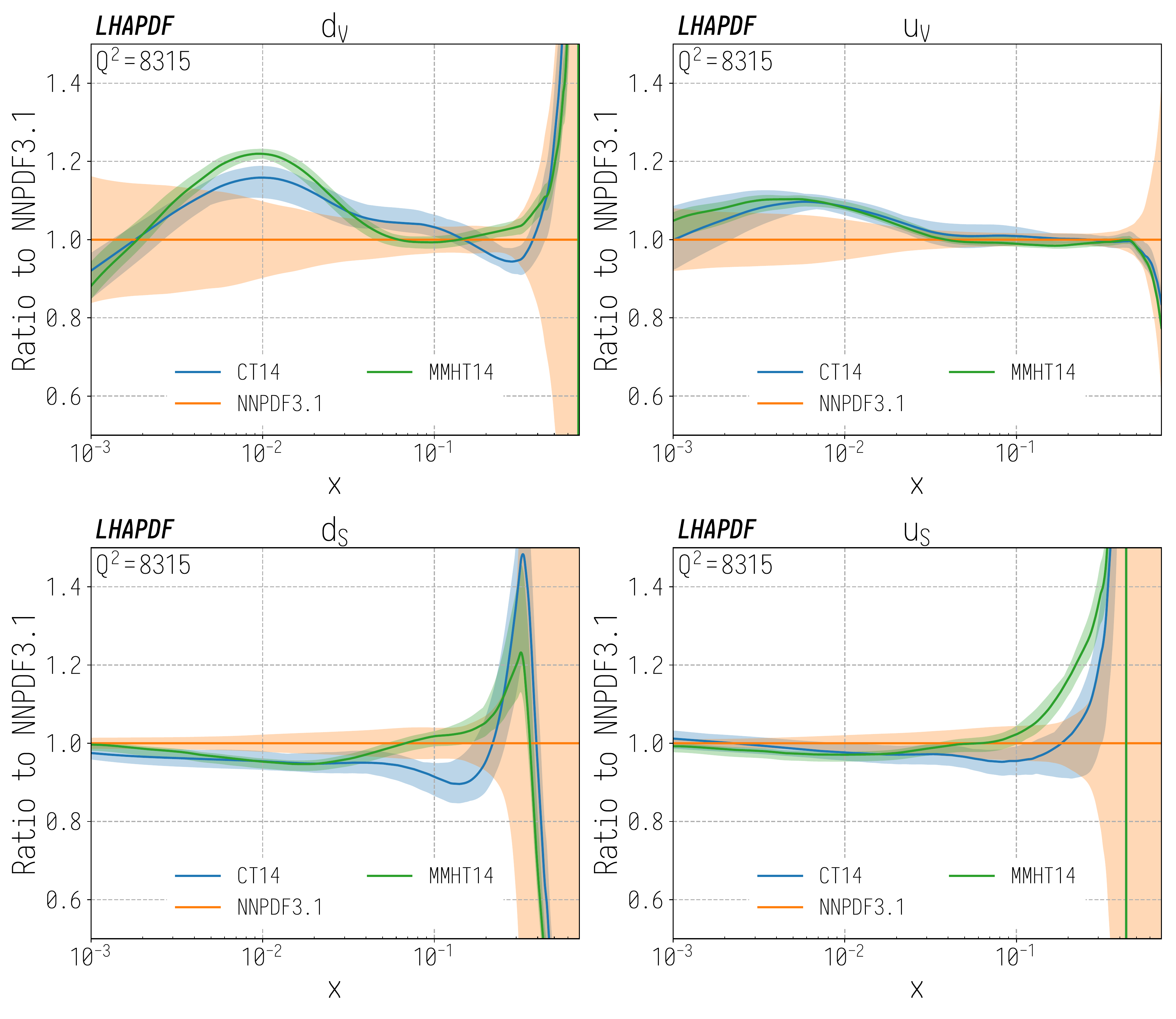}
  \caption{The ratio of \texttt{MMHT14} and \texttt{CT14} PDF sets to the central member of the \texttt{NNPDF3.1} set as a function of Bjorken-$x$.
    From the top left panel clockwise, the panels correspond to the \Pqd valence, \Pqu valence, \Pqu sea-quark, and \Pqd sea-quark
    contributions to the various PDF sets at $Q^2=\mz^2$.
    The given uncertainties are the PDF uncertainties.}
  \label{fig:Z3D_PDF_Bjorken_x}
\end{figure}

For the central members of both \texttt{MMHT14} and \texttt{CT14} sets we see a shape difference across the variation in rapidity, with central rapidities showing a $\mathcal{O}(3\text{--}4\%)$ difference with respect to the \texttt{NNPDF3.1} results, which decreases to $\mathcal{O}(1\text{--}2\%)$ in the forward regions of the measurement.
This can be interpreted primarily as the impact of different sea and valence quark distributions between the three sets, given the dominant incoming partonic sub-process in Drell-Yan production is quark-antiquark annihilation.
These are analysed in more detail in~\cite{Ball:2017nwa}, where the primary driver of differences between the sets occurs in the anti-quark distributions at $Q\sim\mz$, and are visualised in Figure~\ref{fig:Z3D_PDF_Bjorken_x}.
These effects are less pronounced at high $\yll\leftrightarrow x$, where the valence quark contributions dominate over the sea quark and the central members of the PDF sets exhibit better agreement.

This comparison between PDF sets does not, however, account for uncertainties \textit{within} each PDF set, which are parametrised through $\mathcal{O}(30\text{--}100)$ additional Hessian eigenvector or replica sets.
In order to evaluate these using standard \NNLO techniques, one must perform a separate \NNLO calculation for each set member.
Whilst technically possible it is prohibitively expensive computationally.
At \NLO, grid techniques are a well-established solution for dealing with this issue, where the PDF dependence of the (differential) cross section is stored using look-up tables which allow for \textit{a posteriori} convolutions with any PDF~\cite{Carli:2010rw,Britzger:2012bs}.

Whilst grid technologies are being extended at \NNLO for certain processes~\cite{Czakon:2017dip,Britzger:2019kkb,Britzger:2022lbf}, results are not yet widespread and largely still in development.
Standard practise is to reweight NLO results for PDF variation obtained using these look-up tables with $\NNLO/\NLO$ $k$-factors, a technique which is also used within the fitting of the PDFs themselves.
The closure of this method can be checked using dedicated \NNLO runs, either for specific members of a single PDF set, or for central members of different sets, where good agreement is generally found~\cite{Campbell:2019dru}.
The PDF-error uncertainties for each of the different sets used in Figures~\ref{fig:Z3D_PDF_variation}--\ref{fig:Z3D_PDF_variation_CF} have been evaluated in this manner, and they are large enough to accommodate the differences between PDF sets.
As a result, the results produced using the \texttt{MMHT14}, \texttt{CT14} and \texttt{NNPDF3.1} central members are not mutually inconsistent.

\section{Combined \QCD + \EW Predictions}
\label{sect:z3dresults}
In this section, we present a set of fixed-order theory predictions for the Z3D measurement and contrast it with experimental data from ATLAS \cite{Aaboud:2017ffb}.
Results are obtained using the setup described in Section~\ref{sect:numerical_setup} and entail $\mathcal{O}(\alphas^2)$ (\NNLO) \QCD as well as $\mathcal{O}(\alpha)$ (\NLO) \EW and universal \HO \EW corrections at $\mathcal{O}(\alpha^2)$.
In forbidden bins, \NNLO \QCD predictions are supplemented with partial $\mathcal{O}(\alphas^3)$ (\pNthreeLO) \QCD corrections in order to promote these bins to an effective \NNLO accuracy.
Based on the findings of Section~\ref{sect:pdf_variation}, all results are produced using the NNPDF3.1 PDF set.
Before showing the final combined $(\NNLO+\pNthreeLO)_\QCD + (\NLO+\HO)_\EW$ predictions in Section~\ref{sect:combined_predictions}, the effect of including partial \NthreeLO \QCD and \NLO plus \HO \EW corrections is discussed separately in Sections~\ref{sect:partial_nthreelo_corrections} and \ref{sect:nlo_and_ho_ew_corrections}, respectively.

\begin{figure}
  \centering
  \includegraphics[width=0.96\textwidth]{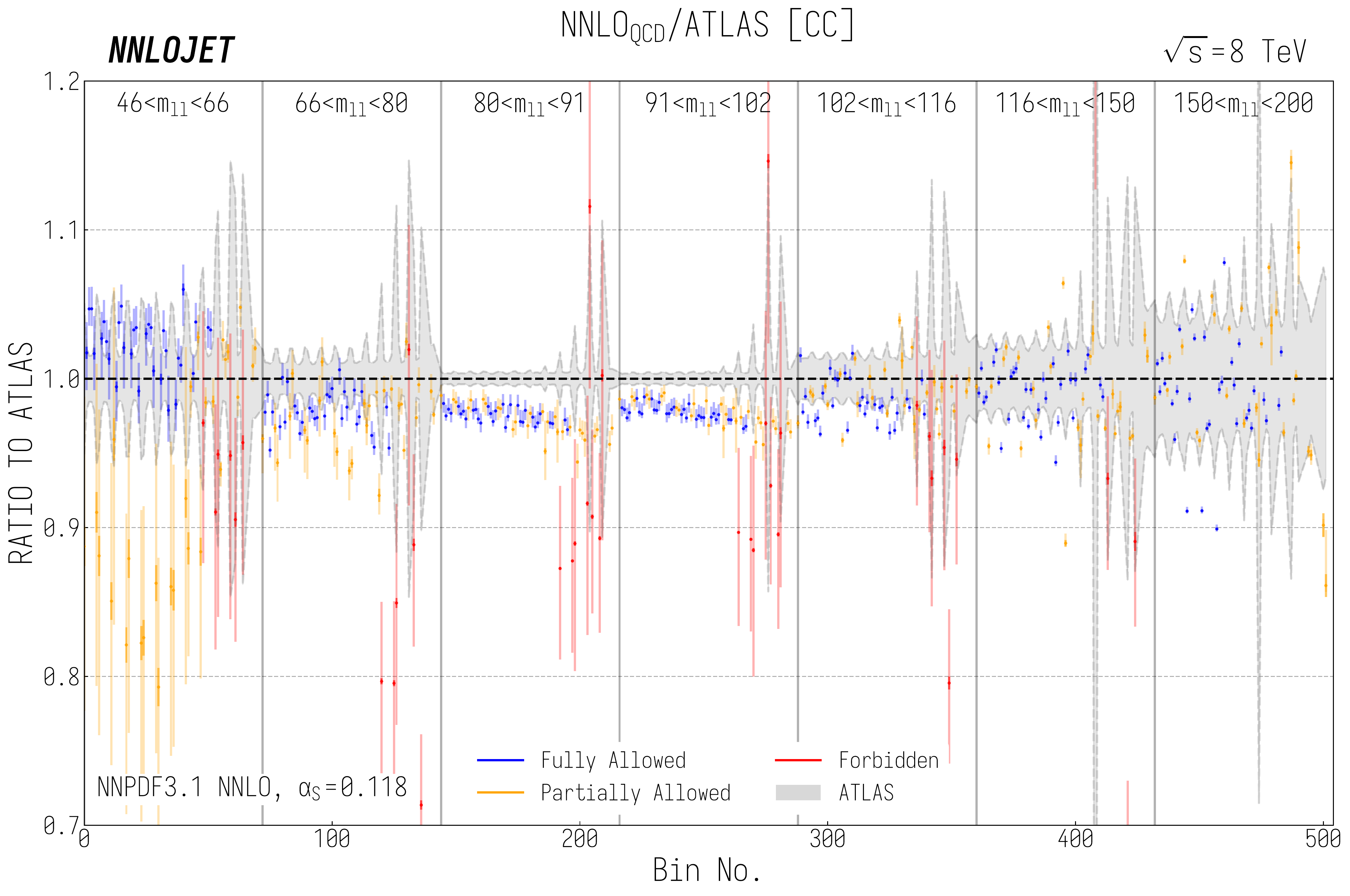}\\
  \includegraphics[width=0.96\textwidth]{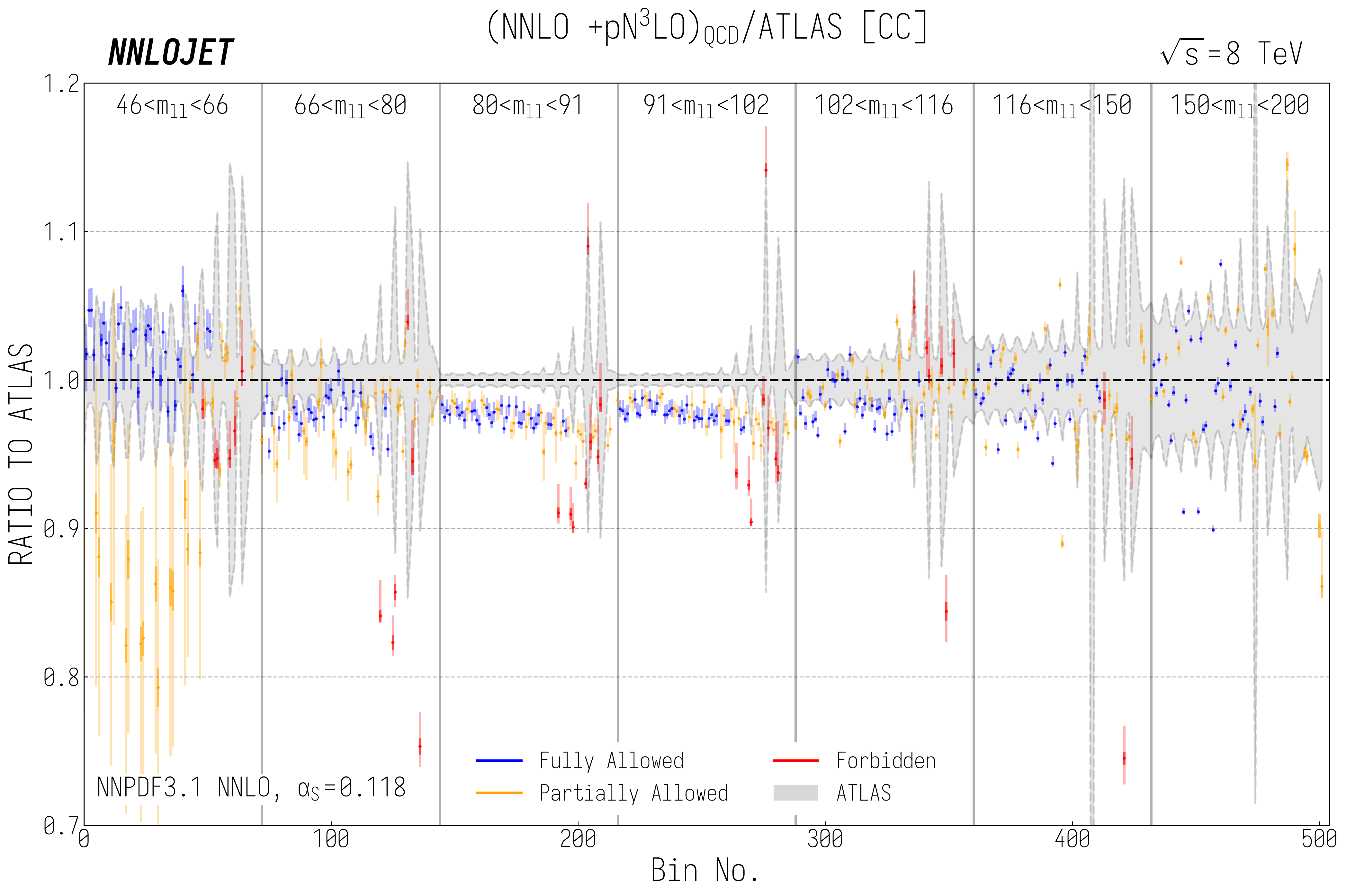}
  \caption{Ratio of \NNLO \QCD predictions to ATLAS data in
    the central-central region of the Z3D analysis. The upper plot
    shows the \NNLO \QCD theory predictions, while
    the lower plot includes \pNthreeLO \QCD contributions for
    forbidden bins. Light error bars indicate scale-variation
    uncertainties and dark error bars correspond to statistical
    uncertainties, whilst the grey shaded area shows the experimental uncertainties
    in the ATLAS measurement. The bin number is as defined in Equation~\eqref{eqn:bin_idx_one}.}
  \label{fig:Z3D_ratio_to_data_CC}
\end{figure}

\begin{figure}
  \centering
  \includegraphics[width=0.96\textwidth]{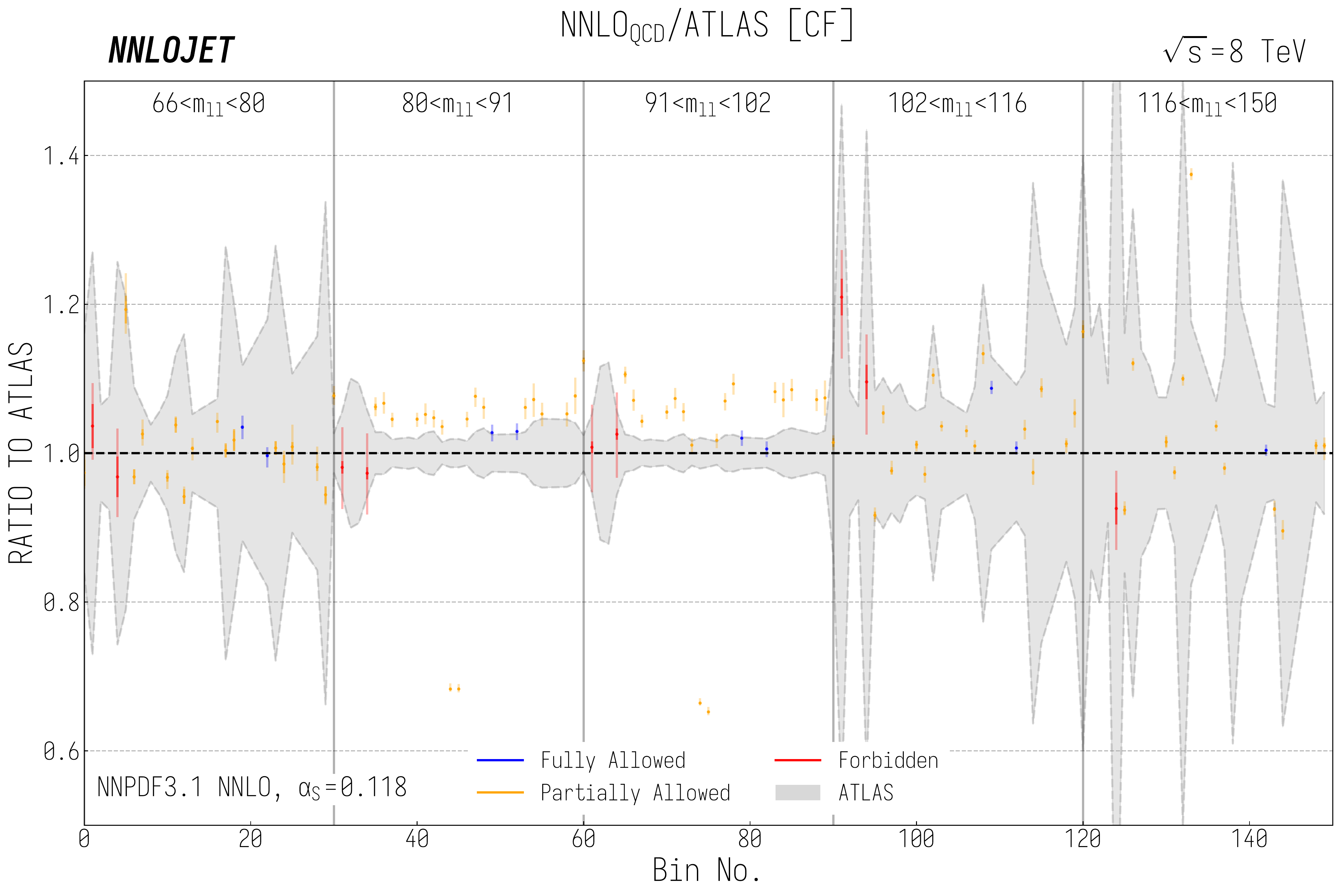}\\
  \includegraphics[width=0.96\textwidth]{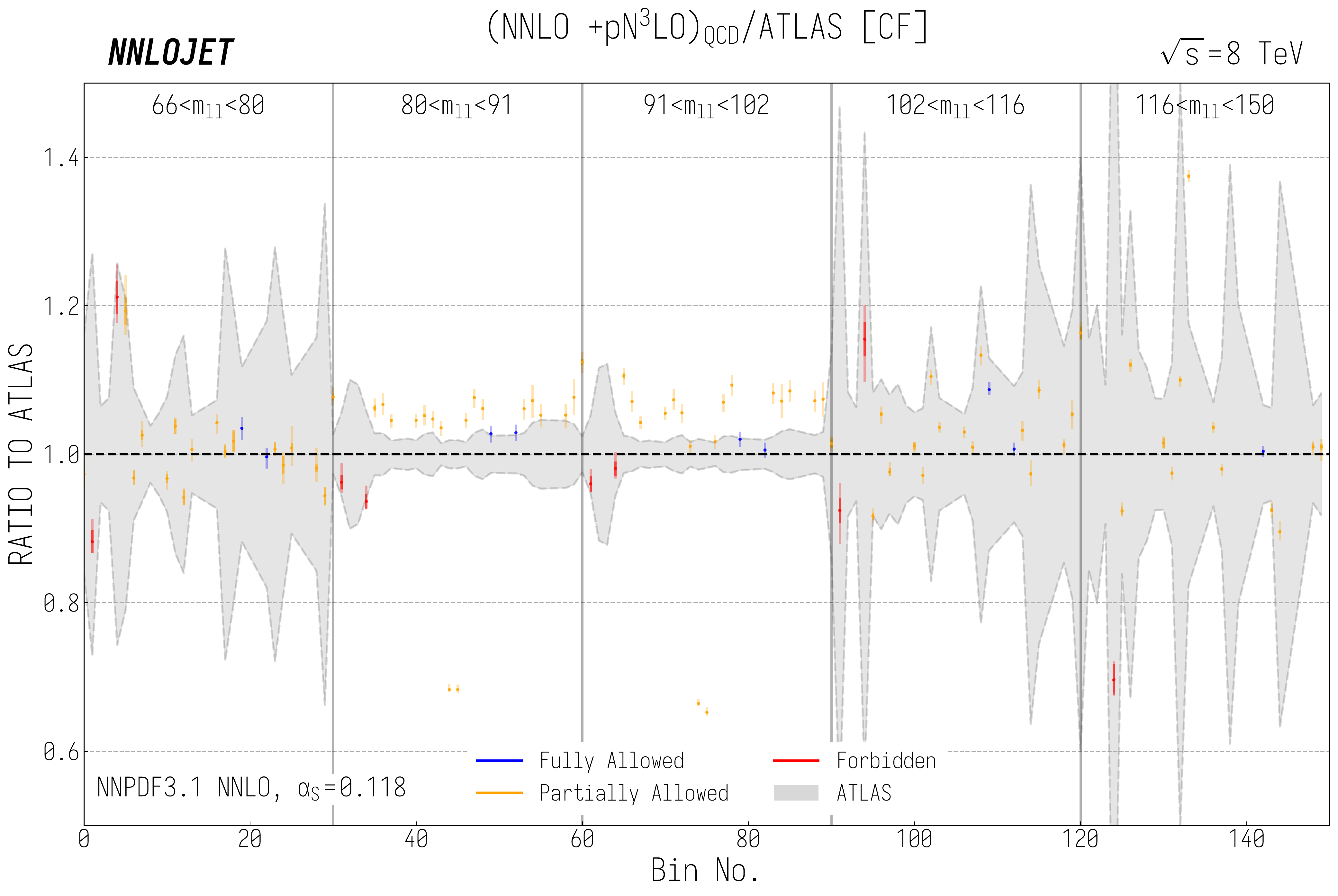}
  \caption{Ratio of \NNLO \QCD predictions to ATLAS data in
    the central-forward region of the Z3D analysis. The upper plot
    shows the \NNLO \QCD theory predictions, while
    the lower plot includes \pNthreeLO \QCD contributions for
    forbidden bins. Light error bars indicate scale-variation
    uncertainties and dark error bars correspond to statistical
    uncertainties, whilst the grey shaded area shows the experimental uncertainties
    in the ATLAS measurement. The bin number is as defined in Equation~\eqref{eqn:bin_idx_CF}.}
  \label{fig:Z3D_ratio_to_data_CF}
\end{figure}

\subsection{Partial \NthreeLO \QCD Corrections}
\label{sect:partial_nthreelo_corrections}
As discussed in Section~\ref{sect:better_than_born}, large theory uncertainties are present in the forbidden region.
These can be remedied by the inclusion of known \NthreeLO \QCD corrections.
We include these corrections by exploiting the existing \NNLO{} \PZJ~calculation in \nnlojet~\cite{Ridder:2016nkl,Ridder:2015dxa}.
The effect of these corrections in the CC and CF region is shown in Figures~\ref{fig:Z3D_ratio_to_data_CC} and \ref{fig:Z3D_ratio_to_data_CF}, respectively, which display the theory predictions (with scale-variation uncertainties) with and without \pNthreeLO corrections to the forbidden bins.
All predictions are normalised to the ATLAS Z3D data.

In the CC region, the \pNthreeLO corrections stabilise the predictions in the forbidden bins and considerably reduce scale-variation uncertainties.
Notably, scale-variation uncertainties are much larger at low invariant masses than at the high invariant masses.
This can be traced back to a steeper gradient of the running of $\alphas$ in this region and a larger factorisation-scale dependence of the PDFs at low scales.
This is particularly evident in the lowest \mll~bin, where the scale uncertainties in the partially-allowed bins are $\mathcal{O}(10\%)$, even at \NNLO.
The inclusion of the \pNthreeLO terms makes the largest impact in this phase-space region, to the point that the scale variation uncertainty in the forbidden bins becomes smaller than in the corresponding fully- and partially-allowed bins.
From this it is reasonable to conclude that, once available, the full triply-differential \NthreeLO \QCD Drell-Yan results will give the largest improvement at low \mll.

For the CF region, the uncertainty on the data is considerably larger than in the CC region.
We see a reasonable description of the data by the theory predictions, albeit with a slight overshoot in the region around the \PZ~pole.
The inclusion of the \pNthreeLO corrections to the forbidden bins again substantially decreases the scale uncertainty in the associated bins.
Due to the extreme kinematic suppression at large \mll, statistical uncertainties of the theory prediction are considerable there.
They are nevertheless sufficient for fitting purposes, as these bins typically also encounter very large experimental uncertainties.

If we now consider the qualitative agreement of the pure QCD predictions with ATLAS data, we see that the theory prediction undershoots the data by $2-3\%$ in the region of the \PZ~pole in the CC region.
In the CF region, theory predictions are instead about $10\%$ larger than the ATLAS data in the two \mll~bins around the Z pole.
This discrepancy can be traced back to two primary causes.
The first is the absence of \NLO \EW corrections, which will be discussed below; the second is the luminosity uncertainty (neglected in the figures), which is estimated to be around $\mathcal{O}(1.8\%)$ and is correlated across all bins.

\begin{figure}
  \centering
  \includegraphics[width=0.95\textwidth]{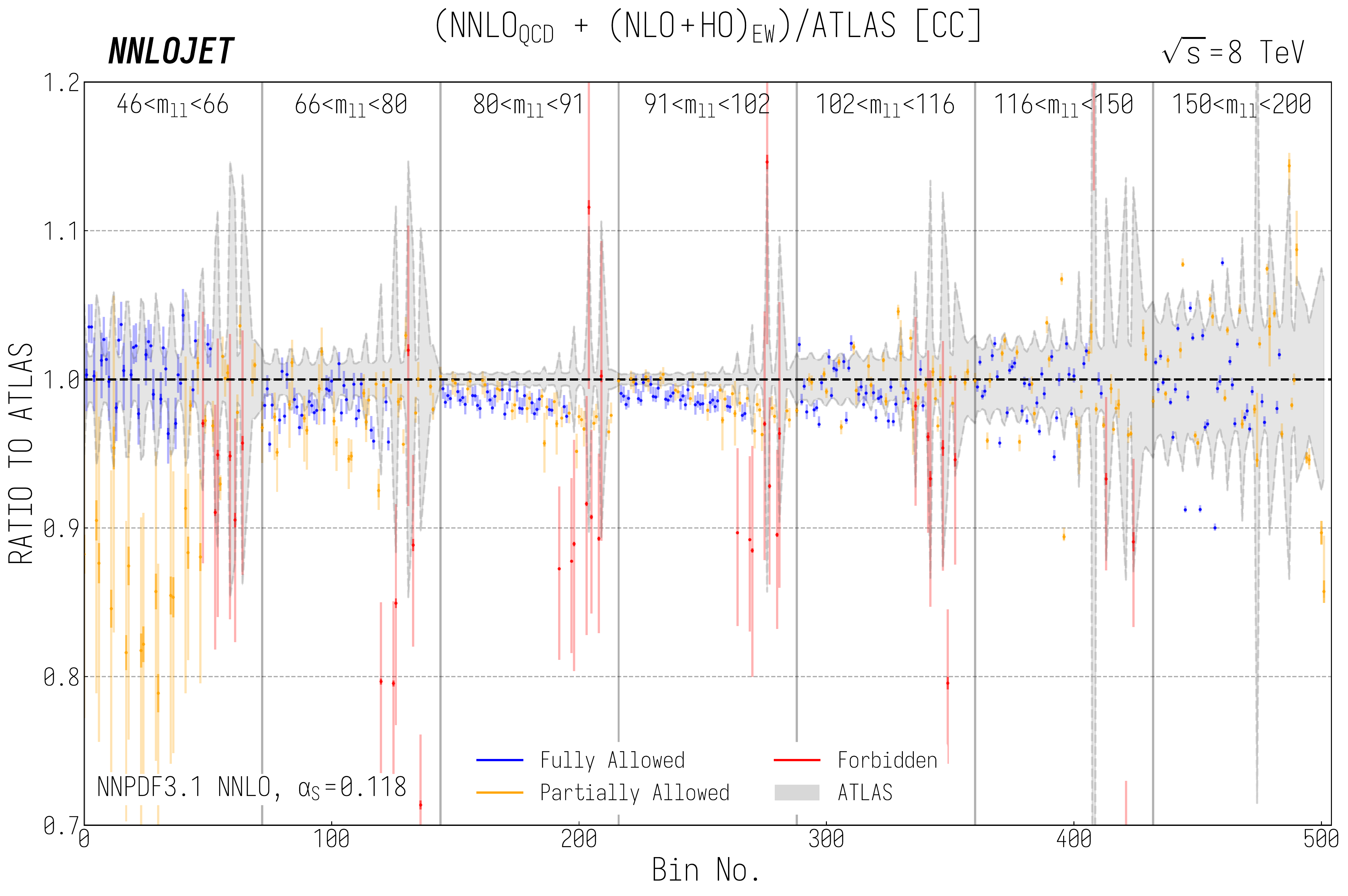} \\
  \includegraphics[width=0.95\textwidth]{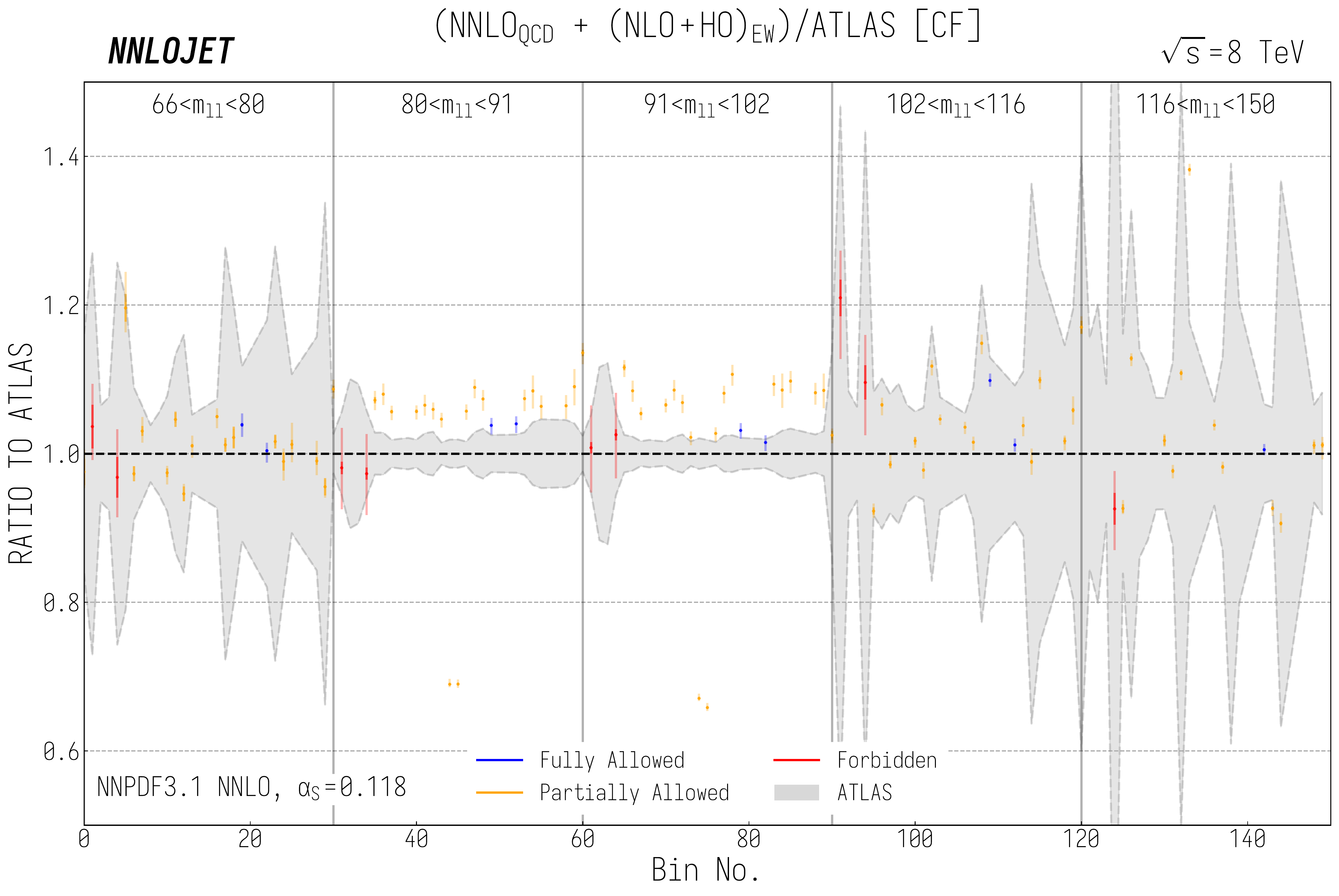}
  \caption{Ratio of $\NNLO_\QCD + (\NLO+\HO)_\EW$ predictions to ATLAS data in the central-central (top panel) and central-forward (bottom panel) region of the Z3D analysis.
    Light error bars correspond to scale-variation uncertainties and dark error bars correspond to statistical uncertainties. The grey shaded region shows the experimental uncertainty of the ATLAS measurement.
    The bin number is as defined in Equation~\eqref{eqn:bin_idx_one} for the CC region and Equation~\eqref{eqn:bin_idx_CF} for the CF region.}
  \label{fig:Z3D_ratio_to_data_NLOEW}
\end{figure}

\subsection{\NLO and Higher-Order EW Corrections}
\label{sect:nlo_and_ho_ew_corrections}
We now turn to the discussion of the importance of \EW corrections for the triple-differential DY measurement.
To this end, we include \NLO as well as some \HO \EW corrections, obtained with \powhegbox as described in Section~\ref{sect:numerical_setup}.
We have validated the equivalence of the \powhegbox and \nnlojet results at LO and found excellent agreement.
The \NNLO \QCD predictions obtained with \nnlojet are amended by purely-weak virtual corrections from \powhegbox according to
\begin{equation}
  \diff\sigma_{\tiny{\NNLO_\QCD+(\NLO+\HO)_\EW}} = \diff\sigma^\text{\nnlojet}_{\tiny{\NNLO_\QCD}} + \diff\sigma^\text{\powhegbox}_{\tiny{(\NLO+\HO)^\text{virt}_\EW}} \, ,
\end{equation}
where the purely-weak virtual corrections, including the higher-order corrections to $\Delta\alpha$ and $\Delta\rho$, are contained in the second term on the right-hand side.
As we include only weak corrections with Born-like kinematics, they can only affect allowed and partially-allowed bins.
To some degree, the effect of these corrections can therefore be understood to be complementary to the one of \pNthreeLO \QCD corrections discussed in the previous subsection.

The ratio of the $\NNLO_\QCD+(\NLO+\HO)_\EW$ predictions to ATLAS data for the CC and CF region is shown in Figure \ref{fig:Z3D_ratio_to_data_NLOEW}.
For the CC region, it is evident that, for allowed and partially-allowed bins, the inclusion of electroweak corrections substantially improves the agreement with data in the central \mll{} bins around the \PZ~pole compared to the pure \NNLO \QCD predictions shown in the top panel of Figure~\ref{fig:Z3D_ratio_to_data_CC}.
For most bins, the remaining discrepancy is around $1\%$ and covered by the experimental and theoretical uncertainties.
In the CF region, the effect is adverse, with \EW corrections slightly increasing the discrepancy between theory predictions and experimental data around the \PZ~pole compared to the pure \NNLO \QCD results shown in the top panel of Figure~\ref{fig:Z3D_ratio_to_data_CF}.
As expected, no improvement can be seen for forbidden bins in either of the two regions, as these are populated only by real radiation, which is omitted in our calculation of \EW corrections, because the data had already been corrected for photon-emission effects.

We wish to close by highlighting the importance of the \HO \EW corrections to the stabilisation of the triple-differential cross sections.
The inclusion of weak one-loop corrections alone is not sufficient to improve the agreement with data in the \mll{} bins around the \PZ~pole.

\begin{figure}
  \centering
  \includegraphics[width=0.95\textwidth]{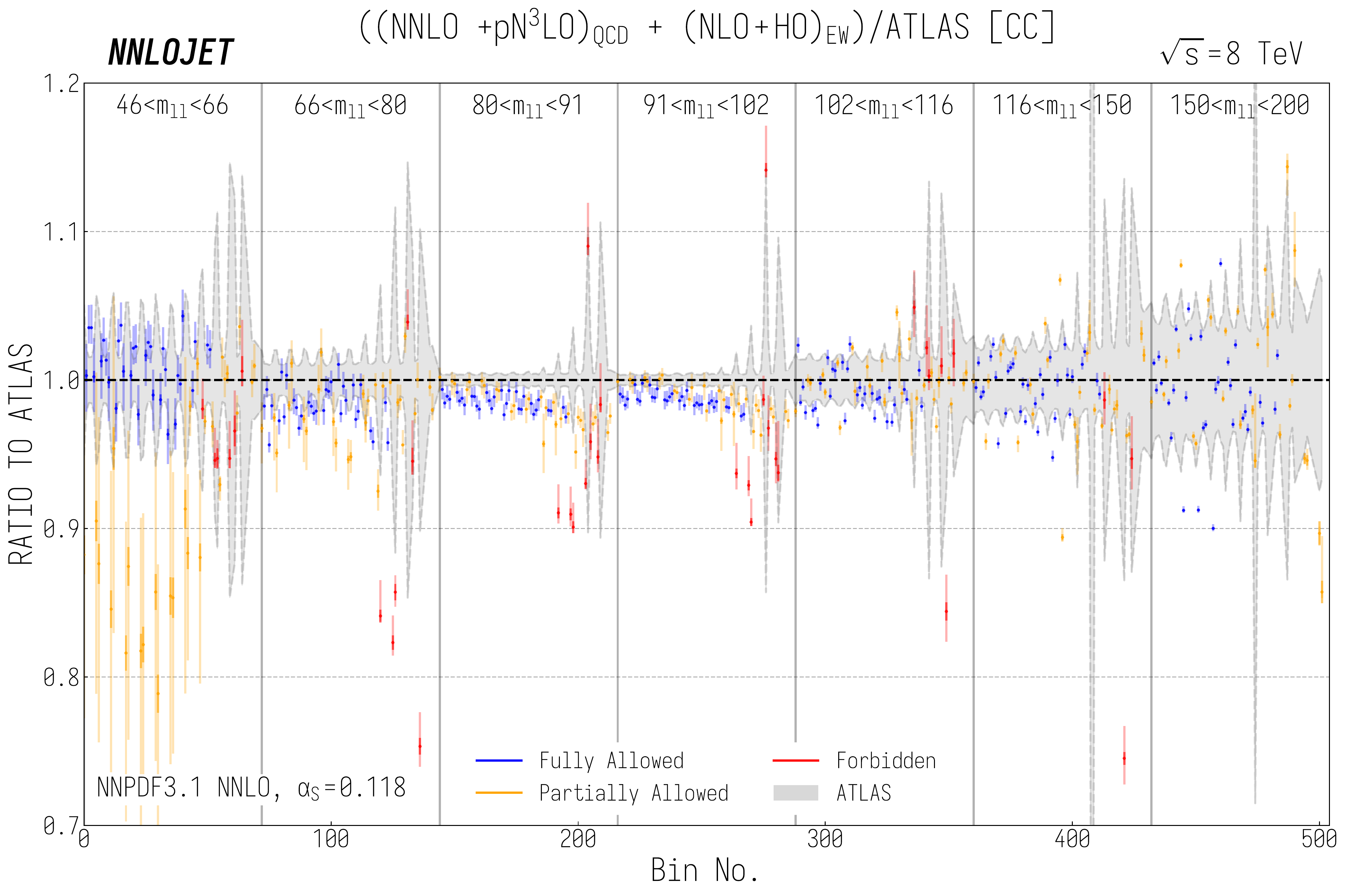} \\
  \includegraphics[width=0.95\textwidth]{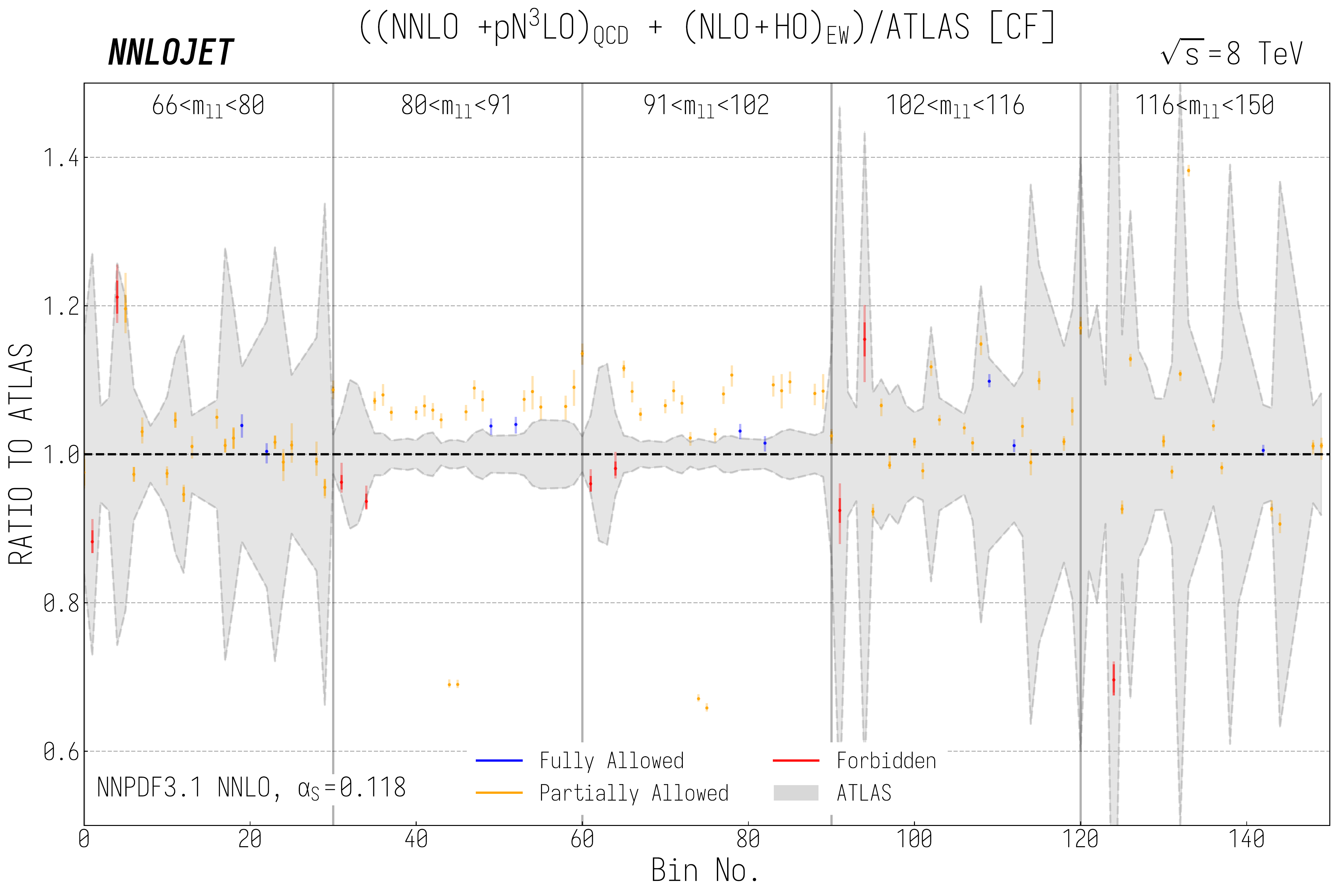}
  \caption{Ratio of $(\NNLO+\pNthreeLO)_\QCD + (\NLO+\HO)_\EW$ predictions to ATLAS data in the central-central (top panel) and central-forward (bottom panel) region of the Z3D analysis.
    Light error bars correspond to scale-variation uncertainties and dark error bars correspond to statistical uncertainties. The grey shaded region shows the experimental uncertainty of the ATLAS measurement.
    The bin number is as defined in Equation~\eqref{eqn:bin_idx_one} for the CC region and Equation~\eqref{eqn:bin_idx_CF} for the CF region.}
  \label{fig:Z3D_ratio_to_data_best}
\end{figure}

\begin{figure}
  \centering
  \includegraphics[width=0.4\textwidth]{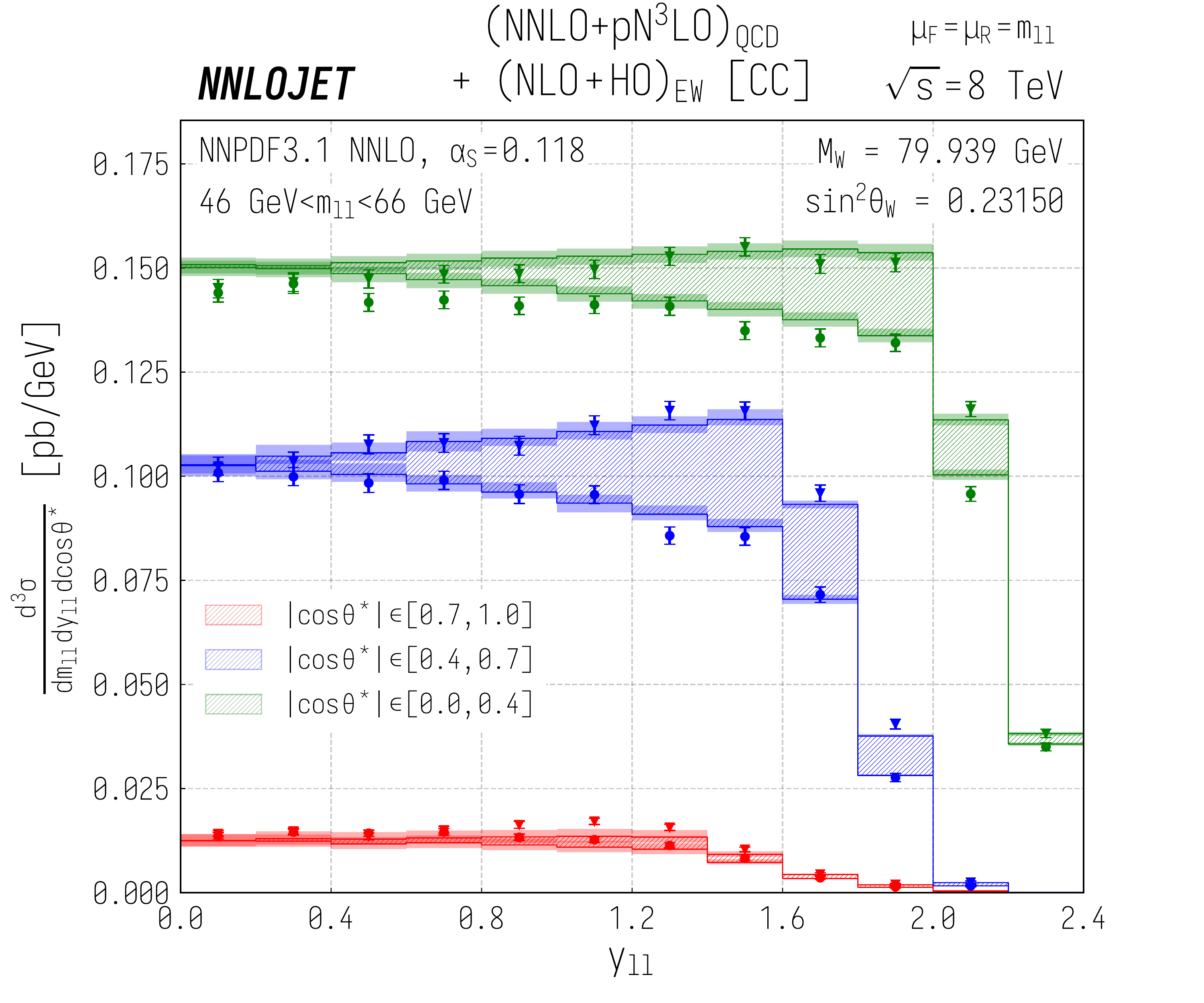}
  \includegraphics[width=0.4\textwidth]{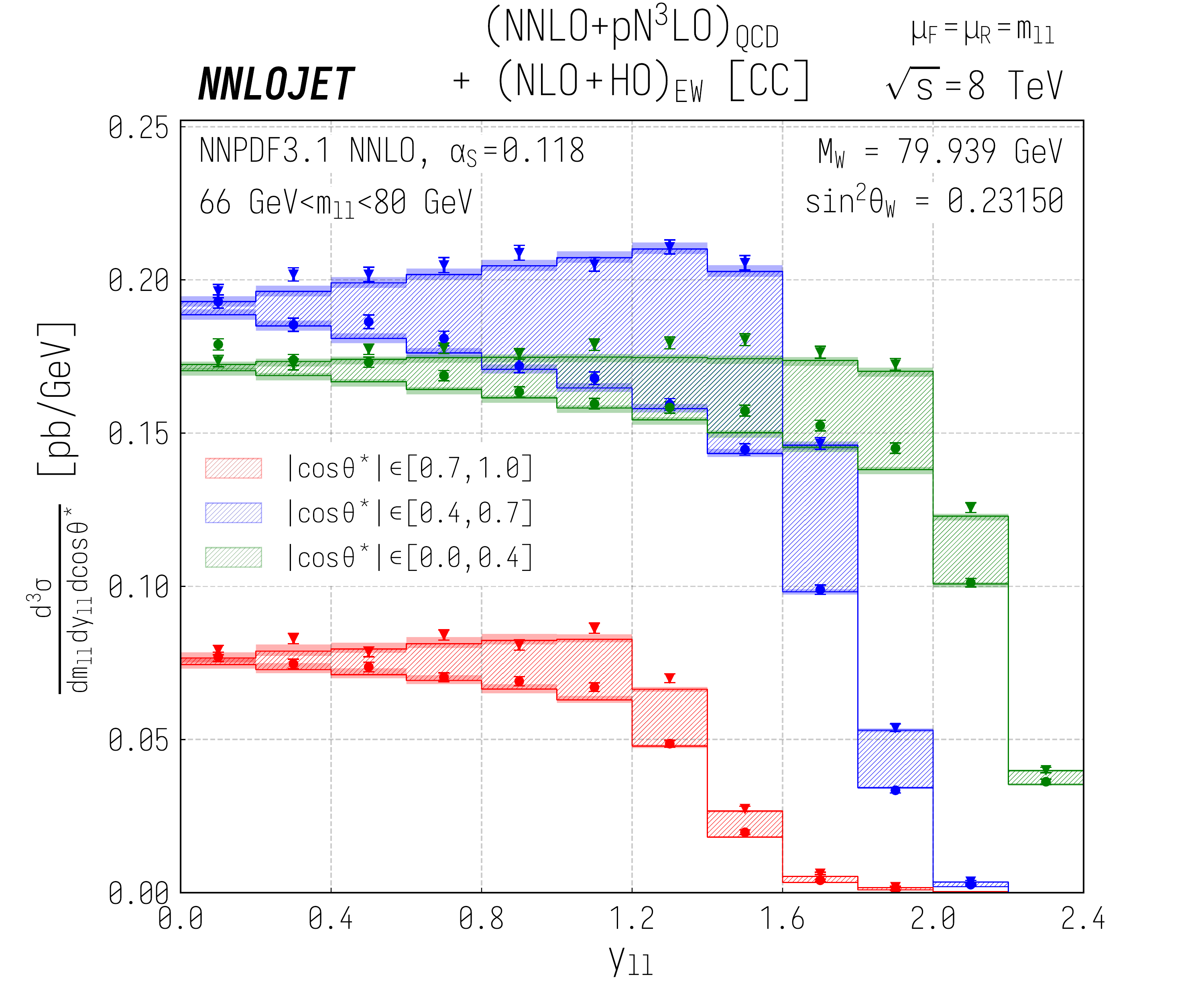} \\
  \includegraphics[width=0.4\textwidth]{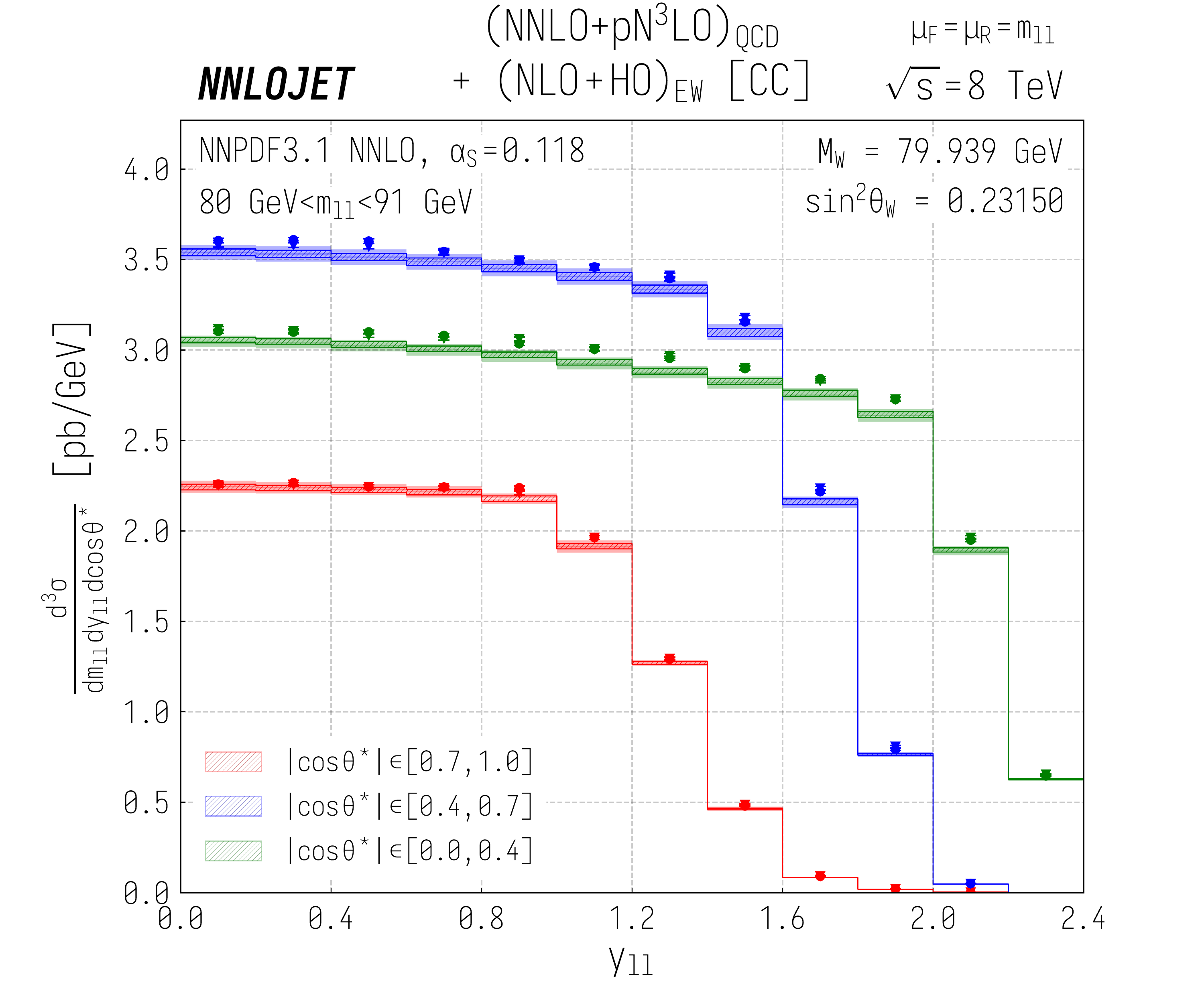}
  \includegraphics[width=0.4\textwidth]{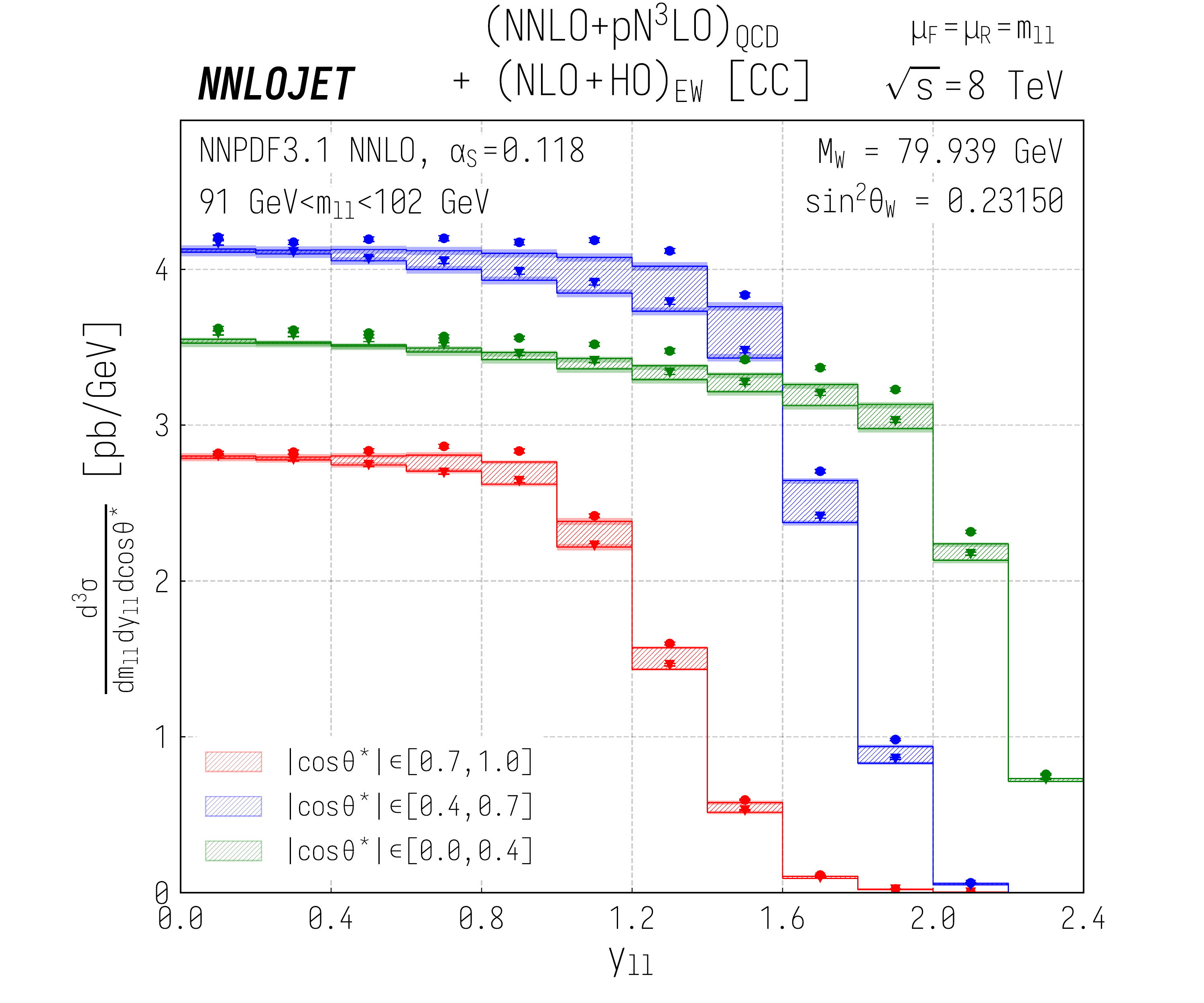} \\
  \includegraphics[width=0.4\textwidth]{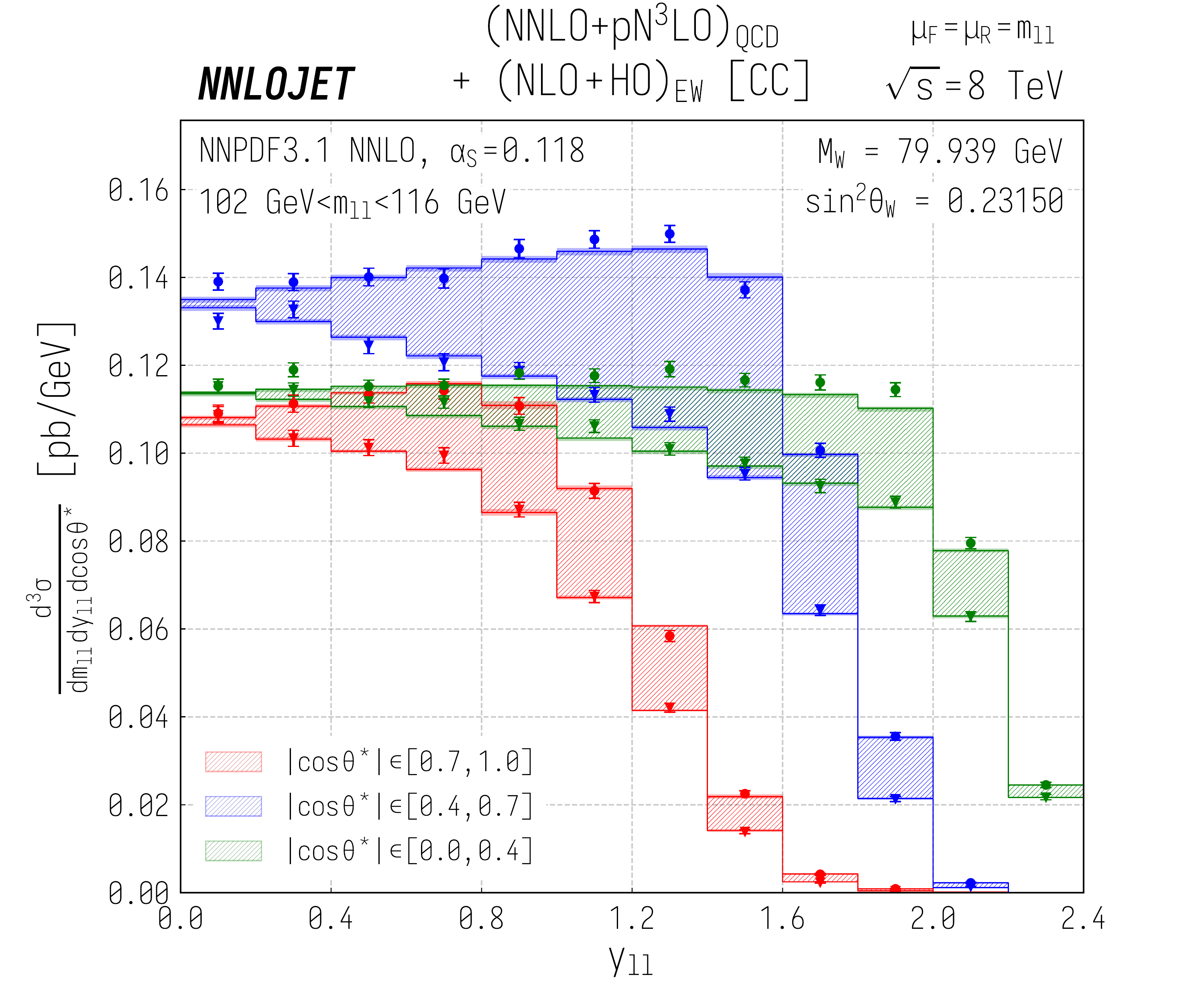}
  \includegraphics[width=0.4\textwidth]{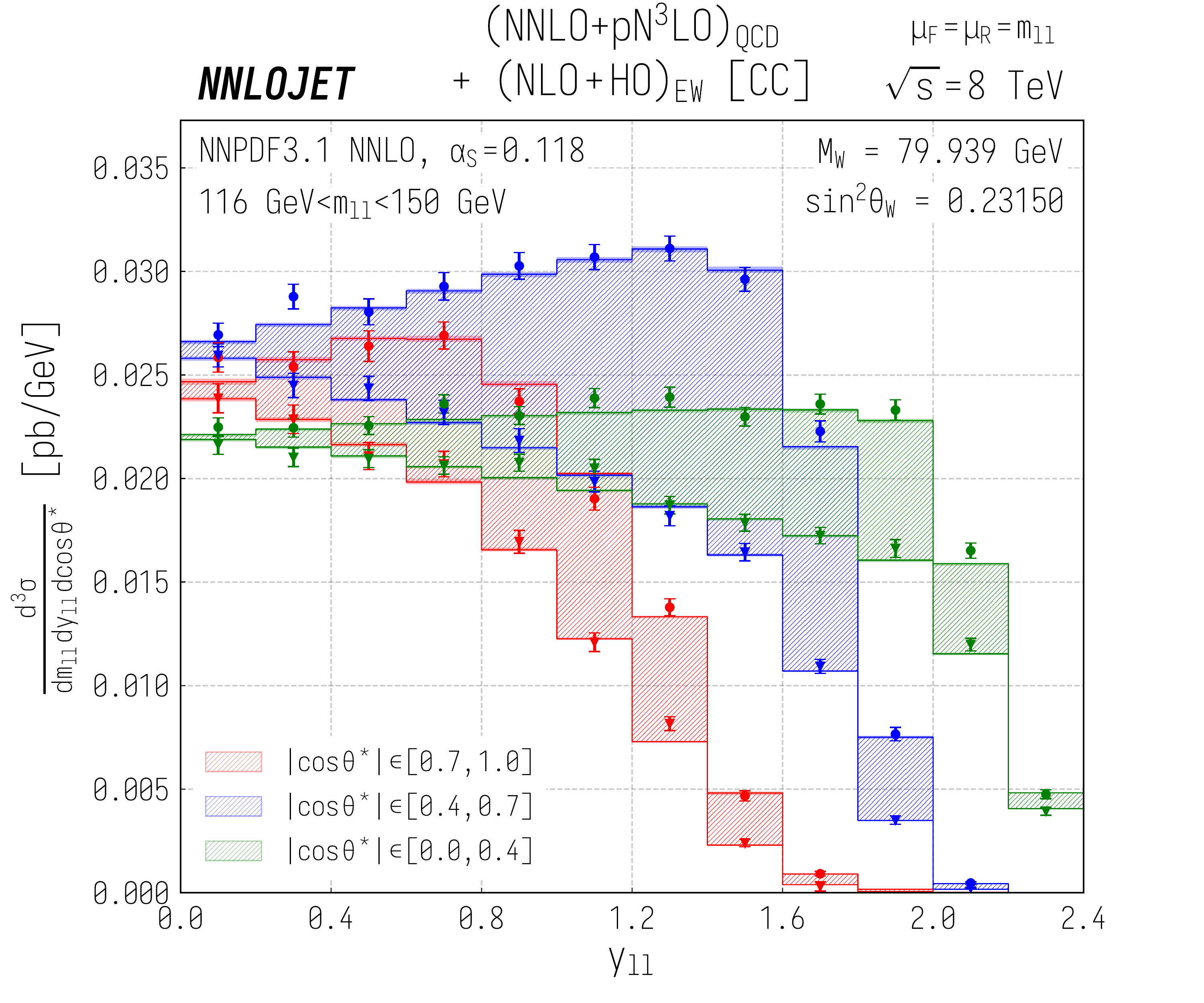} \\
  \includegraphics[width=0.4\textwidth]{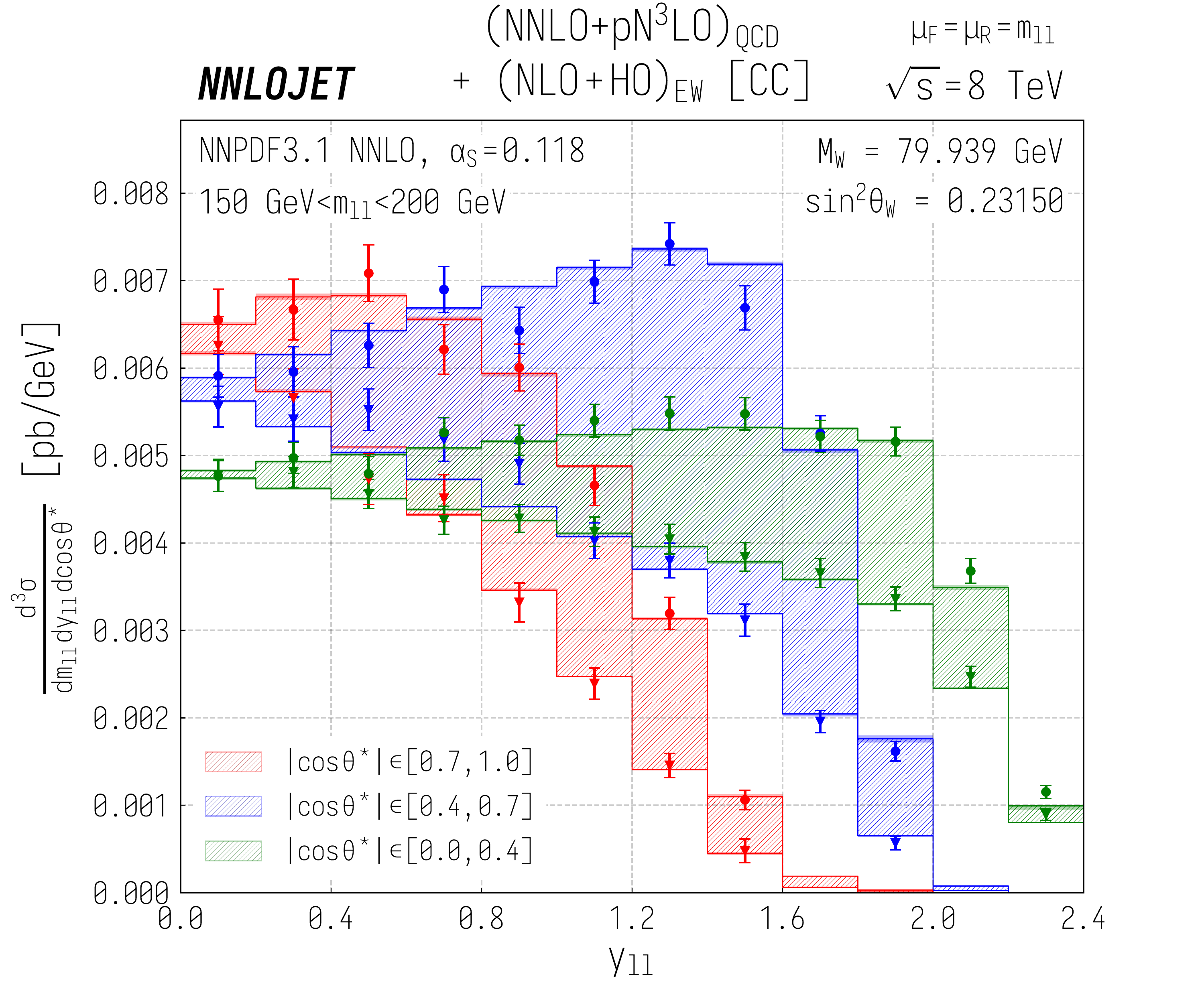}
  \caption{Triple-differential $(\NNLO + \pNthreeLO)_\QCD + (\NLO+\HO)_\EW$
    cross sections in the central-central region of the Z3D analysis in the \gmu~scheme with $\mw=79.939~\GeV$,
    corresponding to $\stw=0.23150$. The solid lines correspond to the theory predictions,
    about which the shaded band corresponds to the scale uncertainty. The markers correspond to the ATLAS
    results and associated uncertainty, and the hatched region gives the asymmetry for each of the three
    regions in $|\cthstar|$. Each panel shows a separate bin in the di-lepton invariant mass \mll.}
  \label{fig:Z3D_predictions_CC}
\end{figure}

\begin{figure}
  \centering
  \includegraphics[width=0.4\textwidth]{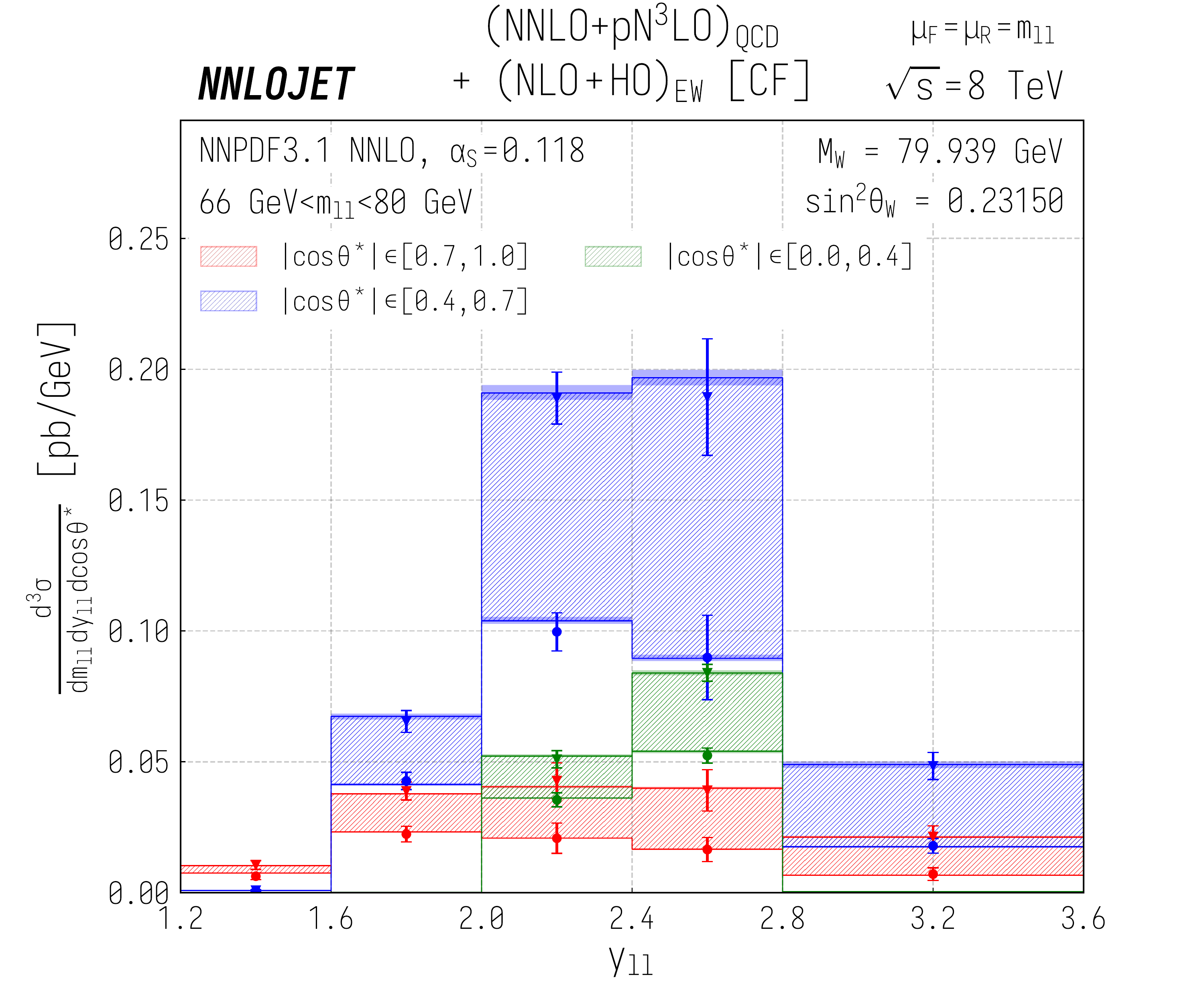}
  \includegraphics[width=0.4\textwidth]{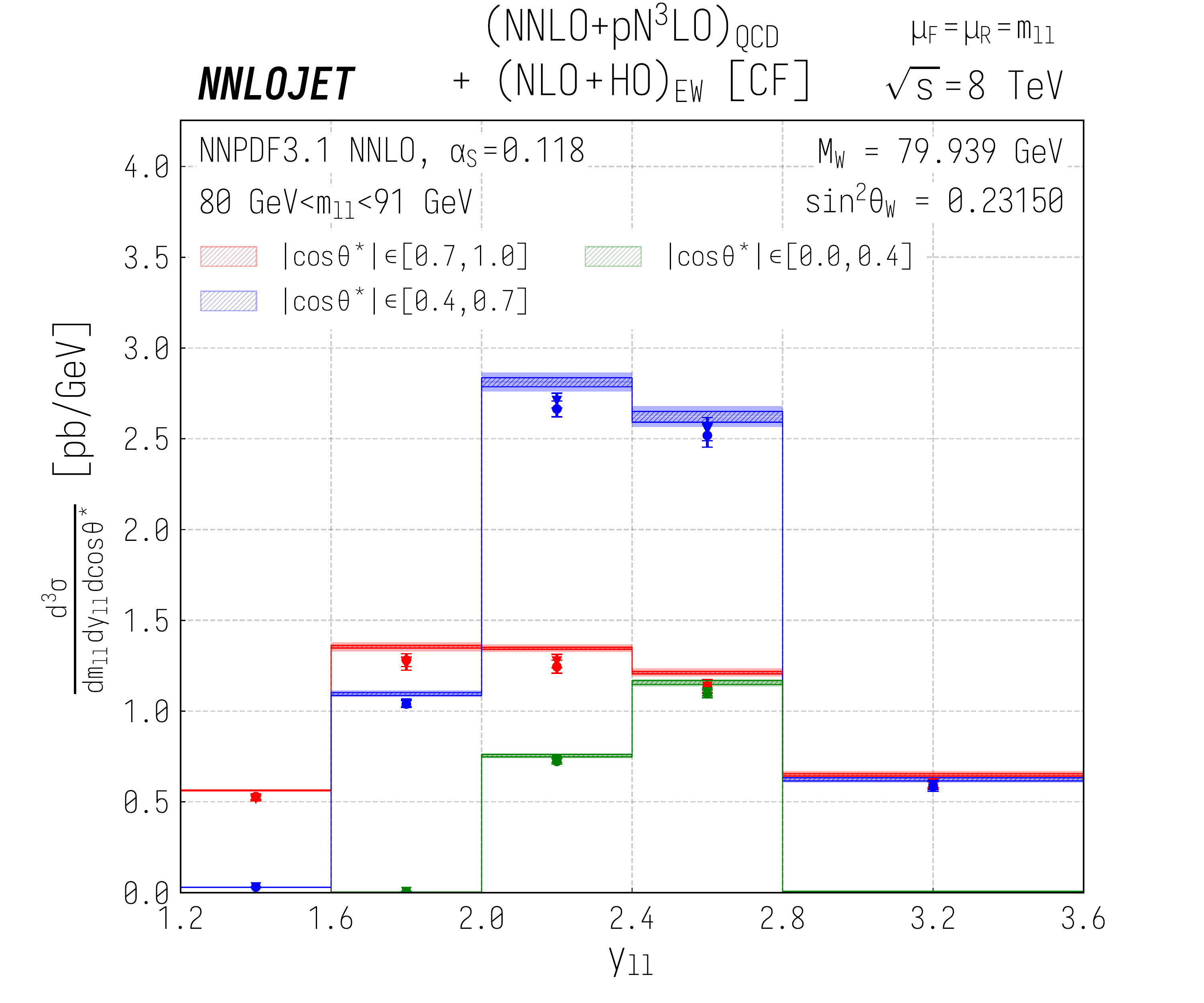}\\
  \includegraphics[width=0.4\textwidth]{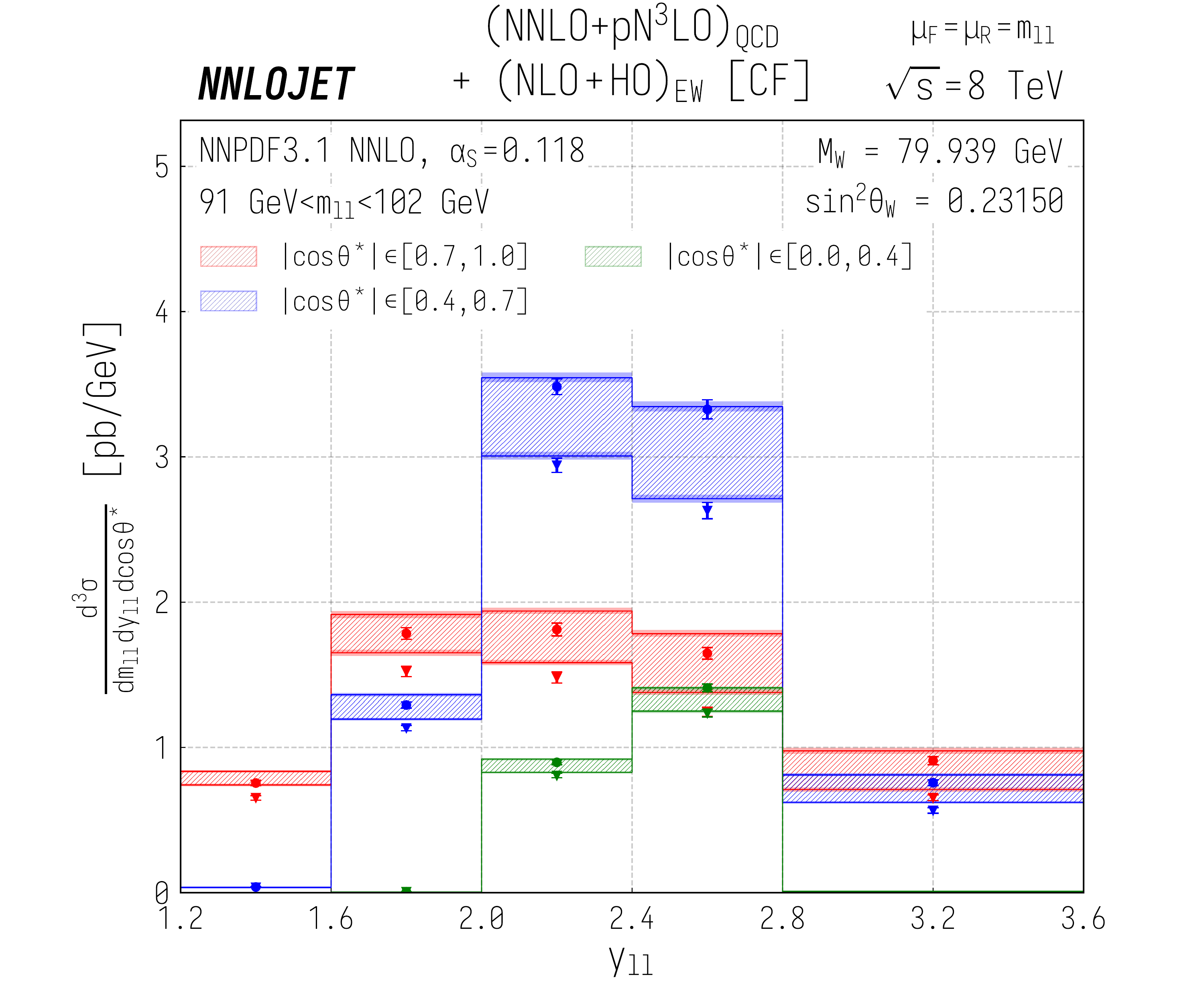}
  \includegraphics[width=0.4\textwidth]{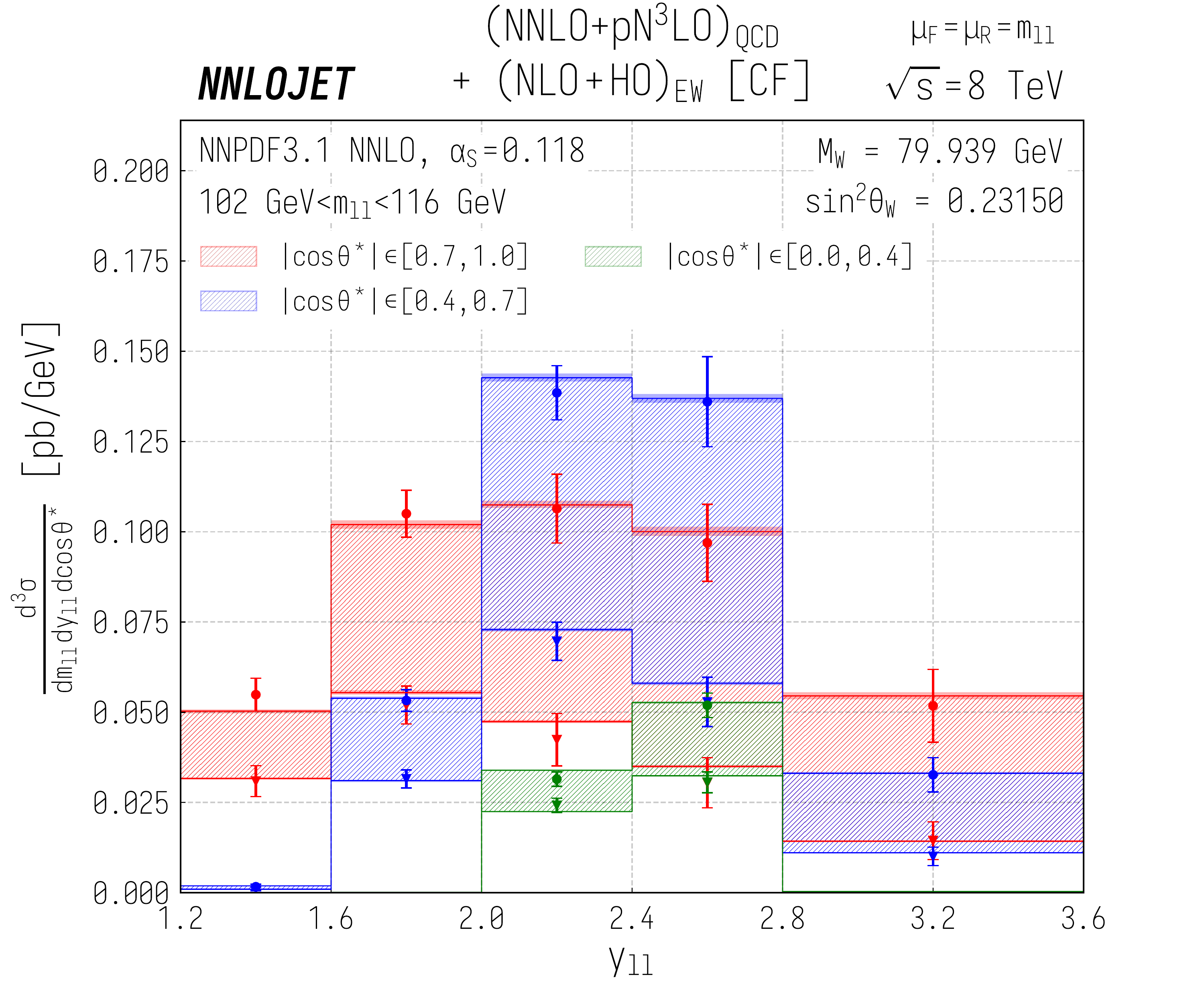}\\
  \includegraphics[width=0.4\textwidth]{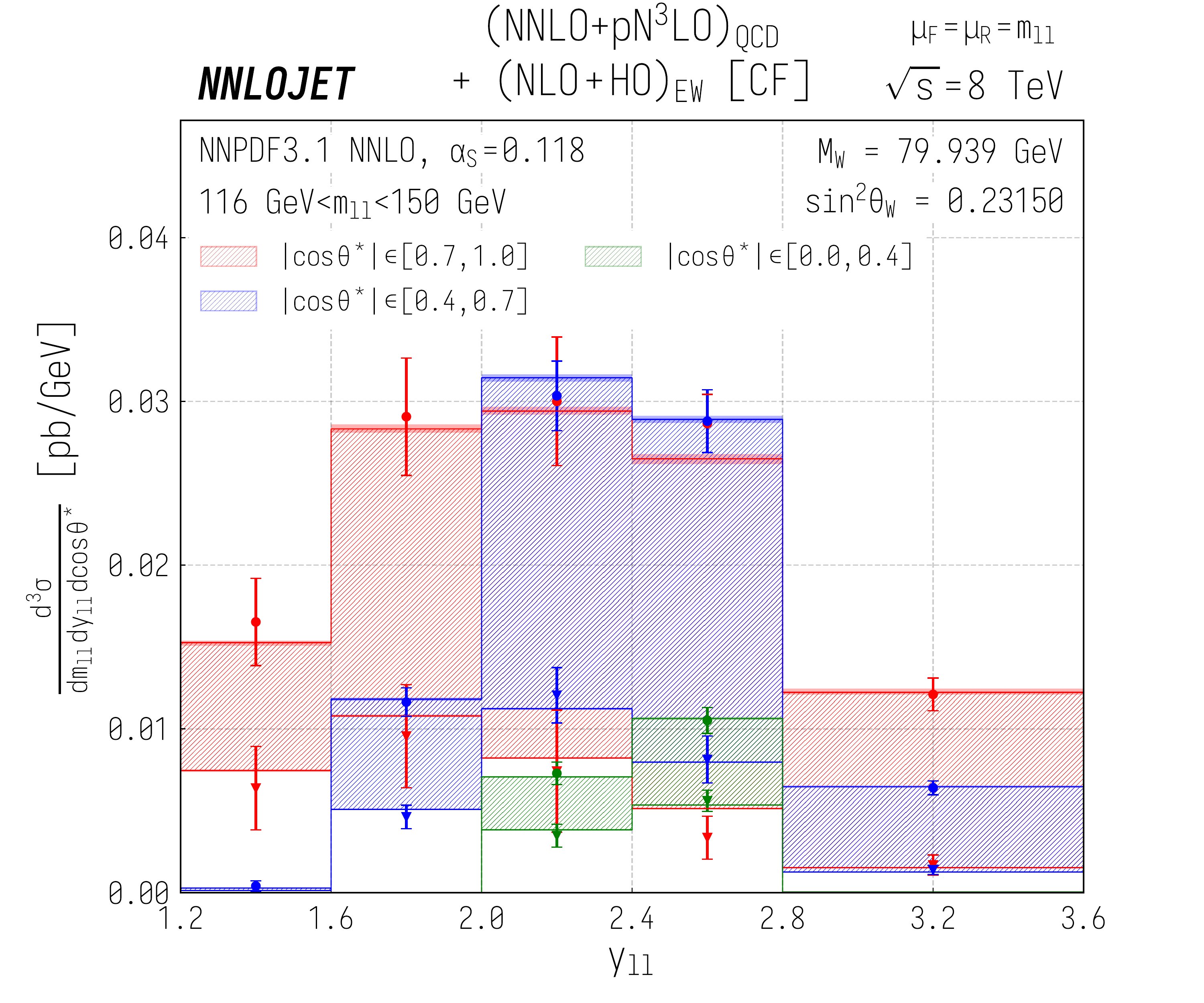}
  \caption{Triple-differential $(\NNLO + \pNthreeLO)_\QCD + (\NLO+\HO)_\EW$
    cross sections in the central-forward region of the Z3D analysis in the \gmu~scheme with
    $\mw=79.939~\GeV$, corresponding to $\stw=0.23150$. The solid lines correspond to the theory predictions,
    about which the shaded band corresponds to the scale uncertainty. The markers correspond to the ATLAS
    results and associated uncertainty, and the hatched region gives the asymmetry for each of the three
    regions in $|\cthstar|$. Each panel shows a separate bin in the di-lepton invariant mass \mll.}
  \label{fig:Z3D_predictions_CF}
\end{figure}

\subsection{Combined \NNLO + \pNthreeLO \QCD and \NLO + \HO \EW Predictions}
\label{sect:combined_predictions}
After discussing the effect of separately including higher-order \QCD and \EW corrections in the previous two subsections, we now assemble the \NNLO \QCD, partial \NthreeLO \QCD, and \NLO + \HO \EW corrections into a set of default predictions for the cross section differential in rapidity for each bin in \cthstar{} and \mll{} using the setup described in Section~\ref{sect:numerical_setup}.
The combined $(\NNLO+\pNthreeLO)_\QCD+(\NLO+\HO)_\EW$ prediction is obtained as
\begin{equation}
  \diff\sigma_{\tiny{(\NNLO+\pNthreeLO)_\QCD+(\NLO+\HO)_\EW}} = \diff\sigma^\text{\nnlojet}_{\tiny{(\NNLO+\pNthreeLO)_\QCD}} + \diff\sigma^\text{\powhegbox}_{\tiny{(\NLO+\HO)^\text{virt}_\EW}} \, ,
\end{equation}
where the second term on the right-hand side corresponds to the purely-weak virtual corrections calculated with \powhegbox, including some higher-order corrections to $\Delta\alpha$ and $\Delta\rho$.

The ratio to data for both the CC and CF regions is shown in Figure~\ref{fig:Z3D_ratio_to_data_best}, where we observe that, in the CC region, the combination of EW and QCD corrections brings the theory closer to data than the QCD predictions of Figure~\ref{fig:Z3D_ratio_to_data_CC} alone.
The remaining discrepancy lies well within the remaining luminosity uncertainty which we do not show.
The same is not true for the CF region, where we see a consistent over-shooting of the data with respect to the theory prediction.
Here, the predictions become very sensitive to the high-$x$ valence quark distribution within the PDFs.

In Figure~\ref{fig:Z3D_predictions_CC}, we compare the combined $(\NNLO+\pNthreeLO)_\QCD+(\NLO+\HO)_\EW$ triple-differential cross sections for the central-central region to ATLAS data.
There, solid lines correspond to the theory predictions, about which the shaded band corresponds to the scale uncertainty.
Markers correspond to ATLAS results with associated uncertainty and the hatched region displays the asymmetry for each of the three regions in $|\cthstar|$.
The difference between the $\cthstar>0$ and $\cthstar<0$ contributions which corresponds to the asymmetry \afb, highlighted by the hatched area, is indeed much smaller around the \PZ~peak than in the extremal \mll~regions.
For each plot, the forbidden and mixed bins lie towards the far right of the \yll{} distribution, with the final four rapidity bins for $0.7<|\cthstar|<1$ (red) and final two rapidity bins for $0.4<|\cthstar|<0.7$ (blue) being forbidden and therefore supplemented with partial \NthreeLO corrections.

Corresponding triple-differential cross sections for the central-forward region are shown in Figure~\ref{fig:Z3D_predictions_CF}, where we observe the same asymmetry pattern with \mll.
Here, the \pNthreeLO-enhanced bins are the two left-most rapidity bins for $0<|\cthstar|<0.4$ (green) and the left-most rapidity bin for $0.4<|\cthstar|<0.7$ (blue).

\section{Summary}
\label{sect:summary}
In this study we have given a detailed overview of Standard-Model theory predictions relevant to the extraction of the effective Weinberg angle \stweff{} using $8~\TeV$ triple-differential Drell-Yan data from the ATLAS collaboration~\cite{Aaboud:2017ffb}.
The intricate kinematics of this process induce a non-trivial interplay between the di-lepton variables used in the triple-differential measurement and the fiducial event-selection cuts applied to each lepton.
This interplay leads to (partially) forbidden kinematical regions at Born-level, and directly influences the higher-order QCD corrections in terms of acceptances and $k$-factors.
It allows for the extension of the theory input to \NthreeLO in regions of phase space which are forbidden at Born level.

We discussed EW scheme considerations for combined $\QCD+\EW$ results, and confronted ATLAS data with $\NNLO + \text{partial }\NthreeLO$ \QCD predictions from \nnlojet in combination with $\NLO + \text{partial higher-order }\EW$ predictions obtained with a modified version of \powhegbox.
These constitute the main theoretical inputs to future \stweff{} fits.

We demonstrated that partial \NthreeLO corrections are necessary to obtain precise results in the so-called forbidden bins, which obtain no contribution from Born-level phase-space points, while \NLO, and in particular partial higher-order \EW corrections, are needed for accurate theoretical predictions around the \PZ~pole.
We have presented a prescription for interpreting the measurement with minimal sensitivity to missing QCD higher-order effects, through a separation of the data into regions of high fiducial acceptance, which are included as cross sections, and regions of low-acceptance, which are included as differential forward-backward asymmetries.
We have also highlighted the discriminating power of the triple-differential data in terms of PDFs, providing important information on valence-quark PDFs and a substantial reduction of PDF uncertainties in \stweff~fits using data from hadron colliders.

The results presented here form a subset of those provided to the ATLAS collaboration for use in a fit of \stweff{} to Z3D data.
Further results not shown here include larger variations of \stweff~for closure tests alongside with results for different values of \alphas{} and central scale choices.
It is anticipated that advances in \NNLO grid technology~\cite{Britzger:2012bs,Britzger:2019kkb} will also allow for a full \NNLO evaluation of PDF uncertainties in the near future.

\acknowledgments
The authors thank Xuan Chen, Juan Cruz-Martinez, James Currie, Marius H\"ofer, Imre Majer, Tom Morgan, Jan Niehues, Joao Pires, and James Whitehead for useful discussions and their many contributions to vector-boson processes in the \NNLOJET code.
The authors are also grateful to Simone Amoroso, Stefano Camarda, Sasha Glazov, Richard Keeler, Tony Kwan, Elzbieta Richter-Was, Eram Rizvi, Andrey Sapronov, and Pavel Shvydkin for many helpful discussions on the ATLAS Z3D measurement.
This research was supported in part by the UK Science and Technology Facilities Council under contract ST/T001011/1, by the Swiss National Science Foundation (SNF) under contracts 200021-197130 and 200020-204200, and from the European Research Council (ERC) under the European Union's Horizon 2020 research and innovation programme grant agreement 101019620 (ERC Advanced Grant TOPUP).
Numerical simulations were facilitated by the High Performance Computing group at ETH Z\"urich and the Swiss National Supercomputing Centre (CSCS) under project ID ETH5f.

\bibliography{Z3D}

\end{document}